\documentclass[12pt]{article}
\usepackage[top=1in,left=1in,bottom=1in,right=1in, paperheight=11in, paperwidth=8.5in]{geometry}

\usepackage{titlesec}
\titleformat{\section}
  {\normalfont\fontsize{14}{3}\bfseries}{\thesection}{1em}{}
\titleformat{\subsection}
  {\normalfont\fontsize{12}{13}\bfseries}{\thesubsection}{1em}{}  
\usepackage{amsmath, bigints, amssymb, amsthm}
\usepackage[mathscr]{euscript}
\usepackage[makeroom]{cancel}
\usepackage{graphicx}
\usepackage{subcaption}
\usepackage{enumerate, verbatim}
\usepackage[font={small,it,singlespacing}]{caption}
\usepackage{lscape}
\usepackage{caption}
\usepackage{hyperref}
\usepackage{cleveref}
\usepackage{natbib2}
\usepackage{parskip}
\usepackage{array}
\usepackage{multirow}
\usepackage{graphicx}
\usepackage{wrapfig}
\usepackage{lscape}
\usepackage{rotating}
\usepackage{epstopdf}
\usepackage[usenames, dvipsnames]{color}
\usepackage{boldline}
\usepackage{url}
\usepackage{filecontents}

\hypersetup{colorlinks,citecolor=blue}
\graphicspath{{./images//}}

\newcommand{\bs}{\boldsymbol}
\newcommand{\R}{\mathcal{R}}
\newcommand{\T}{\mathcal{T}}
\newcommand{\x}{\mathbf{x}}
\newcommand{\z}{\mathbf{z}}
\newcommand{\w}{\mathbf{w}}
\newcommand{\s}{\mathbf{s}}
\newcommand{\e}{\mathbf{e}}

\newcolumntype{L}[1]{>{\raggedright\let\newline\\\arraybackslash\hspace{0pt}}m{#1}}
\newcolumntype{C}[1]{>{\centering\let\newline\\\arraybackslash\hspace{0pt}}m{#1}}
\newcolumntype{R}[1]{>{\raggedleft\let\newline\\\arraybackslash\hspace{0pt}}m{#1}}

\newtheoremstyle{egstyle}
  {6pt} 
  {3pt} 
  {} 
  {} 
  {\bfseries} 
  {.} 
  {.5em} 
  {} 
\theoremstyle{egstyle} \newtheorem{thm}{Theorem}
\theoremstyle{egstyle} \newtheorem{defn}[thm]{Definition}
\theoremstyle{egstyle} 

\begin{document}
\title{\vspace{-0.7in}\normalsize{\textbf{Comparative Study of Differentially Private Data Synthesis Methods}}\vspace{-.8cm}}
\author{\small{\textbf{Claire McKay Bowen$^*$ and Fang Liu$^{\ddag}$}}
\footnote{\noindent Claire McKay Bowen is a postdoctoral research associate at Los Alamos National Laboratory, and Fang Liu is an Associate Professor in the Department of Applied and Computational Mathematics and Statistics, University of Notre Dame, Notre Dame, IN 46556 ($^{\ddag}$E-mail: fang.liu.131@nd.edu). Claire McKay Bowen was supported by the National Science Foundation (NSF) Graduate Research Fellowship under Grant No. DGE-1313583 during part of the development of this paper. Fang Liu is supported by the NSF Grants  \#1546373, \#1717417, and the University of Notre Dame Faculty Research Support Initiation Grant Program.
\newline The publication has been assigned the Los Alamos National Laboratory identifier LA-UR-18-31132.
\newline \textbf{An earlier version of this paper won the Best Student Paper Competition in the 2017 American Statistical Association Survey Research Methods Section (SRMS), Government Statistics  Section (GSS), and Social Statistics Section (SSS).}}
\vspace{-3cm}
}
\date{}
\maketitle{}
\begin{abstract}
\noindent When sharing data among researchers or releasing data for public use, there is a risk of exposing sensitive information of individuals in the data set.  Data synthesis is a statistical disclosure limitation technique for releasing synthetic data sets with pseudo individual records. Traditional data synthesis techniques often rely on strong assumptions of a data intruder's behaviors and background knowledge to assess disclosure risk. Differential privacy (DP) formulates a theoretical approach for a strong and robust privacy guarantee in data release without having to model intruders' behaviors. Efforts have been made aiming to incorporate the DP concept in the data synthesis process. In this paper, we examine current \textbf{DI}fferentially \textbf{P}rivate Data \textbf{S}ynthesis (DIPS) techniques for releasing individual-level surrogate data for the original data, compare the techniques conceptually, and evaluate the statistical utility and inferential properties of the synthetic data via each DIPS technique through extensive simulation studies. Our work sheds light on the practical feasibility and utility of the various DIPS approaches, and suggests future research directions for DIPS.

\noindent \textit{\textbf{keywords}}: differential privacy, DIPS, sufficient statistics,  parametric DIPS, non-parametric DIPS, statistical disclosure limitation
\end{abstract}

\newpage
\section{Introduction}\label{sec:intro}
When sharing data among collaborators or releasing data publicly, a big concern is the risk of exposing the identification and personal information of the individuals who contribute to the data. Even with key identifiers removed, a data intruder may still identify an individual in a data set via linkage with other public information. Some notable examples on individual identification breach in publicly released or restricted access data include the Netflix prize \citep{netflix2},  the genotype and HapMap linkage effort \citep{genotype},  the AOL search log release \citep{AOL}, and the Washington State health record identification \citep{sweeney2013matching}.

Statistical approaches to protecting data privacy are referred to as statistical disclosure limitation. These techniques aim to provide protection for sensitive information while releasing information and data to the public.  Data synthesis is a statistical disclosure limitation technique that focuses on releasing  individual-level data synthesized based on the information in the original data \citep{rubin1993discussion, little1993statistical, liu2003,raghunathan2003multiple, reiter2003,  liu2004, reiter2009, drechsler2011book}. Multiple  synthetic  sets of the identical structure are often released as a way to propagate the uncertainty arising from the synthesis process, a procedure  referred to as multiple synthesis (MS). Methods have been developed to combine the results from multiple synthetic data sets to will yield valid statistical inferences \citep{raghunathan2003multiple,reiter2002, reiter2003}. However, existing disclosure risk assessment approaches for statistical disclosure limitation techniques often depend on the specific values in a given data set as well as various assumptions about the  background knowledge and behaviors of data intruders \citep{reiter2005estimating,hundepool2012statistical, manrique2012estimating}. In some cases, only heuristic arguments are employed without numerical assessment of disclosure risk.

Differential privacy (DP), a concept popularized in the theoretical computer science community,  provides strong privacy guarantee in mathematical terms without making  assumptions about the background knowledge of data intruders \citep{dwork2006calibrating, dwork2008, dwork2011differential}. In brief, if a statistic is released via a $\epsilon$-differentially private mechanism, then,  when the statistic is calculated from two neighboring data sets that differ by one record, the log-difference on the probability to obtain a specific value of that statistic is bounded between $(-\epsilon,\epsilon)$. In layman's terms, DP means the chance an individual will be identified based on the sanitized statistic is  low (the smaller $\epsilon$ is, the lower the probability is) since the statistic would be about the  same  with or without the individual in the database.

DP has spurred a great amount work in developing differentially private mechanisms in general settings  \citep{dwork2006calibrating, mcsherry2007mechanism, mcsherry2009privacy, nissim2015generalization} as well as for specific statistical analysis such as data mining \citep{mohammed2011differentially}, shrinkage regression \citep{Chaudhuri2011, Kifer2012}, principle component analysis \citep{chaudhuri2012near}, genetic association tests \citep{Yu2014}, Bayesian learning \citep{wang2015privacy}, location privacy \citep{xiao2015protecting}, recommender systems \citep{friedman2016differential}, deep learning \citep{abadi2016deep}, among others. Software or web-based interfaces to generate differentially private statistics are also in development, such as RescueDP \citep{wang2016rescuedp}, an online aggregate monitoring scheme that publishes real-time population statistics on spatial-temporal, crowd-sourced data from mobile phone users with DP, and Private data Sharing Interface \citep{gaboardi2016psi}  that aims to allow data sharing among researchers in the social sciences and other fields while satisfying DP.

DP was originally developed and is widely used for releasing aggregate or summary statistics to answering queries submitted to a database. However, query-based data release has several shortcomings. The requirement to pre-specify the level of privacy budget $\epsilon$ often dictates the number and the types of future queries. The curator of a database will refuse to answer any further queries if the prespecified privacy budget is exhausted from answering all previous queries. Additionally, data users would prefer to directly access the individual-level data to perform statistical analysis on their own.

Efforts have also been made to release differentially private individual-level data, which we will refer to as DIPS (\textbf{DI}fferentially \textbf{P}rivate Data \textbf{S}ynthesis). \citet{barak2007privacy} generate synthetic data via the Fourier transformation and linear programming  in low-order contingency tables.  \citet{blum2008learning} discuss differentially private data synthesis from the perspective of  the learning theory.  \citet{abowd2008protective} propose an approach to synthesize differentially private tabular data from the predictive posterior distributions of frequencies, which was applied in the simulations studies in \citet{Charest2010} to explore inferences on proportions from synthesized binary data. \citet{mcclure2012differential} implement a similar technique for synthesizing binary data with a different specification of the differentially private prior. \citet{wasserman2010statistical} propose several paradigms to sample from appropriately differentially private perturbed histograms or empirical distribution functions. They also examined the rate that the probability of empirical distribution of the synthetic data converges to the true distribution of the original data. \citet{zhang2014privbayes} propose PrivBayes to release high-dimensional data from Bayesian networks with binary nodes and low-order interactions among the nodes. \citet{li2014differentially} develope DPCopula for synthesizing multivariate data by using Copula functions to take into account the dependency structure. \citet{liu2016model} proposes a Bayesian technique, model-based DIPS (MODIPS), to release differentially private synthetic data, and explored the inferential properties of the released data. Besides these generic DIPS approaches, there are also DIPS developed for specific type of data such as graphs \citep{proserpio2012workflow},  and mobility data from GPS trajectories \citep{chen2013privacy, he2015}.

The goals of this paper are two-fold. First, it introduces the powerful concept of DP to the statistical community and surveys the current development in DIPS. Second, this paper examines and compares some of the general DIPS approaches based on the statistical and inferential utility of the respective synthesized data; both conceptually and empirically via simulation studies and a  real-life case study.  We aim to, through this comparative examination of different DIPS approaches, demonstrate the useful applications of DP in releasing synthetic data with guaranteed privacy and to provide some guidance on the feasibility of the DIPS methods for practical use.

The remainder of the paper is organized as follows. Section \ref{sec:concept} overviews the basic concepts of DP and some common differentially private mechanisms. Section \ref{sec:methods} presents some currently available DIPS approaches. Section \ref{sec:simulation} compares and examines the utility and inferential properties of the individual-level surrogate data released from some of the DIPS methods introduced in Section \ref{sec:methods} via four simulation studies on different types of data. Section \ref{sec:case} compares some DIPS methods to test the practical feasibility of DIPS on real-world data. Concluding remarks are given in Section \ref{sec:discussion}.

\section{Concepts}\label{sec:concept}
The concepts of DP and the sanitization algorithms were developed originally for releasing results of queries sent to a database. We rephrase the main concepts in DP below in terms of statistics. There is essentially no difference between query results and statistics given that both are functions of data.
Denote the target data for protection by $\x=\{x_{ij}\}$ of dimension ${n\times p}$ ($i=1,\ldots,n;j=1,\ldots,p$). Each row  $\x_i$ represents an individual record with $p$ variables/attributes. 

\subsection{Differential Privacy (DP) and Composition Properties}\label{sec:dp}
\begin{defn}\label{def:dp} \textbf{Differential Privacy} \citep{dwork2006calibrating}:
A sanitization algorithm $\R$ gives $\epsilon$-DP if for all data sets $(\x,\x')$ that is $d(\x,\x')=1$, and all  subsets $Q\subseteq \T$
\begin{equation}\label{eqn:dp}
\left|\log\left(\frac{\Pr(\R( \s(\x)) \in Q)}{\Pr(\R( \s(\x'))\in Q)} \right)\right|\le\epsilon,
\end{equation}
\end{defn}
\noindent where $\T$ denotes the output range of the algorithm $\R$, $\epsilon>0$ is the privacy budget (or loss), and $\s$ denotes the statistics. $d(\x,\x')=1$ implies that $\x'$ differs from $\x$ by only one individual. Mathematically,  Eqn (\ref{eqn:dp}) states that the log-difference on the probability of obtaining a specific value of $\s$ via $\R$ is bounded by $(-\epsilon,\epsilon)$ when it is calculated from two neighboring data sets that differ by one record. If $\epsilon$ is small, then the chance an individual will be identified based on the sanitized query result is low since the query results would be about the  same  with or without the individual in the database. The larger $\epsilon$ is, the sanitized results are more differentiable with vs without an individual in the data set.

Regarding what  value of $\epsilon$ is considered to be appropriate or acceptable for practical use, \citet{dwork2008survey} states the choice of $\epsilon$ is a social question (and ``beyond the scope of'' her paper), but suggests $0.01\sim \ln(3)$, or even up to 3 in some cases, as possible  $\epsilon$ values.  \citet{Clifton2011} state that $\epsilon$ does not easily relate to practically relevant measures of privacy and suggest a formula to calculate $\epsilon$ if the goal is to hide any individual’s presence (or absence) in the database. The formula relies on some assumptions, like query-dependency and also requires knowing the data universe as well as the subset of that universe to be queried. \citet{abowd2015revisiting} examine the question from the economic perspective by accounting for the public-good properties of privacy loss and data utility, and quantify the optimal choice  of $\epsilon$  by formulating a social planner's problem and incorporating $(\epsilon, \delta$)-DP (another relaxation of the pure DP; see Definition  \ref{def:adp}) and $(\alpha,\delta)$-accuracy (the $l_1$ error in released statistics is bounded by $\alpha$ with probability $1-\beta$) to release normalized histograms via the private multiplicative weights method. In two applications, they have examined the optimal $\epsilon$ is 0.067 and 0.044 for $\alpha=0.096$ and $0.073$ (respectively) when  $\beta=0.01$ and $\delta=0.9/N$, where $N$ is the population size for some specific settings on the population size and query set size. If deemed valid, the $\epsilon$ values suggested in the $(\epsilon, \delta)$-DP setting can also be considered for $\epsilon$-DP as the latter implies stricter privacy protection at the same $\epsilon$ value. However, the caveat of possible worse information preservative compared to its relaxed counterpart. Other $\epsilon$ values also came up in the literature.  For example, \citet{machanavajjhala2008privacy} apply DP in the OnTheMap data (commuting patterns of the US population) and used $(\epsilon=8.6,\delta=10^{-5})$-probabilistic DP (a relaxation of the pure DP in Eqn \ref{eqn:dp}; see Definition  \ref{def:pdp}) to synthesize commuter data. \citet{JRSSC} use $\epsilon=3$ and $\epsilon=6$ when synthesizing edges in social networks via a randomized response mechanism with $\epsilon$-edge DP. 
\citet{DPcube} and \citet{DPcopula}  use $\epsilon=1$ in the experiments.

All the work above suggests there are many factors that affect the choice of $\epsilon$, including the type of information released to the public, social perception of privacy protection, statistical accuracy of the release data, among others. Also, that it remains an open question that warrants more research and further investigation. The smaller $\epsilon$ is, the less the privacy loss, but the less accurate the released information. Choosing an ``appropriate'' $\epsilon$  is essentially finding a good trade-off between privacy loss and released information accuracy. We will provide more discussion regarding the choice of $\epsilon$ in Section \ref{sec:discussion}, reviewing what we have learned from the literature and the simulation/case studies.


Often is a data set queried for multiple statistics especially when the data is high-dimensional. Every time the data set is queried, there is a privacy cost (loss) as information is being asked about the same set of individuals. Therefore, the data curator must track all queries and analysis conducted on a data set to ensure the overall privacy spending does not exceed the pre-specified level. Say $r$ queries are sent to the same data set with a total privacy budget of $\epsilon$. The data curator could allocate  $\epsilon/r$  privacy budget to each of the $r$ queries to maintain the total privacy cost at $\epsilon$.   On the other hand, if each query is sent to a disjoint set of data where each set has no overlapping individuals, then the privacy cost does not accumulate.  A typical example is the release of a histogram, where the counts in different bins of the histogram are based on disjoint subsets of data, and each bin is perturbed with the full privacy budget $\epsilon$. These principles are presented in the sequential composition and parallel  composition theorems below.
\begin{thm}\label{thm:comp}
  \textbf{Composition Theorems} \citep{mcsherry2009privacy}: Suppose a differentially private mechanism $\R_j$ provides $\epsilon_j$-DP for $j=1,\ldots,r$.
  \begin{itemize}\setlength{\itemindent}{-10pt}
  \item[a)] \emph{Sequential Composition}: The sequence of $\R_j(\x)$  executed on the same data set $\x$ provides $(\sum_j\epsilon_j)$-DP.
  \item[b)] \emph{Parallel Composition}: Let  $D_j$ be disjoint subsets of the input domain $D$. The sequence of $\R_j(\x\cap D_j)$ provides $\max(\epsilon_j)$-DP.
  \end{itemize}
\end{thm}

\subsection{Relationship between DP and Disclosure Risk Assessment in the Traditional Statistical Disclosure Limitation Setting}
The concept of DP is different from the traditional disclosure risk assessment in statistical disclosure limitation. The former does not rely on any background knowledge or behavioral assumptions of a data intruder while the latter often models what's the data intruder knows and how the disclosure risk is formulated or calculated and could vary significantly, depending on the data and the approaches for assessing disclosure risk, lacking a unified principle. We now illustrate the differences between DP and the traditional disclosure risk assessment using a concrete example.

Suppose a data set contains 11 attributes, one out of the 11 is a sensitive variable, such as HIV status, and the other 10 are pseudo-identifiers such as age, gender, etc. In a typical disclosure risk assessment, the data curator would first make an assumption about what the intruder knows and what the intruder will do to obtain the information she/he is interested in. Therefore, the curator will likely  assume  in this case that: 1) the intruder A wants the information on the sensitive variable on individual B, and A knows that B is in the data set; 2) A knows all 10 pseudo-identifiers of B; 3) A fits a logistic model to calculate the probability of having HIV with the released data set. Suppose the true HIV status of B is T and estimated Pr(HIV=T$|$the 10 attributes) is 5\% from the logistic model based on a synthetic copy of the original data; then from the perspective of the traditional disclosure risk assessment, we would consider B is at a lower risk of getting his/her personal information disclosed. However, how confident are we with this 5\%? What if the data intruder has more information in addition to the released data? What if the data intruder has a more efficient method than the logistic regression to predict the HIV status with high accurate? In other words, the single value 5\% with all the above assumptions could be far from being optimal in reflecting the true disclosure risk.

If the surrogate data set is synthesized via a technique based on the DP framework, then it is guaranteed that any individual (including B) from  the original data has little impact on any statistics calculated from the synthetic data, and the impact is quantified by the probabilities of obtaining the same statistic with vs.\ without any single individual, the ratio between which is bounded by $(e^{-\epsilon}, e^{\epsilon})$. In this example, the statistic $s$ is Pr(HIV$|$the 10 attributes), and the the ratio $\frac{\Pr(s^*|\mathbf{x}^*)}{\Pr(s^*|\mathbf{x'}^*)}\in (e^{-\epsilon}, e^{\epsilon})$, where data $\mathbf{x}^*$ and $\mathbf{x'}^*$ differ by one individual and $s^*$ is the sanitized version of the observed original $s$ based on the synthetic data. Using a small $\epsilon$  leads to a tight neighborhood $(e^{-\epsilon}, e^{\epsilon})$ around 1, and a small privacy loss.

A reviewer asks if DP can be used as an upper bound for disclosure risk assessment. The above example suggests the way the DP bounds the absolute log-ratio of two distribution functions on the sanitized version of $s$ obtained from two neighboring data sets ($\mathbf{x}$ and $\mathbf{x}'$) by $\epsilon$ rather than providing a direct measure on the probability that an individual would be identified or have his/her true value on a sensitive variable disclosed.  \citet{Clifton2011} calculate an upper bound for the posterior probability of a correct guess from an adversary on whether an individual is in a data set given discrete query results sanitized  via the Laplace mechanism under some assumptions. However, the bound is not tight.

In summary, the link between DP and  the traditional disclosure risk assessment is an interesting  topic and open question.  One thing for sure is that DP integrates out all the unknowns (e.g., whether and how data intruder would use that data set, whether an individual in a particular data set or will  participate in any future studies, etc) and covers the worst-case scenario, whether the data curator can think of or not, from the perspective of protecting every individual.

\subsection{Empirical DP and Local DP}\label{sec:local}
Classical DP has inspired other privacy concepts such as the  empirical DP \citep{abowd2013differential} and local DP \citep{Duchi2013}, both of which look for bounding some type of ``privacy'' using a single parameter $\epsilon$.  We will not examine the two concepts further in this discussion for the reasons given below.

Empirical DP was first proposed for privacy protection in Bayesian mixed-effects modeling. In empirical DP, a prior distribution is designed to guarantee that the log difference on the posterior distribution of a parameter with vs. without each of the individuals in the original data is bounded by $(-\epsilon,\epsilon)$. \citet{charest2017meaning} showed that empirical DP is more of an empirical measurement of sensitivity, and relates to the so-called ``local sensitivity'' \citep{localsensitivity} rather than a guarantee or an empirical estimate of DP. In addition, empirical DP is computationally sensitive to how many posterior samples are drawn and how they are binned in its numerical calculation as the analytical form of the posterior distribution is often not available.

For local DP, though its mathematical formulation seems to similar to the classical DP, the two are conceptually different. Local DP focuses on how individual data is collected. Specifically, the true response of the individual goes through a locally $\epsilon$-differentially private randomization mechanism that generates a perturbed response, which is recorded and released. Different from the traditional DP (where the privacy budget $\epsilon$ is possessed by a whole data set) each individual receives a privacy budget $\epsilon$ in local DP, and the log-difference in the probability of generating the same perturbed response from two different individual responses is bounded by $(-\epsilon, \epsilon)$.  Local DP has been applied in practice to collect users' data \citep{erlingsson2014rappor, fanti2016building,tang2017privacy}; but given its conceptual difference from the classical DP, we leave its in-depth investigation and exploration for future research (more discussion on the local DP is provided in Section \ref{sec:discussion}).  

\subsection{Differentially Private Mechanisms}
We introduce two commonly used sanitizers to achieve $\epsilon$-DP: the Laplace mechanism and the Exponential mechanism. A key concept in the Laplace mechanism is the  global sensitivity of $\s$ \citep{dwork2006calibrating}, defined as the following: For all $(\x,\x')$ that is $d(\x,\x')=1$, the global sensitivity of statistics $\s$ is $\Delta_{\s}=\underset{\scriptstyle{\x,\x', d(\x,\x')=1}}{\mbox{max}} \|\s(\x)-\s(\x')\|_1$.
In layman's terms, $\Delta_{\s}$  is the maximum change in terms of $l_1$ norm a person would expect in $\s$ across all possible configurations of $(\x,\x')$ and $d(\x,\x')=1$. The sensitivity is ``global" since it is defined for all possible data sets and all possible ways that two data sets differ by one observation.  The higher $\Delta_{\s}$ is the more disclosure risk there will be on the individuals in the data from releasing the original $\s$.
\begin{defn}\label{def:lap}
  \textbf{Laplace Mechanism} \citep{dwork2006calibrating}:
  The Laplace mechanism of $\epsilon$-DP adds independent noises $\e$ from the Laplace distribution with location parameter 0 and scale parameter $\Delta_{\s}\epsilon^{-1}$ to  each of the elements of the original result $\s$ to generate perturbed result $\s^{\ast}= \s + \e$.
\end{defn}
\noindent By the Laplace distribution, values closer to the raw results $\s$ have higher probabilities of being released than those that are further away from $\s$. The variance of the Laplace distribution is  $2\left(\Delta_{\s}\epsilon^{-1}\right)^2$, implying the smaller the privacy budget $\epsilon$ and/or the larger the $\Delta_{\s}$, the higher the probability that the perturbed result $\s^*$ will be farther way from $\s$ when released. The Laplace mechanism is a quick and simple DP mechanism, but does not apply to all statistics such as statistics that have non-numerical outputs. \citet{mcsherry2007mechanism} introduces a more general DP mechanism, the Exponential mechanism, that applies to all types of queries.

\begin{defn}\label{def:exp}
  \textbf{Exponential Mechanism} \citep{mcsherry2007mechanism}:
  In the Exponential mechanism, a utility function $u$ assigns a score to each possible output $\s^{\ast}$ and  releases $\s^{\ast}$ with probability
  \begin{equation}\label{eqn:exp}
  \frac{\exp\left(u(\s^{\ast}|\x)\frac{\epsilon}{2\Delta_u}\right)}{\int \exp\left(u(\s^{\ast}|\x)\frac{\epsilon}{2\Delta_u}\right) d\s^{\ast} }
  \end{equation}
  to ensure $\epsilon$-DP, where $\Delta_u=\underset{\scriptstyle{\x,\x', d(\x,\x')=1}}{\mbox{max}} \big|u(\s^{\ast}|\x)-u(\s^{\ast}|\x')\big|$
  is the maximum change in score $u$ with one row change in the data (if $\s^*$ is discrete, the integral in Eqn (\ref{eqn:exp}) is replaced with summation).
\end{defn}

\noindent Per the Exponential mechanism, the probability of returning  $\s^{\ast}$ increases exponentially with the utility score.  For example,  if $\mathbf{s}$ is numerical and the goal is to preserve as much original information as possible, metrics measuring the closeness between $\s^{\ast}$ and the original $\s$ are good candidates for $u$ such as the negative $p$-norm  distance $-||\s-\s^{\ast}||_p=-\left( \sum_{j=1}^r \left|s_j- s_j^{\ast} \right|^p \right)^{1/p}$ \citep{liu2016generalized}.  When the $l_1$ norm  is used, the Exponential mechanism in Definition \ref{def:exp} becomes the Laplace mechanism with halved privacy budget \citep{mcsherry2007mechanism,liu2016generalized}. Both the Laplace mechanism and Exponential mechanism are widely applied in developing more complicated mechanisms, such as  the multiplicative weight approach of generating synthetic discrete data iteratively \citep{hardt2010multiplicative} and the median mechanism  for efficiently releasing correlated queries \citep{roth2010interactive}.

Besides the Laplace mechanism and the Exponential mechanism, there are other sanitizers for general settings, such as the Gaussian mechanism that adds Gaussian noise to satisfy a softer version of DP (Section \ref{sec:relax}) \citep{dwork2013algorithmic, liu2016generalized} and the generalized Gaussian mechanisms that include the Laplace mechanism and the Gaussian mechanism as special cases \citep{liu2016generalized}.

\subsection{Relaxations of Pure \texorpdfstring{$\epsilon$}-DP}\label{sec:relax}
One criticism of the pure $\epsilon$-DP in Section \ref{sec:dp} is the potentially large amount of noise being added to query results to achieve a high level of privacy guarantee. This concern has motivated work on relaxing the pure $\epsilon$-DP. We briefly overview three relaxations: approximate differential privacy (aDP), probabilistic differential privacy (pDP), and concentrated differential privacy (cDP).

\begin{defn}\label{def:adp} \textbf{Approximate Differential Privacy} \citep{dwork2006our}:
A sanitization algorithm $\R$ gives $(\epsilon, \delta)$-aDP if for all data sets $(\x,\x')$ that are $d(\x,\x')\!=\!1$,
\begin{equation}\label{eqn:adp}
\Pr(\R( \s(\x)) \in Q)\le \exp(\epsilon) \Pr(\R( \s(\x'))\in Q) + \delta,
\end{equation}
\end{defn}
\noindent where $\delta>0$ is typically chosen based on the sample size of the data set $n$ that  satisfies $\delta(n)/n^r\rightarrow0$ for all $r>0$. The pure $\epsilon$-DP is a special case of aDP when $\delta=0$.

\begin{defn}\label{def:pdp} \textbf{Probabilistic Differential Privacy} \citep{machanavajjhala2008privacy}:
A sanitization algorithm $\R$ gives $(\epsilon, \delta)$-pDP if
\begin{equation}\label{eqn:pdp}
\Pr\left(\R( \s(\x)) \in \text{Disc}(\x,\epsilon)\right)\leq \delta
\end{equation}
\noindent for all data sets $(\x,\x')$ that are $d(\x,\x')\!=\!1$, where Disc$(\x,\epsilon)$ is the disclosure set $\R(\s(x))$ such that
$\left|\ln\left(\frac{\Pr(\R(\s(\x))\in Q)}{\Pr(\R(\s(\x'))\in Q)}\right)\right|>\epsilon.$  Eqn (\ref{eqn:pdp}) can be interpreted as the pure $\epsilon$-DP fails with probability $\delta$.
\end{defn}

\begin{defn}\label{def:cdp} \textbf{Concentrated Differential Privacy} \citep{dwork2016concentrated}:
For all data sets $(\x,\x')$ that is $d(\x,\x')=1$, a sanitization algorithm $\R$ gives $(\mu, \tau)$-cDP if $D_{\text{subG}}(\R(\x)||\R(\x')) \preceq (\mu,\tau)$, where $D_{\text{subG}}$ stands for \emph{subGaussian divergence}, defined as follows: two random variables $Y$ and $Z$ are $D_{\text{subG}}(Y||Z) \preceq (\mu,\tau)$ if and only if  $\mathbb{E}(L_{Y||Z})\leq \mu$ and  the centered distribution of $(L_{Y||Z}-\mathbb{E}(L_{Y||Z})$ is defined and $\tau$-subgaussian, where  $L_{(Y||Z)}=\ln\left(p(Y) /p(Z) \right)$ is the privacy loss random variable.
\end{defn}

\noindent Both pDP and cDP regard privacy loss as random variables, but cDP has some advantages over  pDP. First, cDP has a bounded expected privacy loss whereas pDP has an infinite privacy loss with probability $\delta$. Second, cDP has better accuracy without compromising the privacy loss from multiple inquiries \citep{dwork2016concentrated}.


\section{Differentially Private Data Synthesis  (DIPS)}\label{sec:methods}
We loosely group the currently available DIPS methods into two categories: the non-parametric approach (NP-DIPS) and the parametric approach (P-DIPS). In the NP-DIPS approach, the synthesizer is constructed based on the empirical distribution of the data, while  in the P-DIPS approach it is constructed based on a parametric distribution or an appropriately defined model for the original data.  

\subsection{Non-parametric DIPS (NP-DIPS)}\label{subsec:npdips}
When the original data is categorical, the statistics $\s$ targeted for differentially private sanitization are the cell counts or proportions in some types of cross-tabulation in NP-DIPS, from which the synthetic data will be from generated.  In the case of continuous data, the NP-DIPS techniques can be applied to generate differentially private histograms, kernel density estimators, or empirical distributions. The list of the NP-DIPS covered in section is given in Table \ref{tab:npdips}. Our goal is not to discuss every NP-DIPS method out there in the literature, which would be impossible to achieve in one paper. The list is not exhaustive, but should provide the readers an idea on how DIPS works in the non-parametric setting.
\begin{table}[!htb]
\caption{Summary of NP-DIPS approaches  discussed in Section \ref{subsec:npdips}}
\label{tab:npdips}\centering\small
\begin{tabular}{L{0.8cm}| L{4.0cm}|L{3.6cm}|L{5.1cm}}
\hline
\textbf{Sec}  & \textbf{Method} & \textbf{Pros}  &\textbf{Cons} \\
\hline
\ref{subsec:table} & Laplace sanitizer  & simple; fast & not accurate for large number of queries \\
\hline
\ref{subsec:table}  &Fourier transformation  & preserves low-order marginals accurately & computationally expensive as the number of attributes increases \\
\hline
\ref{subsec:table}  & multiplicative weights Exponential mechanism  & adaptive, preserves consistency of marginals across tables & difficulty of choosing an appropriate iteration number; inaccuracy\\
\hline
\ref{subsec:hist} &perturbed histogram & simple; fast  & discretization of continuous attributes; doesn't preserve correlation well \\
\hline
\ref{subsec:hist} &smoothed histogram & simple; fast & discretization of continuous attributes; worse than perturbed histogram in accuracy\\
\hline
\ref{subsec:hist} & empirical cumulative density function via Exponential mechanism  & flexible; general & computational infeasibility \\
\hline
\ref{subsec:hist} &kernel density estimator with Gaussian process noise & general & works for $(\epsilon,\delta)$-aDP; curse of dimensionality\\
\hline
\ref{subsec:others} &histogram with constrained inferences & better accuracy than perturbed histogram& constrains are publicly known or inherent\\
\hline
\ref{subsec:others} &universal histogram & accuracy  for low-order counts & less accurate for high-order counts\\
\hline
\ref{subsec:others} & DPCube  & multidimensional data& inefficiency in constructing accurate high-dimensional histograms; performs worse than the Laplace sanitizer\\
\hline
\ref{subsec:others} & NoiseFirst and StructureFirst   & outperforms several other DP  methods & low dimensional histograms; non-consistency as $\epsilon\rightarrow\infty$\\
\hline
\ref{subsec:others} &Exponential Fourier perturbation and  P-H Partition  & better than NoiseFirst and StructureFirst & depends on histogram compressibility\\
\hline
\end{tabular}
\end{table}

\subsubsection{Synthesis of Categorical Data}\label{subsec:table}
In a data set with $p$ categorical variables, a straightforward approach in generating synthetic data is to add Laplace noise to the cell counts of $k$-way cross-tabulation of $\x$, where $k \leq p$, and then to generate individual level of data from the sanitized counts.  If  $k=p$, it is the full cross-tabulation of $\x$, and the individual-level data are straightforward to generate from sanitized counts. If $k<p$, there are  $(_k^p)$ $k$-way contingency tables, and the sanitization process needs to be carefully planned so that all $k$-way tables are consistent to yield legitimate marginals and individual-level data.

When $k=p$, denote the original frequencies of the $K$ cells formed by the $p$-way cross-tabulation of $\x$ by $\bs{n} = (n_1,..., n_K)$.  The Laplace sanitizer perturbs the original $\mathbf{n}$ via $n_j^\ast=n_j+e_j$, where $e_j\sim \text{Lap}(0,\Delta_{\s}/\epsilon)$ independently for $j=1,\ldots,K$. $\Delta_{\s}$ is the $l_1$ global sensitivity from releasing the whole cross-tabulation.  $\Delta_{\s}$ can be set at  2  or 1, depending on how $d(\x,\x')=1$ is defined. Specifically, if the  change in one individual refers to the case that $n$ remains the same, but the data in exactly one individual change, then $\Delta_{\s}=2$. If the  change in one individual refers to removal of one individual from the data, then $\Delta_{\s}=1$. When  $n$ is relatively large, say $>30$, the difference in the standard deviations of the Laplace noises  $\sqrt{2}n^{-1}$ between the two versions is $O(10^-2)$. There is no practical difference on which one to use. In the simulation studies and the case study presented later,  we used $\Delta_\s=1$. Given the smallest $n$ examined was 40, we expect the results to remain roughly the same if we had used $\Delta_s=2$.  

When $k<p$, \citet{barak2007privacy} conduct early work on constructing $k$-way differentially private, consistent, and non-negative contingency tables via a Fourier transformation. The approach identifies the complete set of metrics required to reproduce a contingency table, where each cell is perturbed to achieve the same level of accuracy. Another approach to generate individual-level data in the discrete domain  is the multiplicative weights Exponential mechanism approach based on linear queries \citep{hardt2012simple}. The multiplicative weights Exponential mechanism method approximates the original distribution in a differentially private manner through an iterative process. It starts from a uniform distribution over the supports of all the attributes in the original data, and then updates the distribution via multiplicative weighting based on a query sampled via the Exponential mechanism and sanitized via the Laplace mechanism in each iteration. Since every iteration access the original data, the total privacy needs to be divided by the number of iterations.

The Fourier transformation based algorithm depends on the linear programming and could be computationally infeasible when $p$ is large. Though the multiplicative weights Exponential mechanism approach is computationally more efficient, it is difficult to choose an optimal iteration number especially when $p$ is large. A small number of iterations would not be sufficient to capture the information in the original queries, leading to biased synthetic data, while a large number of iterations will introduce too much noise during the data generation process, rendering the synthetic data useless. The inaccuracy of the multiplicative weights Exponential mechanism is documented in \citet{li2016differential,vadhan2017complexity,kowalczyk2017hardness} and is also confirmed by the simulation studies we have conducted. For these reasons, we do not evaluate the Fourier transformation based method and the multiplicative weights Exponential mechanism  in Section \ref{sec:simulation}. 

\subsubsection{Synthesis of Numerical Data}\label{subsec:hist}
A straightforward approach for releasing differentially private numerical data is to first generate differentially private histograms, and then synthesize numerical data by drawing a bin according to the relative sanitized frequencies of the histogram bins, and lastly, sample data from the uniform distributions bounded by the sampled bin endpoints in the previous step.

To form histograms on the original numerical  data, discretization is necessary. And there could be a large number of data bins/cubes if high-order interactions exist among the data attributes and are taken into account when the histogram is generated. Let $K$ be the total number of bins (or squares/cubes in the multidimensional case), $n_k=\sum^{K}_{k=1}I(\x_i\in B_k)$ be the number of observations in $B_k$ for $k=1,\ldots,K$,  $\hat{p}_k=n_k/n$,  and $I()$ be the indicator function ($I(\x_i\in B_k)=1$ if $\x_i\in B_k$; 0 otherwise), a mean-squared consistent density histogram estimator is $\hat{f}_K(\x)=\sum^K_{k=1}K\hat{p}_kI(\x\in B_k)$ \citep{scott2015multivariate}. A differentially private perturbed histogram is a direct application of the Laplace mechanism. The sanitized bin counts and proportions with $\epsilon$-DP are given by $n^*_k=n_k+e_k$ and $\hat{p}^\ast_k=n^*_k/\sum_k n^*_k$, respectively, where $e_k\stackrel{iid}{\sim} \text{Lap}(0,\Delta_\s/\epsilon)$ with $\Delta_\s=1$.
The density histogram estimator that satisfies $\epsilon$-DP is thus
\begin{equation}\label{eqn:density2}
\hat{f}^*_K(\x)=\textstyle\sum^K_{k=1}K\hat{p}^\ast_kI(\x\in B_k).
\end{equation}
Note that sanitized $n^*_k$  can be negative since the Laplace noise $\in(-\infty,\infty)$, especially when $n_k$ is small or $\epsilon$ is small. Commonly used  post-hoc processing approaches include replacing negative $n^*_k$ with 0 \citep{barak2007privacy} or using the truncated or boundary inflated truncated Laplace distributions to obtain legitimate  data \citep{liu2016noninformative}.   To incorporate the uncertainty introduced by the sanitization process, releasing multiple sets of $\tilde{\x}$ is suggested, one set per sanitized $\mathbf{n}^*=\{n^*_k\}_{1:K}$.

Another method to generate differentially private histograms is the smoothed histogram approach. \citet{wasserman2010statistical} provided the formulation of smoothed histograms of $\epsilon$-DP for $\x\in[0, 1]^p$, where $p$ is the number of numerical attributes. It is easy to extend the formulation to the general case when $\x$ is bounded by $[c_{10},c_{11}]\times ... \times [c_{p0}, c_{p1}]$. The differentially private smooth histogram is
\begin{equation}\label{eqn:smooth.hist}
\hat{f}^*_K(\x)=\left(1-\lambda\right)\hat{f}_K(\x)+\lambda\Omega, \mbox{ where  }\Omega=\left(\textstyle\prod_{j=1}^p(c_{j1}-c_{j0})\right)^{-1},
\end{equation}
\begin{equation}\label{eqn:smooth.lambda}
\mbox{and }\lambda \geq \frac{K}{K+n\left(e^{\epsilon/n}-1\right)}
\end{equation}
is a constant between 0 and 1 to satisfy $\epsilon$-DP. When $\epsilon\rightarrow0$, $\lambda\rightarrow1$, the synthetic data are simulated from a uniform-like $\hat{f}_K^*(\x)$ that is too noisy to be of any use. When $\epsilon\rightarrow\infty$, $\lambda\rightarrow0$, $\hat{f}_K^*(\x)\rightarrow \hat{f}_K(\x)$, the synthetic data would have minimal privacy protection from the DP perspective. Since $\lambda$ is a constant given $n,K$ and $\epsilon$, $\hat{f}_K(\x)$ is not subject to randomness either,  it is not necessary to release multiple sets of $\tilde{\x}$ from $\hat{f}^*_K(\x)$ from an inferential perspective.

In addition to the perturbed histogram and smooth histogram approaches, there is also the approach to generating data from differentially private empirical cumulative density functions via the Exponential mechanism \citep{wasserman2010statistical}.  Specifically, surrogate data $\tilde{\x}$ is simulated from
\begin{equation}
h(\tilde{\x})=\frac{g_\x(\tilde{\x})}{\int_{[c_{10},c_{11}]\times\cdots\times[c_{p0},c_{p1}]} g_\x(\z)d\z},\label{eqn:exp.dens}
\end{equation}
$$\mbox{where } g_\x(\tilde{\x})=\exp \left(- u(\hat{F}_\x,\hat{F}_{\tilde{\x}} )\frac{\epsilon}{2\Delta_u}\right),
\Delta_u=\!\!\sup_{\x,\x', \Delta(\x,\x')=1}\! \sup_{\tilde{\x}}\!\left|u (\hat{F}_\x,\hat{F}_{\tilde{\x}})-u (\hat{F}_{\x'},\hat{F}_{\tilde{\x}})\right|,$$
$\hat{F}_\x$ is the original empirical cumulative density function, $\hat{F}_{\tilde{\x}}$ is the empirical cumulative density function's of the sanitized data, $u$ is the utility function that denotes a distance measure between the two cumulative density functions,  and $\Delta_u$ is the sensitivity of $u$. If the Kolmogorov-Smirnov distance is used on $u$, $\Delta_u\leq n^{-1}$ \citep{wasserman2010statistical}. However, releasing $\tilde{\x}$ via the Exponential mechanism defined in Eqn (\ref{eqn:exp.dens}) does not seem to be a viable choice in practice. One difficulty lies in defining the set of all possible candidate  cumulative density functions, the size of which increases rapidly with sample size $n$ and $p$, making the synthesis process computationally challenging and unrealistic for a large data set. Due to the impracticality of this approach, we did not implement this method in our simulation studies.

\citet{Hall2013} propose sanitizing kernel density estimator by adding noise from a Gaussian process to yield  DP, from which synthetic data can be generated. If a Gaussian kernel is used, they show there is no loss of accuracy in the differentially private kernel density estimator to the original one with the optimal bandwidth that minimizes the integrated mean squared error. However, the method is currently only available for $(\epsilon,\delta)$-aDP for $\delta>0$, and suffers the same curse of dimensionality for large $p$  \citep{scott2015multivariate}.

\subsubsection{Other NP-DIPS Methods}\label{subsec:others}
There are also various extensions to the basic Laplace sanitizer and the perturbed histogram approach with the purposes to improve their accuracy. \citet{hay2010boosting} suggested boosting the accuracy of differentially private histograms by sorting the bin values after sanitation if the order of the bin size is known to the public. They also suggested a universal histogram approach by exploring the inherent consistency in a hierarchical histogram, and  proved that the accuracy of lower-order contingency tables/marginals is improved, but at the sacrifice of high-order contingency tables \citep{qardaji2013understanding,hay2016principled}. \citet{xiao2012dpcube} applied a 2-phase partitioning strategy, DPCube, to multidimensional data cubes. \citet{gardner2013share} implemented DPCube in biomedical data to demonstrate its practical feasibility on real-world data sets, but found that DPCube was still inefficient in constructing accurate high-dimensional histograms. Additionally, \citet{hay2016principled} showed that DPCube performed worse than the Laplace sanitizer.  \citet{xu2013differentially} proposed two mechanisms, NoiseFirst and StructureFirst, that performed well against some DP methods, but only applied to low dimensional histograms due to long running time. Moreover, StructureFirst is inconsistent; where the error of the statistics does not tend to 0 as $\epsilon$ increases to infinity \citep{qardaji2013understanding,hay2016principled}. \citet{acs2012differentially} presented two sanitization techniques, the Exponential Fourier perturbation algorithm and the P-H Partition, that  sanitize compressed data to exploit the inherent redundancy of real-life data sets. From the experimental results, the techniques outperformed some DP methods, including NoiseFirst and StructureFirst, but the performance depended on the compressibility of a histogram.

In summary, the accuracy improvements, if any, of the above methods over the basic Laplace sanitizer or the perturbed histogram either utilize some constraints that only exist in certain types of histograms/data, or only benefit low dimensional histograms. For these reasons, we do not explore these extensions in the simulations studies in Section \ref{sec:simulation}. That being said, it is of our interest to further explore these extended methods that provide better accuracy in low-dimensional histograms in the future, by coupling them with efficient and accurate dimensional reduction techniques.

\subsection{Parametric DIPS (P-DIPS)}\label{subsec:pdips}
The synthesizers in the P-DIPS category are based on an assumed distribution or an appropriately defined model given the original data. In what follows, we describe the Multinomial-Dirichlet synthesizer and other methods motivated by the Multinomial-Dirichlet synthesizer for categorical data, the model-based DIPS (MODIPS) approach for general data types based on a Bayesian modeling framework, and sequential regression synthesizers. The list of the P-DIPS covered in section is given in Table \ref{tab:pdips}. The list is not meant to be exhaustive nor does it list the methods that deal with a specific type of data, but it gives readers an idea on how DIPS works in the parametric setting. The PrivBayes method based on Bayesian networks and the DPCopula, though listed in Table \ref{tab:pdips},  will not be covered in full details in this section given that they are not as widely used for routine data analysis. 

\begin{table}[!ht]
\caption{Summary of Parametric DIPS approaches discussed in Section \ref{subsec:pdips}.}
\label{tab:pdips}\centering\small
\begin{tabular}{L{0.35in}| L{0.95in} | L{0.75in}|L{1.3in}|L{1.8in}}
\hline
\textbf{Sec} & \textbf{Method} &  \textbf{Data}& \textbf{Pros}  & \textbf{Cons} \\
\hline
    \ref{subsec:MD}  & Multinomial-Dirichlet, Binomial-Beta & categorical & straightforward; easy to implement & performs poorly on sparse data; perturbation amount increases with $n$; possible biased inferences on proportions\\
\hline
    \ref{subsec:modips}  & Model-based DIPS (MODIPS) & any & general; the DP version of the model-based  multiple synthesis without DP& model-dependent; relies on identification and sanitization of sufficient statistics or likelihood functions\\
\hline
    \ref{subsec:regression} & Sequential Regression Modeling Synthesizers & any & general; models the correlations among all variables & large amount noises for large $p$; \\
\hline
    & PrivBayes & categorical & models dependency among variable; has an inherent model selection component& requires dichotomization on continuous attributes;  depends on a quality function that can be computationally inefficient\\
\hline
    & DPCopula & any & general & same limitations for copula models in general and quadratic time complexity\\
\hline
\end{tabular}
\end{table}


\subsubsection{Multinomial-Dirichlet Synthesizer}\label{subsec:MD}
\citet{abowd2008protective} proposed the Multinomial-Dirichlet synthesizer  to generate differentially private categorical data in the Bayesian framework. The likelihood of proportions $\bs{\pi}$ is constructed from $f(\mathbf{n}|\bs{\pi}) \sim \text{Multinom}(n, \bs{\pi})$, where $\mathbf{n}=(n_1,\ldots,n_K)$ contains the original cell counts in $K$ categories in the original data and $n=\sum_kn_k$. A Dirichlet prior  $f(\bs{\pi})\!=\!\text{D}(\bs{\alpha})$ is imposed on  $\bs{\pi}$, where each element of  $\bs{\alpha}$ is set at $\alpha^\ast_k= n/(e^\epsilon-1)$,  the minimum value that guarantees $\epsilon$-DP,  for $k=1,...,K$. To generate differentially private surrogate data sets, $\bs{\pi}^*$ is first simulated from the posterior distribution $f(\bs{\pi}^\ast|\x)=D(\bs{\alpha}^\ast+\bs{n})$, and then synthetic data is drawn from $f(\tilde{\mathbf{n}}|\bs{\pi^\ast})=\text{Multinom}(n,\bs{\pi}^\ast)$. To ensure valid inferences in the synthetic data, multiple sets of $\tilde{\mathbf{n}}$ can be released; one for each differentially private $\bs{\pi^\ast}$. The Multinomial-Dirichlet synthesizer reduces to the Binomial-Beta synthesizer in the binary case. \citet{mcclure2012differential} proposed a slightly different approach to synthesizing binary data from $f(\tilde{\mathbf{n}}|\mathbf{n})\!=\!\mbox{Binom}\left(n,\frac{n_1+\alpha_1}{n+\alpha_1+\alpha_2}\right)$,
where $\alpha_1\!=\!\alpha_2\!=\!\left(e^{\epsilon/n}-1\right)^{-1}$ to satisfy $\epsilon$-DP, which we refer to as the Binomial-Beta McClure-Reiter approach. The Binomial-Beta McClure-Reiter  differs from the Binomial-Beta synthesizer not only in how the prior on $\pi$ differ, but also that it does not simulate $\bs{\pi}$  from its posterior distribution thus $\tilde{\mathbf{n}}$ synthesized via the Binomial-Beta McClure-Reiter has one less layer variability.

In both the Multinomial-Dirichlet/Binomial-Beta and the Binomial-Beta McClure-Reiter synthesizers, $\alpha^\ast_k$ increases with $n$, implying that when data/observed information increases, the amount of perturbation required to maintain $\epsilon$-DP also increases and can be nontrivial for any $n$. Furthermore, since all $\alpha^\ast_k$'s for $k=1,...,K$ are equal,  when $n_k$'s are not the same across the $K$ categories, the perturbation will bias the synthetic proportions away from their originals.  \citet{Charest2010} modeled explicitly the Binomial-Beta mechanism in a Bayesian framework in the binary data case  to reduce the bias of the inferences in the synthetic binary data, which seems to be effective as long as $\epsilon$ is not too small.

\subsubsection{Model-based DIPS (MODIPS)}\label{subsec:modips}
\begin{figure}[!ht]
\centering
\includegraphics[scale=0.4]{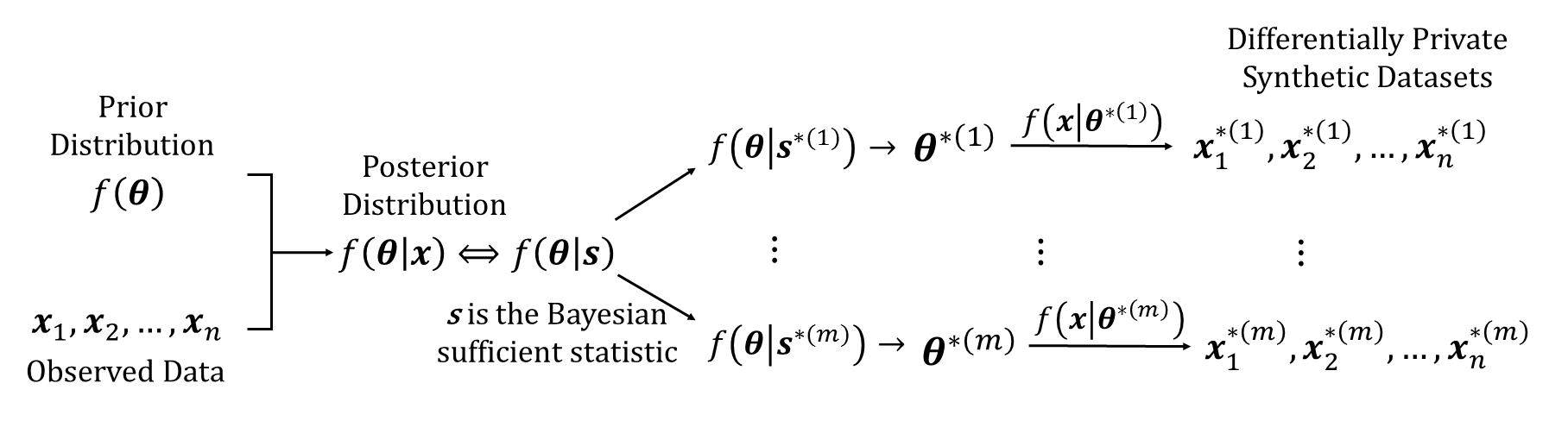}
\caption{The MODIPS algorithm}\label{fig:modips}
\end{figure}

The MODIPS approach is based in a Bayesian modeling framework and releases $m$ sets of multiple  differentially private surrogate data to the original data to account for the uncertainty of the synthesis model \citep{liu2016model}. An illustration of the MODIPS algorithm is given in Figure \ref{fig:modips}.
The MODIPS approach first constructs an appropriate Bayesian model from the original data and identifies the Bayesian sufficient statistics $\mathbf{s}$ associated with the model. The posterior distribution of $\bs{\theta}$ can then be represented as $f(\bs{\theta}|\s)$. The MODIPS then sanitizes $\s$ with privacy budget $\epsilon/m$. Denote the sanitized $\s$ by $\s^*$. Synthetic data $\tilde{\x}$ is simulated given $\s^*$ by first drawing $\bs{\theta}^*$ from the posterior distribution $f(\bs{\theta}|\s^*)$, and then simulating $\tilde{\x}^*$ from $f(\x|\bs{\theta}^*)$. The procedure is repeated $m$ times  to generate $m$ surrogate data sets.

Since the MODIPS approach is model-dependent, the identification and validation of an appropriate model for data $\x$ is critical, and model mis-specification will generate biased samples. If the identification of a suitable model is based on previous knowledge and common practice, then no privacy will need to be spend; however, if the model selection procedure is based on the the data to be released, then the data curator will have to a certain portion of the total privacy budget to the model selection procedure. Differentially private model selection is a separate research topic that is beyond the scope of this paper.   
When there are several plausible models, the model averaging idea can be incorporated into the synthetic data generation and serves as a mitigation measure to model mis-specification and the dependency of the synthetic data on a single synthesis model. The model averaging can be implemented using the Bayesian model averaging method  \citep{hoeting1999bayesian}; but it can be analytically and computationally demanding. An alternative approach, less formal but practically more straightforward manner, can also be applied. Say there are 3 reasonable models -- M1, M2, and M3 -- for the original data, with the ``plausibility'' weights 0.4, 0.3 and 0.3 for each model  (e.g., per Bayes factors).  Suppose 10 sets of synthetic data are to be released; we could then generate 4 sets of synthetic data from M1, 3 sets from M2, and 3 sets  from M3, leading to 10 sets synthesized by 3 different models. The inferences based on the 10 sets (combined using the method given in Section \ref{sec:combinationrule}) will  implicitly integrate out the model uncertainty, and are more robust and less sensitive to the model selection and specification than in the case where all 10 sets are generated from a single model. The downside of the model averaging idea is that it will result in more uncertainty in the synthetic data, a price paid for more robustness. In addition, the weights associated with the set of the models can be subjective, even via the Bayes factor approach, which is known for its dependence on the priors.


\subsubsection{Sequential Regression Modeling Synthesizers}\label{subsec:regression}
Another method to generate DIPS data is through a sequential regression modeling synthesizer approach, which, broadly speaking, can also be regarded as the MODIPS approach.  Specifically, suppose the variables from the data are $X_1,X_2,\ldots,X_p$. The joint distribution of $f(X_1,\ldots,X_p)$ can be decomposed as $f(X_1)f(X_2|X_1)\ldots f(X_p|X_1,\ldots, X_{p-1})$, suggesting data can be generated sequentially by first synthesizing $X_1$ from $f(X_1)$,  then $X_2$ from the conditional model of $X_2$ given $X_1$, and so on. Each of the $p$ regression models needs to differently private and can borrow the existing framework on differentially private empirical risk minimization (ERM) or differentially private regression models.  For instance, \citet{chaudhuri2009privacy} propose directly perturbing the minimizers or perturbing the empirical risk to obtain differentially private minimizers in $l_2$ regularized logistic regression, which is extended to differentially private empirical risk minimization with differentiable and strongly convex regularizers in \citet{Chaudhuri2011} and with non-differentiable regularizer in  \citet{Kifer2012}. 
\citet{zhang2012functional} propose the functional mechanism that adds noise to the objective function using the Laplace mechanism and estimates the global sensitivity through a polynomial representation. The functional mechanism applies to both linear and logistic regression, but the latter is based on the approximation through the Taylor expansion and is susceptible to large amount of noise \citep{zhang2013privgene}. \citet{sheffet2015differentially} examines differentially private inferences (hypothesis testing and confidence interval construction) for ordinary least squares and ridge estimator in linear regression. 


The sequential regression modeling synthesizer accounts for the correlations among the various variables. In addition, the synthesizer can be implemented in most data types as long as we have differentially private versions for the commonly used regression model types in practice. The sequential modeling framework has been successfully implemented in practice for imputing missing data  \citep{raghu2001, IVEaware} and data synthesis \citep{kinney2011towards}). Used as a DIPS methods, there are a few potential drawbacks. First, the total privacy budget needs to be divided into $p$ portions due to the DP sequential composition, and each regression model receives only a single portion. If $p$ is large, this approach could perform poorly in terms of statistical utility of the synthetic data due to the lack of privacy budget per regression. Most of the DP regression techniques mentioned above often output a single point estimate for the parameters involved in each regression models, which can be plugged in to generate synthetic data. To properly propagate the uncertainty around the parameters, we either have to model the synthesis process analytically or release multiple synthetic data sets by drawing and plugging in multiple sets of parameters as in the MODIPS approach. Third, efficient DP regression models are not available for all model types (e.g., the Cox regression for survival data). One possible solution that circumvents regression modeling is the STEPS approach \citep{STEPS}, a nonparametric synthesizer that is also based  sequentially ``modeling'' of the data.  


\subsection{Inferences from Synthetic Data via DIPS}\label{sec:combinationrule}
Synthetic data generated by DIPS approaches are perturbed through the sanitization process with random  noise into the original data. Some P-DIPS approaches (such as the Multinomial-Dirichlet synthesizer and MODIPS) also incorporate the uncertainty around the distribution and model assumed on the original data. There are at least two approaches that account for the sanitization/synthesis uncertainty in the inferences based on the synthetic data. The first approach is to model the sanitization/synthesis process directly, such as in \citet{Charest2010} for synthesizing binary data and in \citet{JRSSC}, where the edges of a social network are synthesized via a randomized response mechanism under $\epsilon$-edge DP in the exponential random graph models and then likelihood-based inference for missing data and Markov chain Monte Carlo techniques (more specifically, Metropolis-Hastings algorithms) are applied to model the synthesis process. This approach can be demanding for data users both analytically and computationally. The second approach is to release multiple sets of synthetic data, which can be regarded as a Monte Carlo version of the former. In the multiple release approach, data users only need to analyze each surrogate data set as if they had the original data set, and then combine the multiple sets of inferences in a legitimate way to yield the final inferences. Suppose the parameter of interest is $\beta$. Denote the estimate of $\beta$ in the $j\textsuperscript{th}$ synthetic data by $\hat{\beta}_j$ and the associated standard error by $v_j$. The final point estimate $\beta$ is
\begin{equation}
\bar{\beta} = \textstyle{m^{-1}\sum_{j=1}^m}\hat{\beta}_j\label{eqn:mean}
\end{equation}
with Var$(\bar{\beta})$ estimated by
\begin{equation}
	T = m^{-1}B + W \label{eqn:var},
\end{equation}
where $B= \sum_{j=1}^m (\hat{\beta}_j-\bar{\beta})^2 /(m-1)$ (between-set variability) and  $W=m^{-1}\sum_{j=1}^m v^2_j$  (average per-set variability); and tests and confidence intervals are based on
\begin{equation}
	(\bar{\beta}-\beta)T^{-1/2} \sim t_{\nu=(m-1)(1+mW/B)^2}.\label{eqn:test}
\end{equation}
The formal proof of the variance combination rule for the MODIPS approach can be found in \citet{liu2016model}. Eqn (\ref{eqn:var}) coincides with the variance combination rule in  \citet{reiter2003} for obtaining inferences from multiply synthetic sets in the context of non-DP setting for partial synthesis, but the components of  the between-set variability $B$ difference in these two cases:  $B$ in MODIPS has one more source of variability from sanitizing $\s$ compared to the non-DP MS approach. Due to this, inferences from synthesized data via the MODIPS approach will be less precise than those from non-DP MS approaches  -- a price paid for the DP guarantee.  Though not formally proved, it is expected  Eqns (\ref{eqn:mean}) to (\ref{eqn:test}) also apply in the Multinomial-Dirichlet synthesizer and other DIPS approaches that use multiple set releases to account for sanitization and synthesis uncertainty, though the sources that compose $B$ might differ.

A reviewer questioned why the variance combination rule $T = (1+ m^{-1})B - W$ for full synthesis \citep{raghunathan2003multiple}  in the non-DP setting does not apply in this case. The reasons is that $T=(1+ m^{-1})B-W$ deals with  the situation where data from non-sampled units in  a finite population are regarded as missing values  and then filled in by multiple imputation, followed by a second step of sampling from the imputed population while in  the context of MODIPS, there is no synthesis of a finite population or the extra step of sampling from the population. As pointed out by \citet{raghunathan2003multiple} themselves, due to ``the random sampling of the units that compose the synthetic samples from each multiply synthetic population'' that is  ``not presented in the usual multiple imputation'', ``the between imputation already reflects the usual within imputation variability'', and thus $T=(1+ m^{-1})B-W$.  In fact, the synthesis of the full sample, as done in the MODIPS approach, can be viewed as a special case of the ``partial synthesis'' examined in \citet{reiter2003} with the selection percentage of synthesis equal to 100\%. We also conducted simulation studies and showed $T = (1+ m^{-1})B - W$ led to underestimate of the variance and undercoverage of the confidence interval, while  Eqn (\ref{eqn:var}) delivered the nominal coverage.

\section{Simulation Studies} \label{sec:simulation}
We assess the utility and  inferential properties of  the sanitized data via some of the DIPS approaches presented in Section \ref{sec:methods} in four simulation studies. We examine the approaches in the setting of the pure $\epsilon$-DP through the application of the Laplace mechanism, but all the DIPS approaches can be applied under softer versions of DP (eg, ($\epsilon,\delta$)-pDP) via the employment of appropriate sanitizers (such as the Gaussian mechanism). The first and second simulation studies focus on univariate categorical data and univariate continuous data, respectively; the third and fourth simulation studies involve a mixture of categorical and continuous variables, but data are generated from different models.

Though the first and second simulation studies seem to be simple, the results on the impacts of DP on the statistical inferences are in fact insightful even in these simple cases, especially considering there is few work out there comparing different DIPS approaches in statistical inferences; they also provide justifications for the choices of the DIPS approaches used in the third and fourth simulation studies. In the fourth simulation, we examined the effect of mis-specification on the synthesis model for the MODIPS approach, and investigated the importance of selection and validation of an appropriate synthesis model. We did not implement the model selection/validation procedures in simulation studies 1 to 2 due to the simplicity of the data and thus the obvious choice of an appropriate model; neither did we in simulation study 3 given that its similarity in the data structure with simulation 4, and it would make no difference whether simulation studies 3 or 4 was used for the purposes of illustrating the model selection/validation procedure in the MODIPS approach.

We varied the privacy  budget  $\epsilon$ from $e^{-4}$ to $e^4$ in each simulation to examine its effect on the statistical inferences. Though $\epsilon$ as large as $e^4$ might not be used in practice due to privacy considerations, it is a useful theoretical exploration on the amount of privacy sacrifice in order to have inferences based on the synthetic data to be close to the original; likewise, $\epsilon$ as small as $e^{-4}$ helps us to understand what level of privacy would ruin the inferences to an unacceptable degree based on the synthetic data. In all the examined DIPS approaches, the sample size of each released synthetic set was the same as the original data, and 5 sets of synthetic data were generated in DIPS approaches except for the smoothed histogram and the Binomial-Beta McClure-Reiter approaches for reasons stated in Section \ref{sec:methods}. For the DIPS approaches that generated 5 synthetic data sets, each synthesis received $1/5$ of the total privacy budget  $\epsilon$ per the sequential composition principle.  The inferences based on the DIPS synthetic data were benchmarked against those based on the original data and the traditional non-DP MS technique.

\subsection{Simulation Study 1: Categorical Data }
The following DIPS  methods are compared in this simulation study: the MODIPS synthesizer, the Laplace sanitizer, the Binomial-Beta McClure-Reiter synthesizer, and the Multinomial-Dirichlet synthesizer. Data was simulated from a Bernoulli distribution $f(x_i)=\mbox{Bern}(\pi)$ for $i=1,\ldots,n$. We examined  9 simulation scenarios for $n\in\{40,100,1000\}$ and $\pi\in\{0.10,0.25,0.50\}$, with 5000 repetitions per scenario.

The non-DP MS and the MODIPS approaches are model-based, and usually model selection and validation should be applied to select an appropriate synthesis model. However, we didn't perform model  and selection and validation given obvious choice of the likelihood with the simplicity of the data in this simulation. With the binomial likelihood and prior $\mbox{Beta}(\alpha, \beta)$ on $\pi$, the posterior distribution of $\pi$ given $\x$ is $f(\pi|\x)=\mbox{Beta}(\alpha+n_1,\beta+n-n_1)$ where $n_1=\#\{x_i=1\}$. We set $\alpha=\beta=1/3$ \citep{kerman2011neutral}.  In the MODIPS approach, we first located the Bayesian sufficient statistics $\s$ associated with the posterior distribution  $f(\pi|\x)$, which is $n_1$ with global sensitivity being 1. The  Laplace mechanism was then employed to obtain $n_1^{\ast}=n_1+e$, where $e\sim\mbox{Lap}(0,\epsilon^{-1})$. Finally, we sampled $\pi^{\ast}$ from $f(\pi^{\ast}|n_1^{\ast})=\mbox{Beta}(\alpha+n_1^{\ast},\beta+n-n_1^{\ast})$, and $\tilde{x}_i$ from $f(\tilde{x}_i|\pi^{\ast})=\mbox{Bern}(\pi^{\ast})$ for $i=1,\ldots,n$ to generate one set of synthetic data. The cycle was  repeated 5 times (from sanitizing $n_1$ to drawing $\tilde{\x}$) to obtain 5 sets of synthetic binary data.  The non-DP MS approach generated synthetic data in a similar manner to the MODIPS approach except that there was no perturbation of $n_1$ and $\pi$  was sampled directly from $f(\pi|\x)=\mbox{Beta}(\alpha+n_1,\beta+n-n_1)$, and then $\tilde{x}_i$ was sampled from $f(\tilde{x}_i|\pi)=\mbox{Bern}(\pi)$.  In the Laplace sanitizer, five sets of sanitized binary data  were directly generated per $n_1^{\ast}=n_1+e$ without any distributional assumption or model fitting. 

In both the Laplace sanitizer and the MODIPS, the sanitized $n^*$ could be out of bounds $[0,n]$ as the Laplace noise is drawn from the real line. To legitimize $n_1^*$, we applied truncation (out-of-bounds $n_1^*$ is thrown away and only in-bounds values are kept), and the boundary inflated truncation (setting  $n_1^*<0$ values at 0 and those $>n$ at $n$). Neither post-hoc processing procedures compromised DP as no new information was leaked from the original data (sample size $n$ is assumed to be insensitive information and can be released) \citep{liu2016noninformative}. Bounding $n^*$ at $[0,n]$ in the MODIPS implies that   $\alpha^*=\alpha+n_1^{\ast}\ge0$ and $\beta^*=\beta+n-n_1^{\ast}\ge\beta$ in $f(\pi^{\ast}|n_1^{\ast})=\mbox{Beta}(\alpha+n_1^{\ast},\beta+n-n_1^{\ast})$. A reviewer suggested bounding $n^*$ at $[-\alpha,\beta + n]$, thus $\alpha^*\ge0$ and $\beta^*\ge0$ and a wider range of $\pi^{\ast}$ can be sampled. We compared this truncation scheme with the above two in this simulation and no significant differences were found. 

Both the Binomial-Beta McClure-Reiter synthesizer and the Multinomial-Dirichlet synthesizer simulated data  $\tilde{\x}$ from $\text{Bern}(p^*)$; however, $p^*$ was fixed at $\frac{n_1+\alpha^*}{n+\alpha^*+\beta^*}$ with $\alpha^*=\beta^*=(e^{\epsilon/n}-1)^{-1}$ for the Binomial-Beta McClure-Reiter synthesizer, and was drawn from $f(p^\ast|\alpha^\ast,\beta^\ast)=\text{Beta}(\alpha^*+n_1,\beta^*+n-n_1)$ for the Multinomial-Dirichlet synthesizer with $\alpha^*=\beta^*=n/\left(e^{\epsilon}-1\right)$.  In the Binomial-Beta McClure-Reiter sanitizer, a single synthetic set was released.  In the Multinomial-Dirichlet synthesizer, five synthetic sets were generated, one  per each sanitized $p^\ast$.

\begin{landscape}
\begin{figure}[!htp]
\begin{subfigure}{.5\textwidth}
\includegraphics[width=1.34\linewidth]{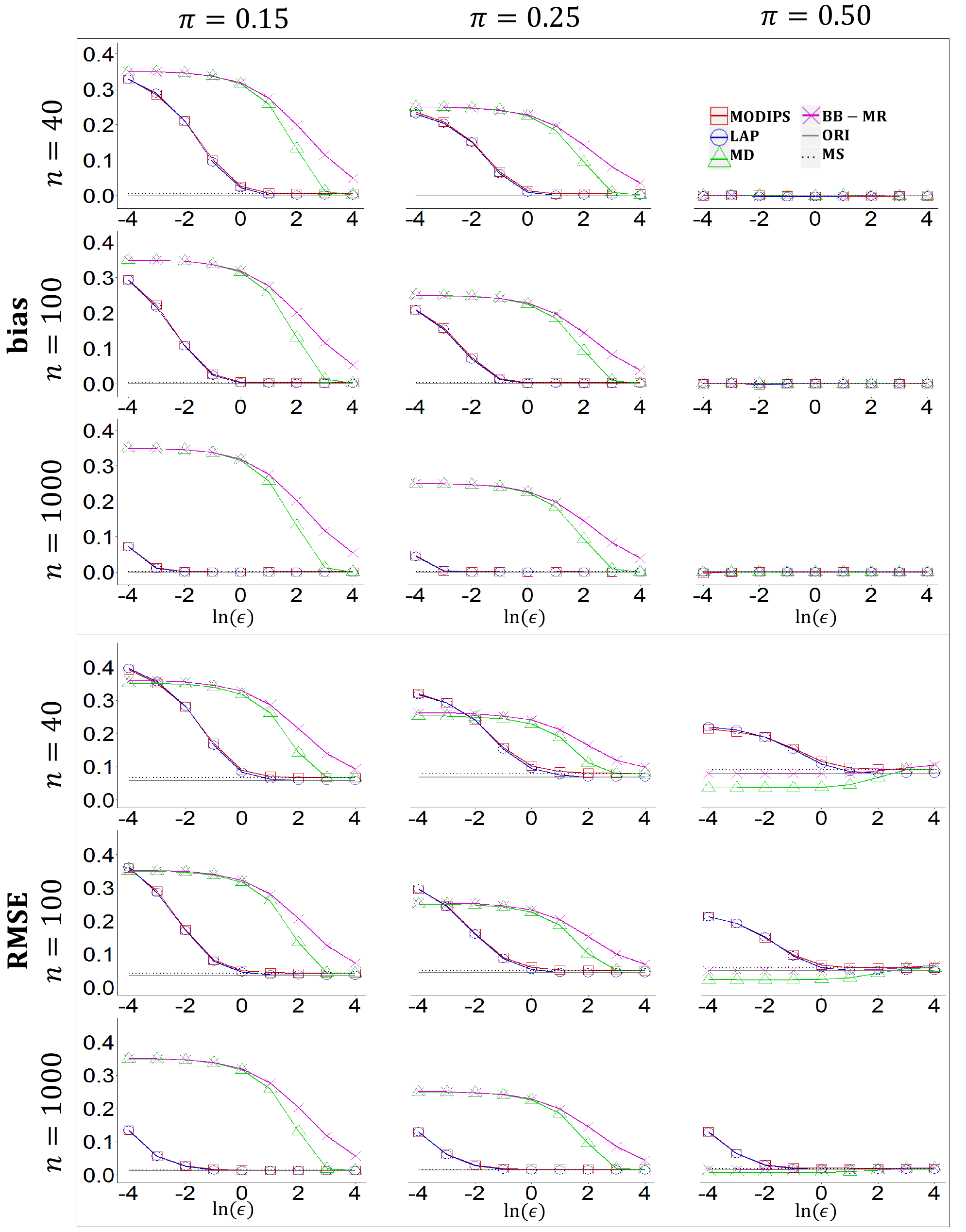} 
\end{subfigure}\hspace{2.5cm}
\begin{subfigure}{.5\textwidth}
\includegraphics[width=1.34\linewidth]{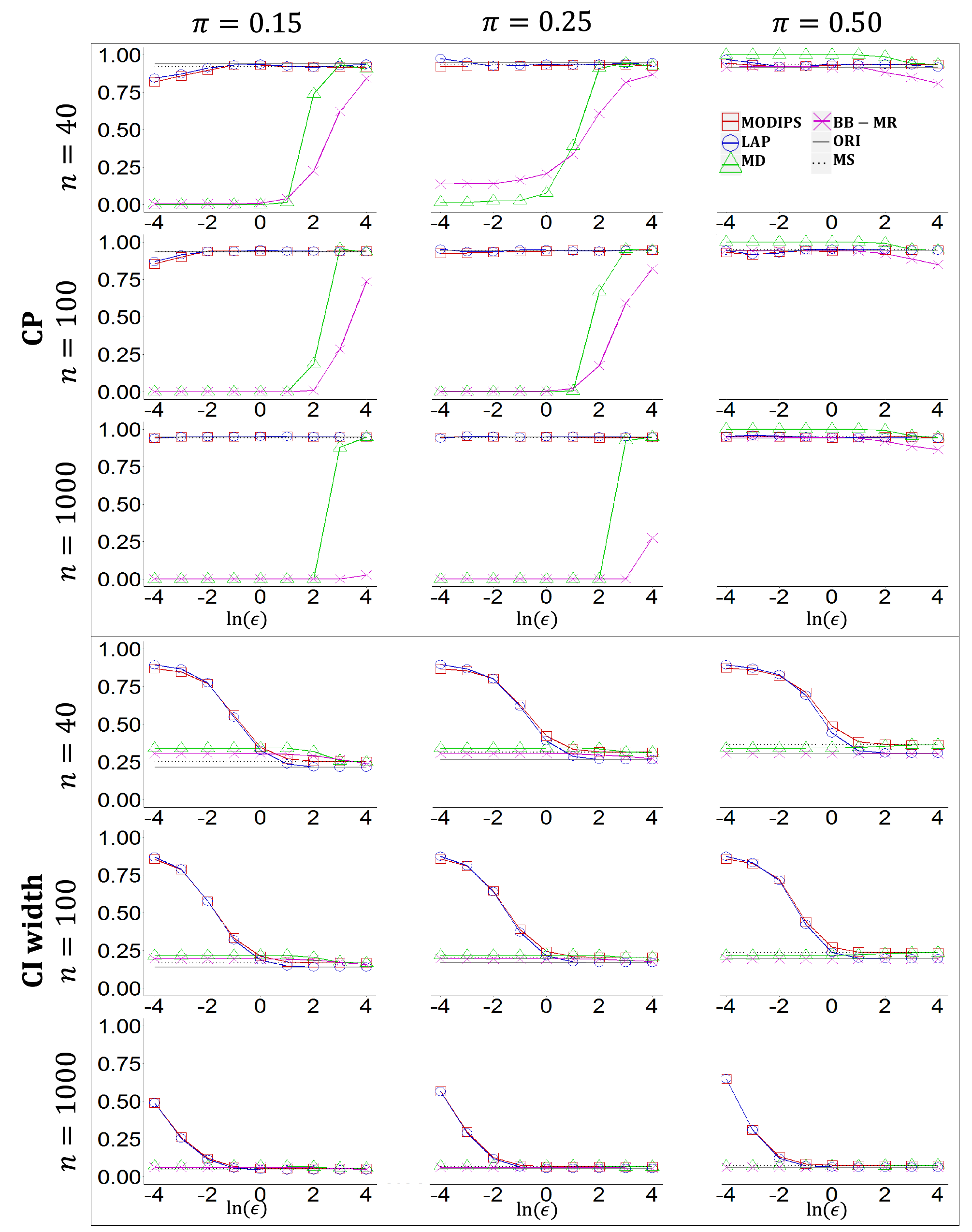}
\end{subfigure}
 \caption{The bias, RMSE,  95\% coverage probability (CP), and 95\% confidence interval (CI) width of $\pi$ in simulation study 1. MODIPS represents the model-based differentially private synthesis, LAP represents the Laplace sanitizer, MD represents the Multinomial-Dirichlet synthesizer, BB-MR represents Binomial-Beta McClure-Reiter synthesizer, Ori is the original results without any perturbation, and MS is the traditional multiple synthesis method without DP.}\label{fig:sim1}
\end{figure}
\end{landscape}

To obtain inferences on $\pi$ from the released data, each of the $5$ sets was analyzed separately in all the above synthesis approach except for the Binomial-Beta McClure-Reiter approach. The point estimate of $\pi$ in the $j$-th ($j=1,...,5$) synthetic data set was the sample proportion $\hat{p}_j$, and its variance was estimated as $v_j=\hat{p}_j (1-\hat{p}_j)n^{-1}$. Eqns (\ref{eqn:mean}) to (\ref{eqn:test}) were then applied to obtain a final estimate of $\hat{p}$ and the associated 95\% confidence interval (CI).  Figure \ref{fig:sim1} depicts the results on the bias and root mean squared error (RMSE), CI width, and the coverage probabilities (CPs) of the 95\% CIs for $\pi$ from each DIPS approach, the non-DP MS approach, and the original data (we present only the results from the boundary inflated truncation post-processing, which were better than the results from the truncation approach).

The results are summarized as follows.  \textbf{1)} The overall performances of the MODIPS and the Laplace synthesizer were similar while those of the Multinomial-Dirichlet and Binomial-Beta McClure-Reiter synthesizers were similar; in general, the inferences in the former two were better than the latter two. \textbf{2) } There was noticeable bias, large RMSE, and some undercoverage especially when $\epsilon <1$ and $n$ was small across all DIPS approaches. The inferences improved as $\epsilon$ increased  (more privacy budget and thus less perturbation), eventually
 approaching the original results for the Multinomial-Dirichlet and Binomial-Beta McClure-Reiter synthesizers, and approaching the non-DP MS results for the MODIPS and the Laplace sanitizer. \textbf{3)} In the MODIPS and Laplace sanitizer, the amount of noise remained constant regardless of $n$; in other words, the noise became less significant for larger $n$.  In the Multinomial-Dirichlet and the Binomial-Beta McClure-Reiter synthesizers, the perturbation introduced through the prior information increased monotonically with $n$. As a result, there was little improvement in inferences with larger $n$, which was a significant drawback for the Multinomial-Dirichlet and the Binomial-Beta McClure-Reiter synthesizers.  \textbf{4)} Since the prior mean of $\pi$ was 0.5 for the Multinomial-Dirichlet and the Binomial-Beta McClure-Reiter synthesizers, when the sample proportion was not 0.5, the two sanitizers would introduce bias into the released data. Therefore,  the inferences were the best when $\pi=0.5$ for the Multinomial-Dirichlet and Binomial-Beta McClure-Reiter synthesizers given the consistency between the prior information and the data. \textbf{5)} For the Laplace sanitizer and the MODIPS approach, the inferences were also the best when $\pi = 0.5$ since $0.5$ was the mid point for the range of a proportion, truncating at 0 or 1 did not skew the distribution of $\pi$  as much as when $\pi$ was close to 0 or 1. 
\textbf{6)} The RMSE values from the MODIPS and Laplace sanitizers were much smaller than those from the Binomial-Beta McClure-Reiter and Multinomial-Dirichlet synthesizes for most $\epsilon$ values when $\pi\!\ne\!0.5$; when $\pi\!=\!0.5$, the Binomial-Beta McClure-Reiter and Multinomial-Dirichlet synthesizers offered smaller RMSE values for small  $\epsilon$; actually, the values were even smaller than the original RMSE values for small $\epsilon$ values and decreased when there was more perturbation ($\epsilon$ decreased). Again, this was due to the consistency of the prior information and the data when $\pi\!=\!0.5$. As $\epsilon$ decreased, the prior in the Multinomial-Dirichlet and Binomial-Beta McClure-Reiter priors became more ``informative'', and injected more ``useful'' prior information about $\pi$ that was consistent with the data, leading to smaller RMSE. \textbf{ 7)} The MODIPS and Laplace sanitizers produced  close-to-nominal  coverage (0.95) across all the $n$ and $\pi$ values, except for some undercoverage at small $\epsilon$ and $n$ due the relatively large bias with the truncation at 0 and 1 for sanitized proportions.  Eventually all CPs  converged to the nominal level as $\epsilon$ increased in all the sanitizers except for the Binomial-Beta McClure-Reiter synthesizer. \textbf{8)}  On the other hand, the CIs for the Laplace and the MODIPS sanitizers were much wider when $\epsilon<e^{-1}$ than those from the Binomial-Beta McClure-Reiter and Multinomial-Dirichlet synthesizes. 

\subsection{Simulation Study 2: Continuous Data}\label{sec:cont}
The following methods are compared in this simulation study: the MODIPS synthesizer, the NP-DIPS synthesizers via the perturbed histogram and the smoothed histogram approaches.  Data was simulated from $\mbox{N}(\mu,\sigma^2)$. We manually truncated the simulated data at  bounds $\left[c_0=\mu-3\sigma,c_1=\mu+4\sigma\right]$ around $\mu$ to generate bounded data so that global sensitivity for the sample mean and variances are finite and calculable.  Since there was minimal probability mass ($0.0013$) outside the $[\mu-3\sigma,\mu+4\sigma]$, the normal assumption was hardly affected with the truncation (note that the bounds we used are asymmetric around the true $\mu$, which is  more representative of real life data than symmetric bounds). We also examined symmetric bounds, but present the results in the Supplementary materials. We examined  9 simulation scenarios for $n=\{20,100,1000\}$ and $\sigma^2=\{1,4,9\}$, with 5000 repetitions per scenario. Without loss of generality, $\mu$ was set to 0 in all scenarios.

With the obvious choice of the likelihood given the simplicity of the data, we did not perform model selection and validation in this simulation. Let the prior be $f(\mu,\sigma^2)\propto \sigma^{-2}$,  the posterior distributions were $f(\sigma^2|\x)=\mbox{Inv-Gamma}\left[(n-1)/2, (n-1)S^2/2\right]$ and $f(\mu|\x,\sigma^2)=\mbox{N}(\bar{x},n^{-1}\sigma^2)$, where $\bar{x}$ and $S^2$  were the sample mean and variance, respectively. In the non-DP MS,  a synthetic set was generated by first drawing $\sigma^2$ and $\mu$ from their posterior distributions, and then drawing $\tilde{\x}$ from the normal distribution given the drawn $\mu$ and $\sigma^2$. The process repeated $5$ times to generate 5 sets of synthetic data to release.

The MODIPS procedure started with sanitizing sufficient statistics $\s$ via the Laplace  mechanism, which were, in the posterior distribution $f(\mu,\sigma^2|\x)$, $\s= (\bar{x},S^2)$.  To calculate the global sensitivity for $\bar{x}$ and $S^2$, we would need the global bounds of $X$. We assumed the bounds of the data were publicly known knowledge, which is a realistic assumption in general. It is very likely an attribute in a data set was not studied previously, and thus its bounds are  known (e.g., human height, income, or published biomarkers, etc).  A reviewer questioned the possible conservativeness of the bounds. If the bounds are conservative compared to the local data, then it would not be a concern as DP protects against the worst case scenario and the global bounds are what is needed instead of the local data. If the bounds are conservative at the global level, this implies there is insufficient information on the attribute. In this case, it would be better to be conservative rather than not from a privacy protection perspective though it means more than necessary noises are injected. Future studies are expected to help gain more understanding on the attribute and tighten the bounds. Note that using the local bounds directly would violate privacy even if one is willing to spend some privacy budget to perturb the bounds before using them. However, how to perturb the minimum and maximum can be difficult without knowing the global bounds in the first place.

The global sensitivity was $(c_1-c_0)n^{-1}$ for $\bar{x}$  and $(c_1-c_0)^2n^{-1}$ for $S^2$, where $(c_1-c_0)=7\sigma$ \citep{liu2016model}. Since the data was bounded, so were $\bar{x}$ and  $S^2$. Specifically, the bounds for $\bar{x}$ were $[c_0,c_1]$, and that of $S^2$ were $\left[0,(c_1-c_0)^2/4 \cdot n/(n-1)\right]$ \citep{macleod1984bounds}. If a sanitized statistic was outside its range, it was post-processed by the boundary inflated truncation procedure. Given the sanitized $\s^*=\{\bar{x}^*,S^{2\ast}\}$, the MODIPS technique drew $\sigma^{2\ast}$ from $\mbox{Inv-Gamma}\left[(n-1)/2, (n-1)S^{2\ast}/2\right]$ and $\mu^{\ast}$ from $\mbox{N}(\bar{x}^{\ast},n^{-1}\sigma^{2\ast})$. Finally,  $\tilde{x}^*_i$ was simulated from $\mbox{N}(\mu^{\ast},\sigma^{2\ast})$ for  $i=1,\ldots,n$ to generate one synthetic set. The whole procedure was repeated $5$ times to generate $5$ surrogate data sets. $\epsilon/5$ of the total budget was spent per synthesis. In addition, since there were two statistics, $(\bar{x},S^2)$, to sanitize over the same set of data, the  $\epsilon/5$ budget per synthesis was further split in half between the sanitization of $\bar{x}$ and $S^2$. 

In deciding the number of bins for the histograms for the perturbed and smoothed histogram approaches,  we applied the Scott's Rule after comparing it with the Sturges' rule and the Freedman-Diaconis rule \citep{scott2015multivariate}. Specifically, the bin width was set at $\hat{h}=3.5Sn^{-1/3}$, where $S$ was the sample standard deviation of $\x$ and $n$ was the sample size. The median number of bins was 7, 10, and 21 for $n=20,100,$ and 1000, respectively, across all simulations (Table \ref{tab:histstat} in Supplemental Materials). In the perturbed histogram,  all bin counts were perturbed via the Laplace mechanism with $\Delta_{\s}=1$ to obtain the perturbed density histogram (Eqn \ref{eqn:density2}). The procedure was repeated 5 times to obtained 5 sets of differentially private $\hat{\mathbf{p}}^*_j$ (the perturbed bin counts), based on the 5 sets of synthetic data that were simulated. For the smoothed histogram, we first calculated $\lambda$ for a given $\epsilon$ using Eqn (\ref{eqn:smooth.lambda}), and then constructed the smoothed histogram by applying Eqn (\ref{eqn:smooth.hist}), from which a single set of synthetic data was generated and released.

To obtain the inference on $\mu$ and $\sigma^2$ from the multiple released data sets via the MODIPS, the perturbed histogram sanitizers, and the non-DP MS approach, each synthetic set $j$ was analyzed  to obtain point estimates of $\mu$ and $\sigma^2$, which were $\bar{x}_j$ (the sample mean) and $s^2_j$ (the sample variance), respectively; the associated within-set variance estimates were $s_j^2/n$ and $\left(s^2_j)^2(2(n-1)^{-1}\!+\! \kappa_j n^{-1}\right)$, respectively, where $\kappa_j$ was the excess kurtosis in the $j\textsuperscript{th}$   set.  Eqns (\ref{eqn:mean}) to (\ref{eqn:test}) were then applied to obtain the final estimates and 95\% CIs.

\begin{landscape}
\begin{figure}
\begin{subfigure}{.5\textwidth}
\includegraphics[width=1.35\linewidth]{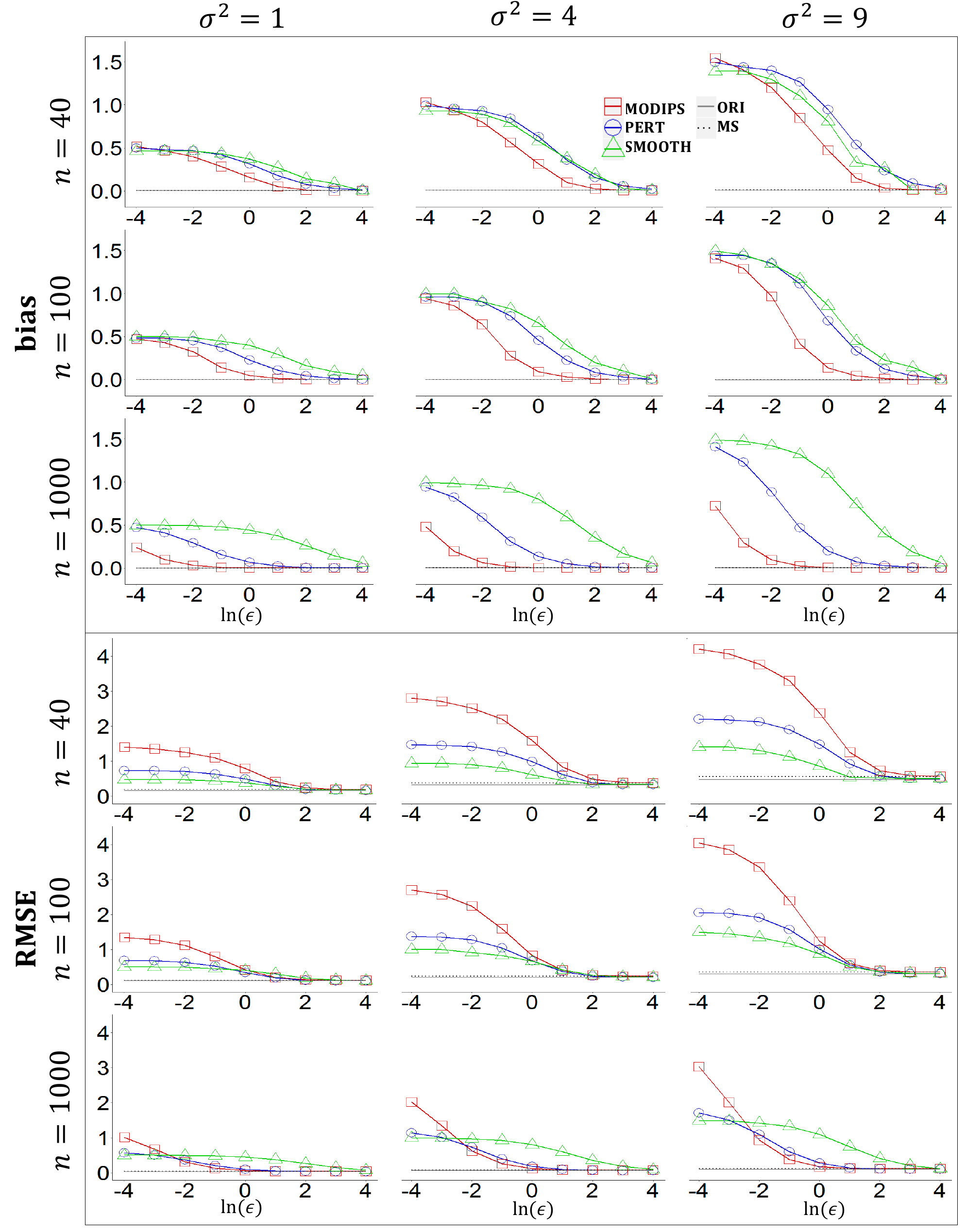}
\end{subfigure}\hspace{2.5cm}
\begin{subfigure}{.5\textwidth}
\includegraphics[width=1.38\linewidth]{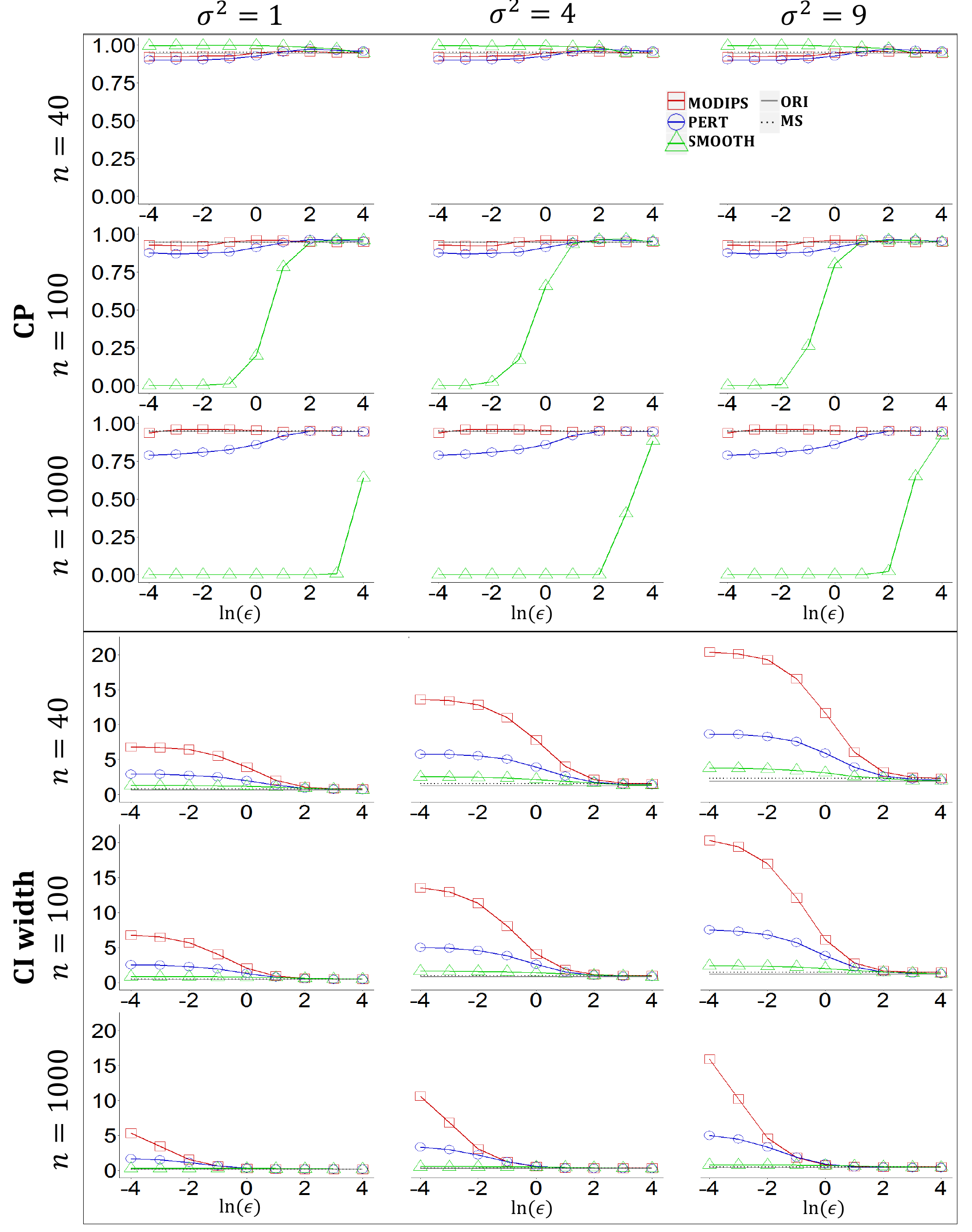}
\end{subfigure}
 \caption{The bias,  RMSE, 95\% CP and 95\% CI width of $\mu$ in simulation study 2. MODIPS represents the model-based differentially data private synthesis, PERT represents the perturbed histogram method, SMOOTH represents the smoothed histogram method, MS is the traditional multiple synthesis method without DP, and Ori is the original results without any perturbation.}\label{fig:sim2mu}
\end{figure}
\end{landscape}

\begin{landscape}
\begin{figure}
\begin{subfigure}{.5\textwidth}
\includegraphics[width=1.38\linewidth]{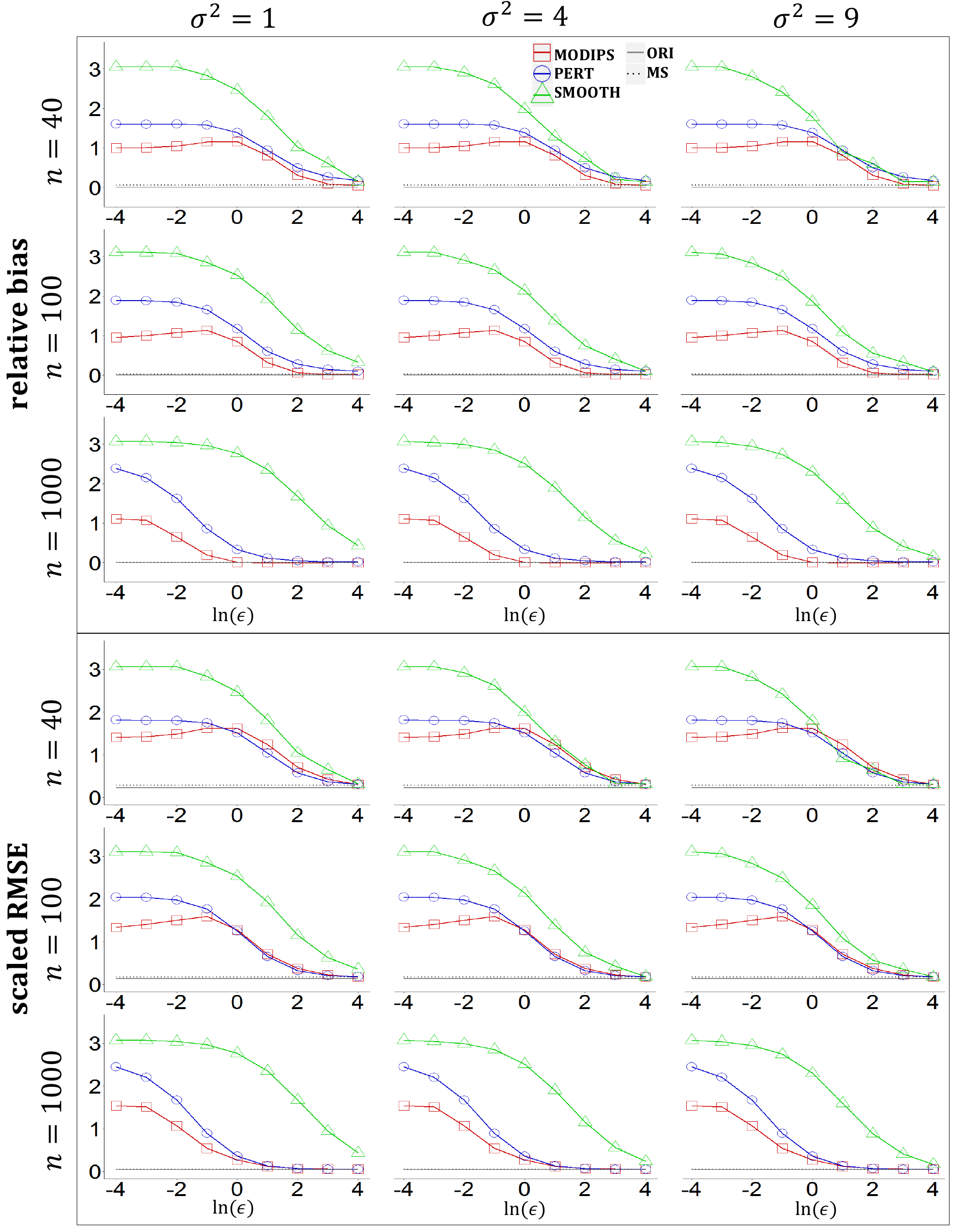} 
\end{subfigure}\hspace{2.75cm}
\begin{subfigure}{.5\textwidth}
\includegraphics[width=1.4\linewidth]{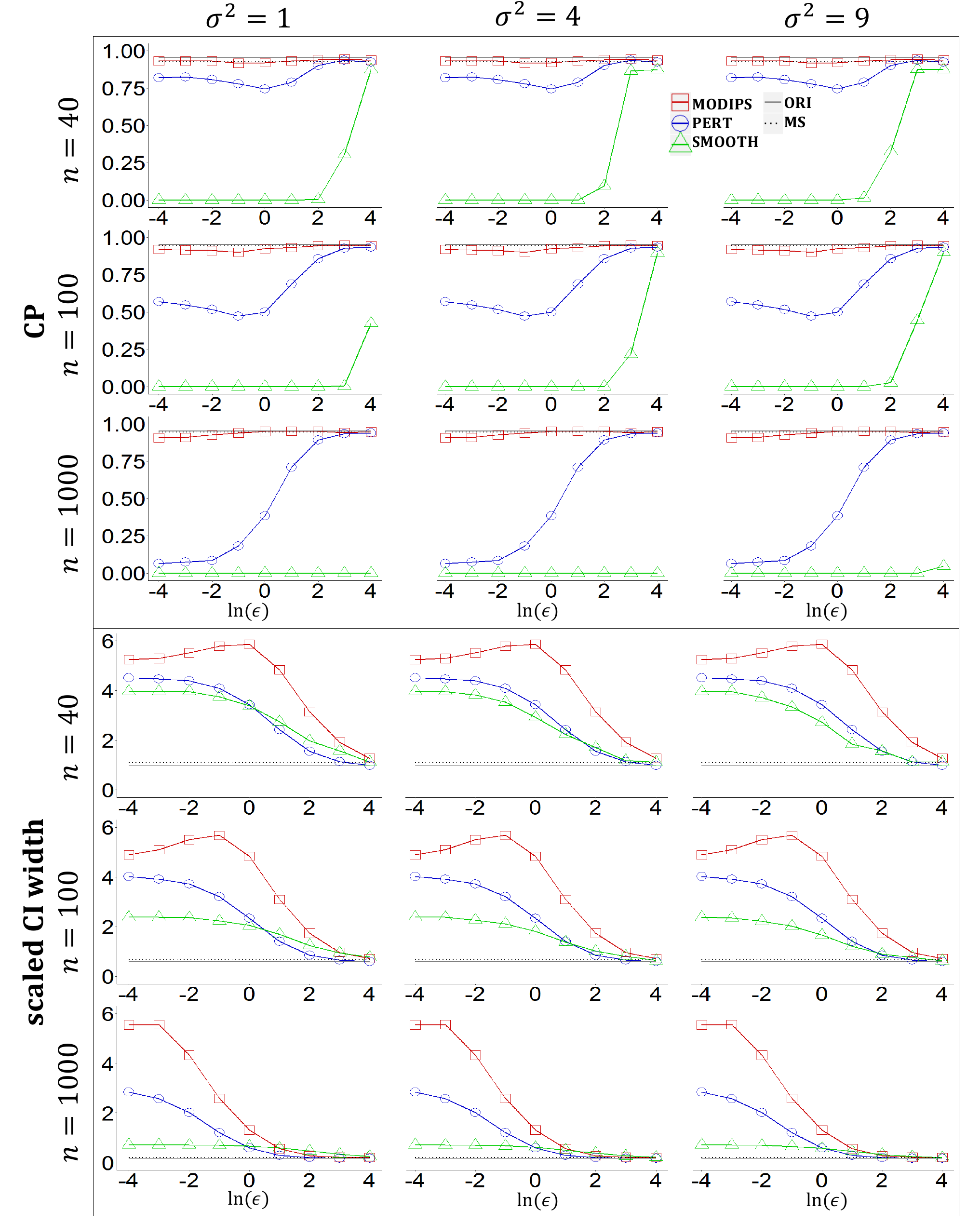}
\end{subfigure}
 \caption{The bias,  RMSE, 95\% CP and 95\% CI width of $\sigma^2$ in simulation study 2. MODIPS represents the model-based differentially data private synthesis, PERT represents the perturbed histogram method, SMOOTH represents the smoothed histogram method, MS is the traditional multiple synthesis method without DP, and Ori is the original results without any perturbation.} \label{fig:sim2sig}
\end{figure}
\end{landscape}

Figures \ref{fig:sim2mu} and \ref{fig:sim2sig} depict the bias, RMSE, 95\% CI width, and the  CP of the 95\% CI for $\mu$ and $\sigma^2$ based on the synthetic data of $\mu$ and $\sigma^2$ based on the synthetic data via the 3 DIPS approaches, respectively. For the purposes of comparability across different values of $\sigma^2$, the bias, RMSE, and CI width for  $\sigma^2$ were scaled by the true $\sigma^2$, referred to as the relative bias, scaled RMSE and scaled CI width, respectively.

The results are summarized as follows. \textbf{1)} For all approaches, there were some noticeable biases and large RMSE at small $\epsilon$ for both $\mu$ and $\sigma^2$, which improved as $\epsilon$ increased and eventually approached the original or the non-DP MS results.  Overall, the perturbed histogram seemed to offer the best trade-off between bias and variance for the inferences based on the synthetic data. \textbf{2)}  For the MODIPS and the perturbed histogram approaches, the amount of injected noise became immaterial as $n$ increased, and the inferences improved. In the the smoothed histogram,  $\lambda$ in Eqn (\ref{eqn:smooth.lambda})  got larger  and approached $K/(K+\epsilon)$ as $n$ increased. As a result, increasing $n$ did not help the inferences in the smoothed histogram. \textbf{3)}   The positive bias in $\mu$ can be explained by the asymmetric bounds  $[\mu-3\sigma, \mu+4\sigma]$ of data $\x$ around $\mu$. When sanitized $\bar{x}^*$ or synthesized data were out of bound, they were set at the boundary values per the boundary inflated truncation procedure. Since the left bound $\mu-3\sigma$ was closer to $\mu$, there were more values at $\mu-3\sigma$ than at $\mu+4\sigma$, resulting in overestimation. The observed positive bias in $\sigma^2$ was expected due to the randomness introduced through synthesis and sanitization. \textbf{4)} In terms of RMSE,  the histogram-based approaches produced smaller RMSE for $\mu$ than the MODIPS for most of $n$ and $\epsilon<1$, but the situation switched for $\sigma^2$ with the smallest RMSE coming from the MODIPS. \textbf{5)}  In terms of CP, the MODIPS produced close-to-nominal level coverage in all examined scenarios for both $\mu$ and $\sigma^2$ at the cost of wide CIs for $\epsilon<1$; the perturbed histogram had moderate to mild undercoverage with much narrower CIs; and the smooth histogram had unacceptable severe undercoverage at small $\epsilon$ for large $n$.

The results when the data bounds $[\mu-4\sigma, \mu+4\sigma]$ were symmetric around the true mean are presented in Figures \ref{fig:sim2mu2} and \ref{fig:sim2sig2_cp} in the Supplemental Materials. As expected, there were minimal biases on $\mu$ in all the DIPS approaches (there was some fluctuation in MODIPS for small $\epsilon$), and the CP in all approaches were at nominal-level.  The histogram-based approaches delivered more precise estimates than MODIPS in the inferences of $\mu$ (smaller RMSE and narrower CIs). However, the histogram-based approaches did not perform as well as MODIPS in the inferences of $\sigma^2$. Both the bias and RMSE were large and there was severe undercoverage at small values of $\epsilon$.

\subsection{Simulation Studies 3 and 4: Mixture of Continuous and Categorical Data}\label{sec:mixture}
In simulation studies 3 and 4, we compared the MODIPS synthesizer and the NP-DIPS synthesizer in data with mixed Gaussian variables $\x$ and categorical variables $\w$.  The data were generated from the GLOM (\textbf{G}eneral \textbf{LO}cation \textbf{M}odel)  based on $f(\x|\w)f(\w)$ in simulation 3, and from the SLOMAG model  (\textbf{S}equential \textbf{LO}gistic regression with \textbf{MA}rginal \textbf{G}aussian distribution) $f(\w|\x)f(\x)$ in simulation 4. We also investigated the impact of mis-specification of the synthesis models.   For NP-DIPS in both simulations, we applied the Laplace synthesizer on $\w$ coupled with the perturbed histogram for $\x$. We did not implement the Multinomial-Dirichlet synthesis or the smoothed histogram approach given their inferior performances to the the Laplace sanitizer, the perturbed histogram, and the MODIPS based on the results from simulation studies 1 and 2.

\subsubsection{Simulation Study 3: GLOM Model}\label{sec:ggm}
Data $\x$ comprised three categorical variables $\mathbf{w}=(\w_1,\w_2,\w_3)$ with 2, 3, and 4 levels, respectively, and  continuous variables $\mathbf{z}$.  Let $n_k$ denote the count in cell $k$ in the full cross-tabulation of $\mathbf{w}$ for $k=1,...,24$. First, the counts $\mathbf{n}=\{n_k\}$ in the 24 cells were simulated from a Multinomial distribution with parameter $\bs{\pi}=\{\pi_k\}$ (which are summarized in Table \ref{tab:sumstatcell} from the Supplemental Material); $\z_{ik,}=(z_{ik1},z_{ik2})'$ was then simulated from $\text{N}_{(2)}(\bs{\mu}_k,\Sigma)$ for $i=1,\ldots,n_k$ and $k=1,...,24$, where $\bs{\mu}_k=(\mu_{k1},\mu_{k2})'$ was the mean of $\z$ in cell $k$, and $\Sigma$ was the covariance matrix that was set to be the same across all 24 cells. The summary of the parameter values of $\bs{\mu}_{1}$, $\bs{\mu}_{2}$, $\bs{\pi}$ across the 24 cells are provided in Table \ref{tab:sumstat} in the Supplemental Materials. We set $n=1000$, the variances of $z_{ik1}$ and $z_{ik2}$, $\sigma^2_1=\sigma^2_2=1$, and their correlation  at $\rho=0.50$ with 5000 repetitions.  $z_{ikj}$  in cell $k$ (where $j=1,2$) was truncated at $[c_{0,kj}=\mu_{kj}-4\sigma_j,c_{1,kj}=\mu_{kj}+4\sigma_j]$ to generate bounded data. Additionally, in the Supplemental Materials, Table \ref{tab:sumstatcell} depicts the summary statistics for the number of observations in the 24 cells across the 5000 repetitions.

For the non-DP MS approach, we imposed Dirichlet prior $\bs{\pi}$, $f(\bs{\pi})=\text{D}(\bs{\alpha})$, where $\bs{\alpha}=\{\alpha_1,...,\alpha_{24}\}= 1/2$, and priors $f(\bs{\mu}_1,\ldots,\bs{\mu}_{24}, \Sigma)\propto |\Sigma^{-1}|$ . The posterior distributions were $f(\bs{\pi}|\w)=\text{D}(\bs{\alpha}')$, $f(\Sigma|\z,\w) =\text{Inv-Wishart}(n-K, \mathbf{S})$, and $f(\bs{\mu}_k|\Sigma,\z,\w)\!=\!\text{N}_{(2)}(\bar{\mathbf{z}}_{k}, n^{-1}_k\Sigma)$,  where $\bs{\alpha}'\!=\!\bs{\alpha}+\bs{n}$,
$\mathbf{S}\!=\!n^{-1}\!\sum^K_{k=1}\sum_{i=1}^{n_k}(\mathbf{z}_{ik}-\bar{\mathbf{z}}_k)(\bs{\mathbf{z}}_{ik}-\bar{\mathbf{z}}_k)'$, and $\bar{\z}_k$ contained the sample means of $\z$ in cell $k$. Synthetic data were simulated from the posterior predictive distribution $f(\tilde{\mathbf{z}}_i,\tilde{\w}_i|\z,\w)$
 by  first drawing
$\bs{\pi}\sim f(\bs{\pi}|\w)$ $=D(\bs{\alpha}+\bs{n})$,
$\Sigma$ from $f(\Sigma|\z,\w)=\text{Inv-Wishart}(n-K,\hat{\Sigma})$, and
$\bs{\mu}_k$  from $f(\bs{\mu}_k|\Sigma,\z,\w)=\text{N}_{(2)}(\bar{\mathbf{z}}_k, n^{-1}_k\Sigma)$; then sampling
$\tilde{\w}$ from $f(\tilde{\w}|\bs{\pi})\!=\!\text{Multinom}(n, \bs{\pi})$, and
$\tilde{\mathbf{z}}_i$ from $f(\mathbf{z}_i|{\tilde{\w}_i},\Sigma)\!=\!\text{N}_{(2)}(\bs{\mu}_k,\Sigma)$ for $i\!=\!1,...,\tilde{n}_k$, where $\tilde{n}_k$ was the count in cell $k$ based on the synthesized $\tilde{\mathbf{w}}$. The drawing process repeated 5 times to generate 5 synthetic sets.

The Bayesian sufficient statistics from the above Bayesian model was $\s =(\mathbf{n}, S, \bar{\mathbf{z}})$; $\bar{\mathbf{z}}$ contained the 24 pairs of cell  means of $\z_1$ and $\z_2$. The MODIPS procedure started with sanitizing $\s$ via the Laplace mechanism to obtain $\s^\ast=(\mathbf{n}^*, S^*, \bar{\mathbf{z}}^*)$ (the  $l_1$ global sensitivity was $1$ for $\mathbf{n}$, $(c_{1,kj}-c_{0,kj})n_k^{-1}$ for $\bar{z}_{kj}$,  and $(c_{1,kj}-c_{0,kj})^2(n-1)(n(n-K))^{-1}$ for each entry in $S$  \citep{liu2016noninformative}, where $c_{1,kj}-c_{0,kj}=8\sigma$ for $k=1,\ldots,24$ and $j=1,2$).  Given $\s^\ast$, the MODIPS method first drew $\bs{\pi}^\ast$ from $f(\bs{\pi}^\ast|\mathbf{n}^\ast)=D(\bs{\alpha}+\mathbf{n}^\ast)$, $\tilde{\w}^\ast$ from $f(\tilde{\w}^\ast|\bs{\pi}^\ast)=\text{Multinom}(n,\bs{\pi}^\ast)$, $\Sigma^*$ from $f(\Sigma^*|\mathbf{S}^*)=\text{Inv-Wishart}(n-K,\mathbf{S}^\ast)$, $\bs{\mu}^*_k$  from $f(\bs{\mu}^*_k|\Sigma^\ast,\bar{\mathbf{z}}^*,\w)=\text{N}(\bs{\bar{z}}^{\ast}_k, n^{-1}_k\Sigma^\ast)$; and then  $\tilde{\mathbf{z}}_{i}$ was simulated from $f(\mathbf{z}_{i}|\bs{\mu}^{\ast}_{\tilde{k}},\Sigma^\ast)=\text{N}(\bs{\mu}^{\ast}_{\tilde{k}},\Sigma^\ast)$ for $i=1,...,,\tilde{n}^*_k$ to generate one set of surrogate data, where $\tilde{n}^*_k$ was the count in cell $k$ based on the synthesized $\tilde{\mathbf{w}}^*$,  and $\tilde{k}$ indicates the cell which the simulated case $i$ belonged to given the synthesized $\tilde{w}^*_i$. The procedure was repeated $5$ times to generate 5 synthetic sets with $\epsilon/5$ privacy budget each. Since $\s$ contained 6 components: $\mathbf{n}$, $\mathbf{\bar{z}}_1$, $\mathbf{\bar{z}}_2$,  two variance terms and one covariance term from $S$, each received $1/6$ of $\epsilon/5$ budget allocated to each synthesis (there was no need to split $\epsilon/30$ further among the 24 elements in $\mathbf{n}$ per the parallel composition as they were formed over  non-overlapping subsets; similar for $\mathbf{\bar{z}}_1$, $\mathbf{\bar{z}}_2$, respectively).

In the NP-DIPS approach, we applied the Laplace sanitizer to sanitize 24 cell counts $\mathbf{n}$ formed by the full cross-tabulation of $\w$, and the perturbed histogram method to sanitize continuous $\mathbf{z}$ within each of the 24 cells. Since $\mathbf{z}$ was 2-dimensional, each bin of the histogram of $\z$ was a square rather than an interval. The number of bins were determined using the Scott's rule,  and the medians ranged from 16 to 49 across the 5000 repeats in the 24 cells (Supplemental Materials Table \ref{tab:histstatsim3}). The process was repeated 5 times to create 5 sets of sanitized $\tilde{\mathbf{n}}$ and 24 perturbed histograms, from which 5 sets of synthetic data were generated. Each synthesis was allocated $1/5$ of the total privacy budget, which was further split between sanitizing the 24 cells formed by $\mathbf{w}$ and sanitizing the histogram formed by $\mathbf{z}$ in a 1:1 ratio.

We examined the inferences on $\bs{\mu}_{1}, \bs{\mu}_{2}$, $\sigma_1^2,\sigma_2^2,\rho$, and the marginal probabilities $\bs{\Pi}=\{\Pr(w_1=1),\Pr(w_2=1),\Pr(w_2=2),\Pr(w_3=1),\Pr(w_3=2),\Pr(w_3=3)\}$ based on the synthetic data sets. In each synthetic set $l$ ($l=1,\ldots,5$), $\bs{\Pi}$ was estimated by the corresponding sample marginal probability $\hat{\mathbf{P}}_l$;  $\bs{\mu}_{1}$ and $\bs{\mu}_{2}$ were estimated by the sample cell means $\bar{\z}_{1,l}$ and $\bar{\z}_{2,l}$; and $\Sigma$ was estimated by the pooled variance-covariance $S_l$. The within-set variance was estimated by $\hat{\mathbf{P}}_l(1-\hat{\mathbf{P}}_l)n^{-1}$ for $\hat{\mathbf{P}}_l$, $S^2_{j,l}\bs{n}^{-1}$ for $\bar{\z}_j$ ($j=1,2$), $\left(S^2_{j,l})^2(2(n-1)^{-1} + \kappa_l n^{-1}\right)$  for $S_{k,l}^2$, and $(1 - r_l^2)(n - 2)^{-1}$ for the correlation between $Z_1$ and $Z_2$, respectively, where $S^2_{1,l}$ and $S^2_{2,l}$ were the diagonal elements of $S_l$, $\kappa_l$ was the excess kurtosis, and $r_l$ was derived from $S_l$. Eqns (\ref{eqn:mean}) to (\ref{eqn:test}) were then applied to obtain the final estimates of the parameters and the 95\% CIs.

\begin{figure}[!htb]
\makebox[\textwidth][c]{\includegraphics[width=6.2in]{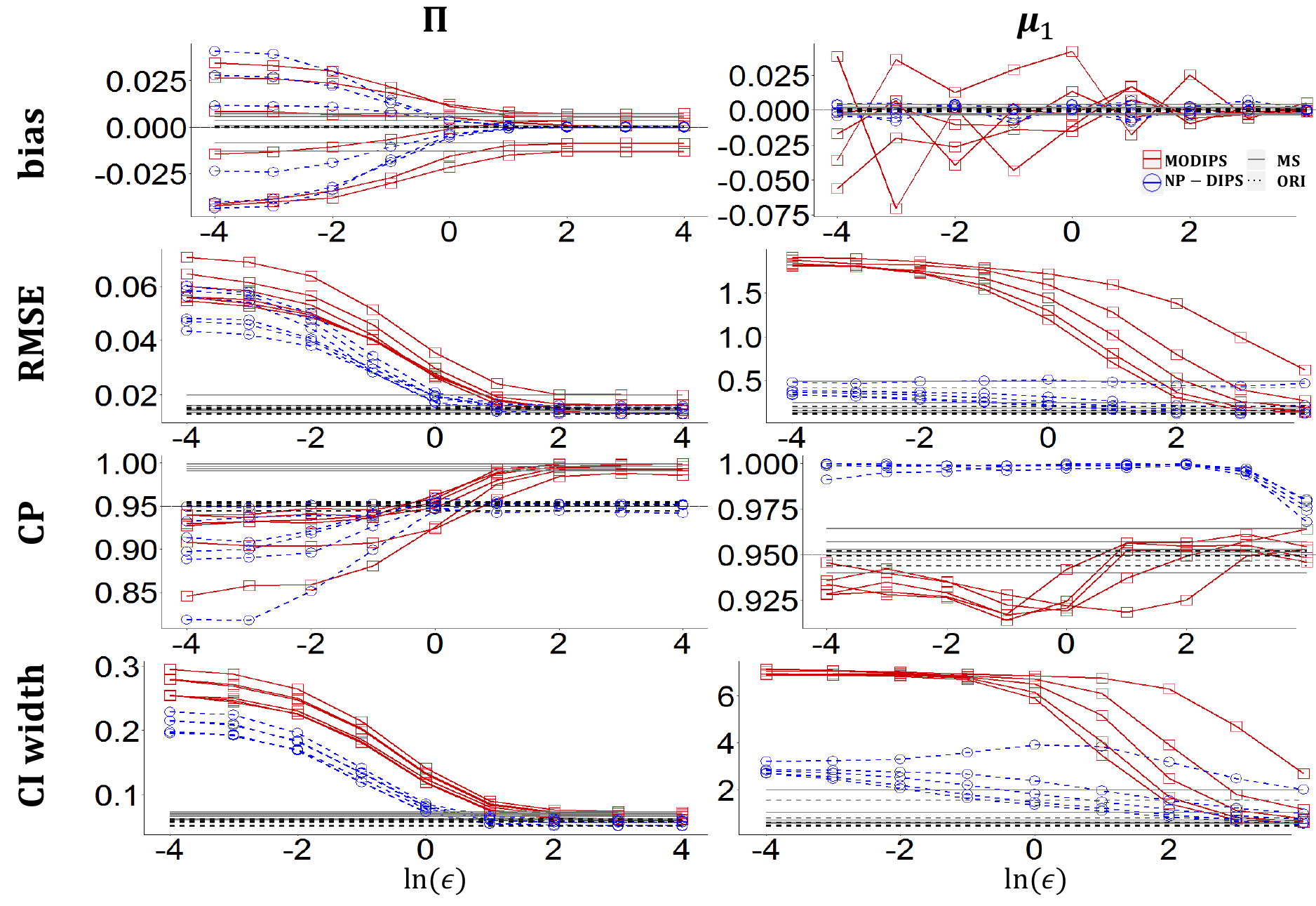}}
 \caption{The bias, RMSE, CP, and 95\%  CI width of $\bs{\Pi}$ and $\bs{\mu}_1$ in simulation study 3. In the plot of $\bs{\Pi}$, each line presents a different marginal probability.  In the plot of $\bs{\mu}_1$, the lines represent the min, med, max, Q1, Q3 of the true 24 cell means, respectively. MODIPS represents the model-based differentially private data synthesis, NP-DIPS represents the Laplace sanitizer+perturbed histogram method, MS is the traditional MS method without DP, and Ori is the original results without perturbation.}
\label{fig:sim3muprop}
\end{figure}

Figure \ref{fig:sim3muprop} shows the results on the bias, RMSE,  CP, and the 95\% CI width of $\bs{\mu}_1$ and $\bs{\Pi}$. The results on  $\bs{\mu}_2,\sigma_1^2,\sigma_2^2,\rho$ are provided in Figures \ref{fig:sim3mu2rho} and \ref{fig:sim3sigma} in the Supplemental Materials. The results are summarized as follows.  \textbf{1)}  NP-DIPS performed better than MODIPS in the inferences of $\bs{\mu}_1,\bs{\mu}_2$ and $\bs{\Pi}$ with smaller bias, similar or smaller RMSE, closer-to-nominal coverage and slightly narrower CIs for $\epsilon > e^{-2}$.  \textbf{2)} On the other hand, the MODIPS  outperformed the NP-DIPS approach in the inferences on $\sigma_1^2,\sigma_2^2,\rho$ with much smaller bias and RMSE when $\epsilon>e^{-1}$ and delivered nominal CP with reasonable CI widths when $\epsilon>1$; the NP-DIPS approach experienced severe undercoverage in all 3 variance/covariance components and never reached the nominal level of 95\% at all level of $\epsilon$. The severe undercoverage in the NP-DIPS at large $\epsilon$ (where the injected noise is supposed to be small) is due to the discretization in forming the histogram bins. The performance of the MODIPS was based on the correct specification of the synthesis model (the GLOM). Mis-specification of the synthesis model is expected to lead to worse results, which we will explore in simulation 4.

\subsubsection{Simulation study 4: SLOMAG Model}\label{sec:lgm}
In this simulation, we first simulated $\z$  from the bivariate normal distribution $f(\mathbf{Z})=\text{N}_{(2)}(\bs{\mu},\Sigma)$ and then generated the categorical variables $\w$ from a sequence of logistic regression models. We set  $\bs{\mu} =(\mu_{1},\mu_{2})' =\mathbf{0}$, the variances of $Z_1$ and $Z_2$ at $\sigma^2_1=\sigma^2_2=1$, and their correlation at $\rho=0.50$. $Z_j$ ($j=1,2$) was truncated at $[c_{0j}=\mu_j-4\sigma_j,c_{1j}=\mu_j+4\sigma_j]$ to generate bounded data. $\w$ contains 3 categorical variables $W_1,W_2,W_3$ with 2, 2, and 3 levels, respectively, and was generated from   $W_{1}|Z_1,Z_2\!\sim\text{Bern}(\pi_{1})$ with $\pi_{1}=e^{(1,Z_{1},Z_{2})\bs{\beta}_1}(1+e^{(1,Z_{1},Z_{2})\bs{\beta}_1})^{-1}$, $W_{2}|Z_1,Z_2,W_1\sim\text{Bern}(\pi_{2})$ with $\pi_{2}=e^{(1,Z_{1},Z_{2},W_{1})\bs{\beta}_2}(1+e^{(1,Z_1,Z_2,W_1)\bs{\beta}_2})^{-1}$, and $W_3|Z_1,Z_2,W_1,W_2\!\sim\!\text{Multinom}(1, (\pi_{31},\pi_{32},\pi_{33}))$, where $\pi_{31}=(1+A+B)^{-1},\pi_{32}\!=\!A(1+A+B)^{-1},\pi_{33}\!=\!B(1+A+B)^{-1}$, $A\!=\!e^{(1,Z_{1},Z_{2},W_{1},W_{2})\bs{\beta}_3}$, $B\!=\!e^{(1,Z_1,Z_2,W_1,W_2)\bs{\beta}_4}$; and $\bs{\beta}_1=(\beta_{01},\beta_{11},\beta_{21})' = (-1,0.5,-1)'$, $\bs{\beta}_2=(\beta_{02},\beta_{12},\beta_{22},\beta_{32})'= (-2,\!-1,1.5,0.5)'$, $\bs{\beta}_3  =  (\beta_{03},\beta_{13},\\ \beta_{23}, \beta_{33},\beta_{43})'=(0,\!-2.5,1,0.5,0.4)'$, and $\bs{\beta}_4=(\beta_{04},\beta_{14}, \beta_{24},\beta_{34},\beta_{44})'=(0.1,-1,-0.5,0,1.5)'$. We ran 1000 repetitions, each sized at $n=1000$. 

The implementation of the NP-DIPS approach was straightforward. $\z$ was first discretized via the Scott's rule to form a 2-way histogram (the number of histogram bins are presented in Table \ref{tab:histstatsim4}), which was then combined with  $(\w_1, \w_2,\w_3)$ to form a 5-way cross-tabulation. The counts from which were sanitized via the Laplace sanitizer with global sensitivity is $1$, from which 5 sets of synthetic data were generated.

The non-DP MS and the MODIPS and reply on the specification on a synthesis model. In the case of the non-DP MS, an appropriate model can be identified without having to worry about privacy costs. For MODIPS, if the identification of a suitable model is based on previous knowledge and common practice, then no privacy is needed to be spent; however, if the model selection procedure is based on the the data to be released, then the data curator will have to allocate a certain portion of the total privacy budget to the model selection procedure. Differentially private model selection is a separate research topic that is beyond the scope of this paper. For simplicity, we assumed the correct SLOMAG model was identified beforehand without using the current data set.

In the SLOMAG model, we employed  priors $f(\bs{\mu}, \Sigma)\!\propto\!|\Sigma^{-1}|$ on $\bs{\mu}$ and $\Sigma$ and assumed  $f(\bs{\beta}_1, \bs{\beta}_2,\bs{\beta}_3,\bs{\beta}_4)\!=\!f(\bs{\beta}_1)f(\bs{\beta}_2)f(\bs{\beta}_3,\bs{\beta}_4)$. The joint posterior distribution of the parameters can be factorized as
$f(\Sigma|\z)f(\bs{\mu}|\Sigma,\z)f(\bs{\beta}_1|\z,\w_1)f(\bs{\beta}_2|\z,\w_1,\w_2) f(\bs{\beta}_3,\bs{\beta}_4|\z,\w_1,\w_2,\w_3)$, where $f(\Sigma|\z) = \text{Inv-Wishart}(n, \bs{S}), f(\bs{\mu}|\Sigma,\z)=\text{N}(\bar{\z}, n^{-1}\Sigma)$ ($\bar{\z}$ was the sample means of $\z$, and $\bs{S}=n^{-1}\sum_{i=1}^{n}(\mathbf{z}_{i}-\bar{\mathbf{z}})(\bs{\mathbf{z}}_{i}-\bar{\mathbf{z}})'$ was the sample covariance matrix of $\z$), and
\begin{align}
& f(\bs{\beta}_1|\w_{1},\z)
\propto f(\bs{\beta}_1)\prod^{n}_{i=1}\frac{e^{w_{i1}(1,z_{i1},z_{i2})\bs{\beta}_1}}{1+e^{(1,z_{i1},z_{i2})\bs{\beta}_1}}\label{eqn:beta1}\\
&f(\bs{\beta}_2|\w_{2},\w_{1},\z)
\!\propto f(\bs{\beta}_2)\prod^{n}_{i=1}\!\frac{e^{w_{i2} (1, z_{i1},z_{i2},w_{i1})\bs{\beta}_2}}{1\!+\!e^{(1, z_{i1},z_{i2},w_{i1})\bs{\beta}_2}}\label{eqn:beta2}\\
& f(\bs{\beta}_3,\bs{\beta}_4|\w_3,\w_{2},\w_{1},\z)\propto
\prod^{n}_{i=1}\left\{\!\left(\!\frac{1}{1+A_i+B_i}\!\right)^{\!I(w_{i3}=1)}
\!\!\!\!\!\!\!\times\!\!\left(\!\frac{e^{w_{i3}(1,z_{i1},z_{i2},w_{i1},w_{i2})\bs{\beta}_3}}{1+A_i+B_i}\!\right)^{\!I(w_{i3}=2)}\right.\notag\\
&\quad\times\!\!\left.\left(\!\frac{e^{w_{i3}(1,z_{i1},z_{i2},w_{i3},w_{i2})\bs{\beta}_4}}{1+A_i+B_i}\!\right)^{\!I(w_{i3}=3)}\!\right\}f(\bs{\beta}_3,\bs{\beta}_4)
=\frac{e^{a_i\bs{\beta}_3+b_i\bs{\beta}_4}}{\prod^{n}_{i=1}(1+A_i+B_i)}f(\bs{\beta}_3,\bs{\beta}_4),\label{eqn:beta3}
\end{align}
\noindent where $a_i=(\sum^n_{i=1}\!I(w_{i3}\!=\!2),\sum^n_{i=1}\!z_{i1}I(w_{i3}\!=\!2),\sum^n_{i=1}\!z_{i2}I(w_{i3}\!=\!2),\sum^n_{i=1}\!w_{i1}I(w_{i3}\!=\!2),\\ \sum^n_{i=1}\!w_{i2}I(w_{i3}\!=\!2)), b_i=(\sum^n_{i=1}\!I(w_{i3}=3),\sum^n_{i=1}\!z_{i1}I(w_{i3}=3),\sum^n_{i=1}\!z_{i2}I(w_{i3}=3), \\ \sum^n_{i=1}\!w_{i1}I(w_{i3}\!=\!3), \sum^n_{i=1}\!w_{i2}I(w_{i3}\!=\!3)), A_i\!=\!e^{(1,z_{i1},z_{i2},w_{i1},w_{i2})\bs{\beta}_3}$, and $B_i\!=\!e^{(1,z_{i1},z_{i2},w_{i1},w_{i2})\bs{\beta}_4}$.

To synthesize $\tilde{\z}_i$  for $i=1,...,n$ in the traditional non-DP MS approach via the SLOMAG model,we first drew $\Sigma$ from $f(\Sigma|\z)=\text{Inv-Wishart}(n,\bs{S})$, and $\bs{\mu}$ from $f(\bs{\mu}|\Sigma,\z)=\text{N}(\bar{\z},n^{-1}\Sigma)$, and then simulated $\tilde{\z}_i$ from $f(\tilde{\z}_i|\bs{\mu},\Sigma)=\text{N}(\bs{\mu},\Sigma)$ given the drawn $(\Sigma,\bs{\mu})$. To synthesize $\tilde{\w}_i=(\tilde{w}_{i1}, \tilde{w}_{i2}, \tilde{w}_{i3})$, we assumed $f(\bs{\beta}_1)f(\bs{\beta}_1)f(\bs{\beta}_3,\bs{\beta}_4)\propto$ constant, and applied the Metropolis  algorithm to sample $\bs{\beta}_1,\bs{\beta}_2,\bs{\beta}_3, \bs{\beta}_4$ from their posterior distributions  and, after checking on the convergence of the MCMC chains  (2 chains, a burn-in period of 1500, a thinning period of 10, and 10,000 iterations to yield a total of 7,650 samples), simulated $\tilde{w}_{i1}, \tilde{w}_{i2}$ and $\tilde{w}_{i3}$  from $f(\tilde{w}_{i1}|\bs{\beta}_1,\Sigma,\tilde{\z}), f(\tilde{w}_{i2}|\bs{\beta}_2,\tilde{w}_{1i},\tilde{\z}_i)$, and  $f(\tilde{w}_{i3}|\bs{\beta}_3,\bs{\beta}_4,\tilde{w}_{i2},\tilde{w}_{i1},\tilde{\z}_i)$, respectively. We calculated  the potential scale reduction factor (psrf) using the \verb;R; package \verb;coda; to check on the convergence of the MCMC chains. In the Supplemental Materials, we provide the  MCMC trace plots from a random sample out of the 1000 repeats on $\bs{\beta}_1$ as an example.

For the MODIPS approach, there were total 8 sets of quantities to be sanitized: $\bar{\mathbf{z}},\mathbf{S}$, and 3 sets of estimated regression coefficients $\hat{\bs{\beta}}_1, \hat{\bs{\beta}}_2$ and ($\hat{\bs{\beta}}_3,\hat{\bs{\beta}}_4)$, implying the total privacy budget per synthesis ($\epsilon/5)$ should be divided by $8$.  $\bar{\z}$ (2 components) and $\mathbf{S}$ (3 components) were the Bayesian sufficient statistics associated with the posterior distributions of $\Sigma$ and $\bs{\mu}$. Since $c_{1,j}-c_{0,j}=8\sigma $ for  $j=1,2$, the $l_1$ global sensitivity was $8\sigma n^{-1}$ for $\bar{z}_j$ and $(8\sigma)^2n^{-1}$ for each entry in $\mathbf{S}$. To obtain differentially private samples from the  posterior  distributions of  $\bs{\beta}_1, \bs{\beta}_2,\bs{\beta}_3$ and $\bs{\beta}_4$ in Eqns (\ref{eqn:beta1}) to  (\ref{eqn:beta3}), we implemented Algorithm 1 in \citet{chaudhuri2009privacy}. Specifically, denote by  $\hat{\bs{\beta}}$ the optimizer of the loss function from a logistic regression model with $l_2$ regularization (tuning parameter $\lambda$) and normalized  predictors $\mathbf{x}$  for all $i=1,\ldots,n$ (per Euclidean norm $||\mathbf{x}||_i\leq 1$), then the differentially private coefficient estimates are given by $\hat{\bs{\beta}}^*=\hat{\bs{\beta}}+\mathbf{e}$, where the distribution of the noises $\e$ is $f(\e)\propto \exp\left(-n\lambda\epsilon||\mathbf{e}||/2\right)$. In this simulation study, we added the noise $\e$ to a random draw from the posterior distribution of the $\bs{\beta}$ instead of to the optimizer, which would be the posterior mode. This is because the injected noise only depends on $n$, $p$ and is independent of the specific values in the data set  as along as the specification of $\lambda$ does not use the local data.  $\lambda$ is often chosen by cross-validation if there is no privacy concern, but would cost privacy otherwise. \citet{Chaudhuri2011} suggest two ways of selecting $\lambda$; using a separate public data set, which does not cost privacy budget; or subsetting the data and then apply the Exponential mechanism with the prediction accuracy as the scoring function to choose $\lambda$  (Algorithm 4 in \citet{Chaudhuri2011}). Here we assumed there existed a public data set (which was simulated from the same joint distribution of $\mathbf{X}$ and $\mathbf{W}$ and attempted five $\lambda$'s ($0.01,0.05,0.1,0.5$ and 1) in each regression on this public data set. We found $\lambda_1,\lambda_2,\lambda_3$ around $0.5$ seemed to perform the best in terms of prediction accuracy, which was used as the final $\lambda$'s in the MODIPS approach.

In summary, the steps for generating the differentially private data from the SLOMAG model are as follow. 
1) sanitize $\bar{\mathbf{z}}$ and $\mathbf{S}$, draw $\bs{\mu}$ and $\Sigma$ from their posterior distributions with the sanitized $\bar{z}^*$ and $\mathbf{S}^*$, and simulate $\tilde{\mathbf{z}}^*$ from its posterior predictive distribution given $\bs{\mu}^*$ and $\Sigma^*$. 
2) Fit the $l_2$ regularized logistic regression on $\mathbf{w}_1$ in the Bayesian framework with the normalized predictor $\mathbf{z}'$ and prior $f(\bs{\beta}_1)\overset{\mbox{\scriptsize{ind}}} {\sim}N(0,\lambda_1^{-1})$; draw $\bs{\beta}_1$ from its posterior distribution and sanitize it as outlined above; simulate $\tilde{\mathbf{w}}^*_1$ given the sanitized $\bs{\beta}^*_1$ and the normalized sanitized $\tilde{\mathbf{z}}^{*'}$ from the first step. 
3) Fit the $l_2$ regularized logistic regression on $\mathbf{w}_2$ in the Bayesian framework  with the normalized predictor $(\mathbf{z}',\mathbf{w}'_1)$ and prior $f(\bs{\beta}_2)\overset{\mbox{\scriptsize{ind}}} {\sim}N(0,\lambda_2^{-1})$; draw $\bs{\beta}_2$ from its posterior  distribution and sanitize it as outlined above; simulate $\tilde{\mathbf{w}}^*_2$ given the  sanitized $\bs{\beta}^*_2$ and the normalized $(\tilde{\mathbf{z}}^{*'},\tilde{\mathbf{w}}^{*'}_1)$  from the first two steps. 4) Fit the $l_2$ regularized multinomial logistic regression on $\mathbf{w}_3$ in the Bayesian framework with the normalized predictor $(\mathbf{z}',\mathbf{w}'_1,\mathbf{w}'_2)$ and prior $f(\bs{\beta}_3,\bs{\beta}_4)\overset{\mbox{\scriptsize{ind}}} {\sim}N(0,\lambda_3^{-1})$; draw $(\bs{\beta}_3,\bs{\beta}_4)$ from their posterior distributions and sanitize them as outlined above; simulate $\tilde{\mathbf{w}}^*_3$ given the sanitized $(\bs{\beta}^{*}_3,\bs{\beta}^{*}_4)$ and the normalized $(\tilde{\mathbf{w}}^*_1,\tilde{\mathbf{w}}^*_2,\tilde{\mathbf{z}}^*)$  from the previous three steps.  Similar to the non-DP MS case, we calculated the psrf to check on the convergence of the MCMC chains. In the Supplemental Materials, we provide the MCMC trace plots from a random sample out of the 1000 repeats on $\bs{\beta}_1$ as an example. 

To check on the impact of the mis-specification of synthesis models in MODIPS on the inferences from the synthetic data, we also synthesized data from a mis-specificated model.  There are many possible models for a mixture of categorical and continuous variables, for example, assuming independence among ($\x,\w$), dropping a predictor  in one of the logistic regression equations above, switching the order of the logistic regression sequence, or applying the GLOM, which are all mis-specifications in this simulation. For the purposes of checking on the impact of the mis-specification of synthesis models, it would make no essential difference which misspecified  model to be compared with the correct specification. Therefore, we used the GLOM as a representation for the model misspecification. The Bayesian GLOM in this case was similar to that in Sec \ref{sec:ggm}, with the multinomial distribution on $(W_1,W_2,W_3)$  (12 cells) and the bivariate Gaussian assumption on $(Z_1,Z_2)$ in each of the 12 cells, to generate 5 synthetic data sets. To examine how bad the misspecification could be in terms of inferences, the supplementary materials present a posterior predictive check from fitting the SLOMAG and the GLOM models to the original data. Though the GLOM led to biased synthetic data, this mis-specification brought some potential side-benefit in terms of privacy protection. Specifically, the misspecification could offer more privacy guarantee as it provides another type of noise that deviated the synthetic information from the original. However, the additional privacy protection, if there is any, can be difficult to quantify or even untraceable in the framework of DP. As a result, the comparison in statistical utility below from the synthetic data might be unfair to the  GLOM model as the actual privacy it provided could be more than the nominal $\epsilon$-DP, which the other synthesizer used.

We examined the inferences on $\Sigma,\bs{\mu}$, and $\bs{\beta}_1, \bs{\beta}_2,\bs{\beta}_3,\bs{\beta}_4$. $\mu_1$ and $\mu_2$ in synthetic data set $l$ ($l=1,\ldots,5)$ were estimated by the sample means $\bar{z}^{(l)}_1$ and $\bar{z}_2^{(l)}$, and $\Sigma$ was estimated by the sample covariance $\bs{S}^{(l)}$. The corresponding within-set variance was estimated by  $(S^2_k)^{(l)}n^{-1}$ for $\bar{z}_k^{(l)}$ ($k=1,2$), $\left((S^2_k)^{(l)})^2(2(n-1)^{-1} + \kappa^{(l)} n^{-1}\right)$  for the marginal variances of $\z_1$ and $\z_2$, and $(1 - (r^{(l)})^2)(n - 2)^{-1}$ for the correlation between $Z_1$ and $Z_2$, respectively, where $(S^2_1)^{(l)}$, $(S^{2}_2)^{(l)}$ were the diagonal elements of  $\bs{S}^{(l)}$, $\kappa^{(l)}$ was the excess kurtosis, and $r_l$ was derived from $\mathbf{S}^{(l)}$.  The regression coefficients $\bs{\beta}$ were estimated using the \texttt{logistf} function with the Firth's bias reduction method in  the \texttt{R} package \texttt{logistf} along with the corresponding estimated variance estimates. Eqns (\ref{eqn:mean}) to (\ref{eqn:test}) were then applied to obtain the final estimates of the parameters and the 95\% CIs in each DIPS approach. 

\begin{figure}[!htb]
\begin{center}
\hspace*{-0.2in}\includegraphics[width=6.7in]{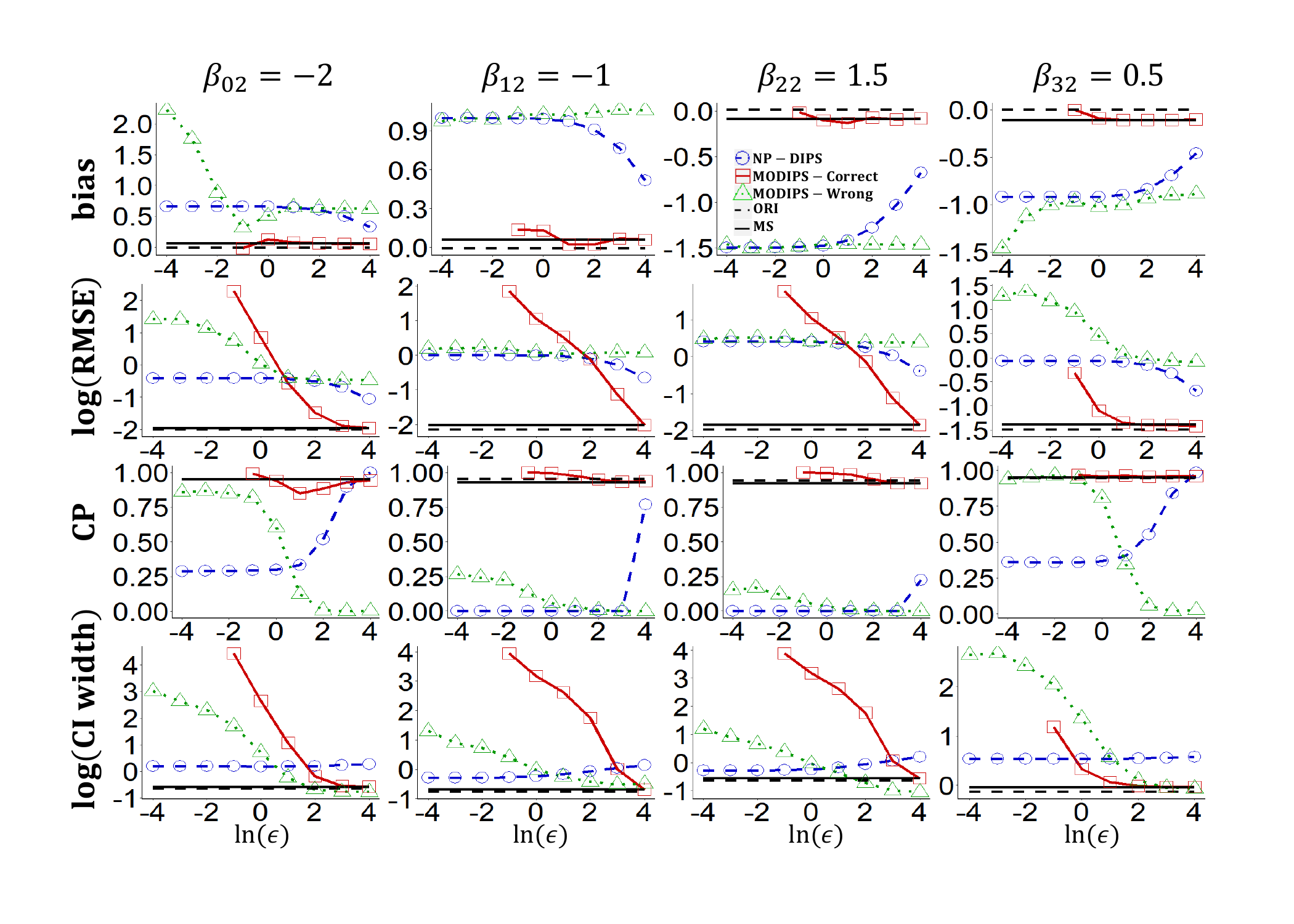}
\caption{The bias, log(RMSE), 95\% CP, and log(95\% CI width) for $\bs{\beta}_2$ in simulation study 4. MODIPS represents the model-based differentially private synthesis, NP-DIPS represents the Laplace sanitizer + perturbed histogram method, MS is the traditional multiple synthesis method without DP, and Ori is the original results without any perturbation.} \label{fig:sim4reg1}
\end{center}
\end{figure}

\begin{figure}[!htb]
\begin{center}
\hspace*{-0.1in}\includegraphics[width=6.4in]{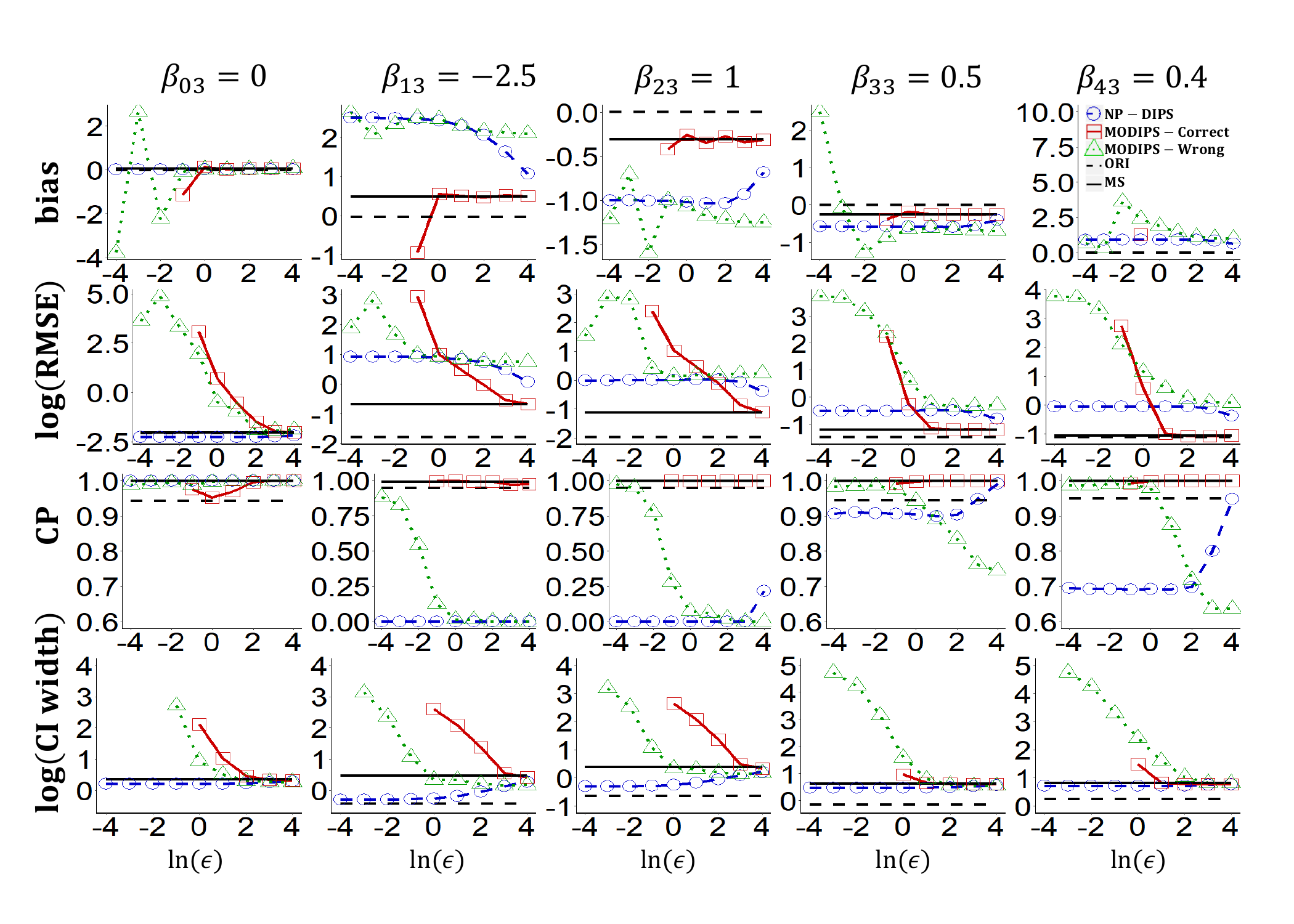}
\caption{The bias, log(RMSE), 95\% CP, and log(95\% CI width) for $\bs{\beta}_3$ in simulation study 4. MODIPS represents the model-based differentially private synthesis, NP-DIPS represents the Laplace sanitizer + perturbed histogram method, MS is the traditional multiple synthesis method without DP, and Ori is the original results without any perturbation.}\label{fig:sim4reg2}
\end{center}
\end{figure}

Due to space limit, we present the results on the bias,  log(RMSE), CP, and log(95\% CI width) for $\bs{\beta}_2$ and $\bs{\beta}_3$ in Figures \ref{fig:sim4reg1} and \ref{fig:sim4reg2}; the results on  $\mu_{1}$, $\mu_{2}$, $\sigma^2_1$, $\sigma^2_2$, $\rho$, $\bs{\beta}_1$, and $\bs{\beta}_4$ are available in the Supplemental Materials. For very small values of $\epsilon=e^{-4}$ to $e^{-2}$, the logistic regression based on the synthetic data from the MODIPS approach fail to converge thus the results were not available for plotting. The results are summarized as follows. \textbf{1)}  First, as expected, MODIPS-Wrong failed to capture the original information due to the model mis-specification (large bias and undercover coverage). \textbf{2)}  Overall, the biases for MODIPS-Correct got smaller and were close zero after $\epsilon>e^{-1}\approx 0.368$, whereas the biases from the NP-DIPS approach did not seem to diminish even at large $\epsilon$. However, the bias in the MODIPS-Correct was unstable and larger than the other methods when $\epsilon$ was small (not plotted).  \textbf{3)}  The RMSE values in general were large compared to the original RMSE values across all DIPS methods.  \textbf{4)} The MODIPS-Correct approach produced coverage at or above the nominal level of 95\% for  $\epsilon>e^{-1}$ at the cost of wide CIs. The CP results from the NP-DIPS approach varies across parameters: some experienced severe under-coverage across all values of $\epsilon$ or only at small values of $\epsilon$, some had close to 95\% coverage across the board. The CI width varied little with $\epsilon$ in the NP-DIPS approach.  \textbf{5)} For $\mu_1$ and $\mu_2$, the bias, RMSE, and CI width of the estimates were smaller in the NP-DIPS approach than those in the MODIPS-Correct approach for $\epsilon<e$, and were similar for $\epsilon>e$; and both provided about 95\% CP. Although the RMSE and CI width decreased as $\epsilon$ increased for MODIPS-Wrong, the bias and CP deviated significantly from the original values. \textbf{6)}  For $\Sigma$, the NP-DIPS and MODIPS--Wrong approaches experienced severe under-coverage in all 3 components ($\sigma_1^2, \sigma_2^2$, and $\rho$) regardless of $\epsilon$. The former suffered for the same reason given in simulation study 3 (discretization and uniform sampling from each histogram bin). The MODIPS--Correct provided nominal CP for $\sigma_1^2, \sigma_2^2$, and $\rho$ and had smaller bias and RMSE than the NP-DIPS approach for $\epsilon>e^{-2}$, at the cost of wide CIs   when $\epsilon<e$. 

Compared to simulation study 3 which also had a mixture of continuous and categorical variables, the results in simulation study 4 were generally worse for both the NP-DIPS and the MODIPS approaches, but in different ways. The identification of statistics to sanitize with the SLOMAG model was less obvious and the inferences based on the synthetic data were less stable in simulation 4 for the MODIPS approach, probably due to the direct sanitization of the likelihood functions. For the NP-DIPS approach, the discretization and sanitization procedure was the same between simulations 3 and 4, but seemed to affect the inferences from the SLOMAG model more than those from the GLOM. The different results from the two simulation studies suggest that even though the NP-DIPS approach was nonparametric, inferences based on the synthesized data in certain models can be more sensitive than others.  

\section{Case Study} \label{sec:case}
We apply several DIPS  approaches to a real-life data set to assess the feasibility these approaches in generating useful synthetic data sets in practice. We used the fertility data set from \citet{gil2012predicting} in a study of 100 student volunteers at the University of Alicante. Each participant  provided a semen sample after 3 to 6 days of sexual abstinence, and answered a questionnaire about their life habits and health status.  The attributes in the data are summarized in Table \ref{tab:datlist} (there were originally 35 variables in \citet{gil2012predicting}, but only 10 variables  were  publicly available on the UCI Machine Learning Repository). The data set is useful for studying risk factors possibly associated with the fertility rate. On the other hand, sharing the data set publically could cause privacy concerns as some of the variables such as ``diagnosis of seminal quality'' are generally regarded as sensitive information. 
\begin{table}[!ht]
\centering 
\small \begin{tabular}{L{2.5in} | L{3.0in}  }
\hline
	\textbf{Variable} & \textbf{Values } \\
\hline
    Season of the analysis & Winter,  Spring,  Summer,  Fall \\
\hline
    Age at the time of analysis (years) & $18\sim 36$   \\
\hline
	Childish diseases & Yes,  No \\
\hline
  	Accident or serious trauma & Yes,  No \\
\hline
	Surgical intervention & Yes, No \\
\hline
	High fevers in the last year & $<3$ months ago,  $>3$ months ago, no  \\
\hline
	Frequency of alcohol consumption & several times a day, every day, several times a week, once a week, hardly ever or never \\
\hline
    Smoking habit & never, occasional, daily  \\
\hline
	Number of hours sitting per day & $1\sim 16$ \\
\hline
	Diagnosis of seminal quality & normal, altered\\
\hline
\end{tabular}
\caption{Variables from the fertility data in \citet{gil2012predicting}.}\label{tab:datlist}
\end{table}

The main goal of \citet{gil2012predicting}  was to compare the performance of three machine learning techniques (decision trees, Multilayer Perception, and Support Vector Machine/SVM) in predicting the seminal quality given various predictors. The authors found that Multilayer Perception and SVM outperformed decision trees with SVM  slightly more accurate. Therefore, we  only employed the SVM in this case study. Specifically, we first randomly split the original data into a training set of 80 subjects and a test set of 20 subjects.  The same test set was then used to evaluate the predictive power of the SVMs constructed from the synthetic data via different DIPS approaches (so to avoid testing the SVM on a test data that was synthesized via the same DIPS approach for generating the training data).  Since this analysis did not involve statistical inferences, we generated a single synthetic data set with $\epsilon=e^1=2.72$, a practically small and reasonable privacy budget. 

We employed the  Laplace sanitizer (ND-DIPS) and the MODIPS (P-DIPS) approach. In the Laplace sanitizer, we first discretized the two continuous variables (age and hours sitting) into a 2D histogram, then sanitized the cell counts $\mathbf{n}$ from the full cross tabulation of the 10 variables with the additive noise from Lap$(0,\epsilon^{-1})$. For the MODIPS approach, the first step was to select an appropriate synthesis model. There were 8 categorical variables with some of them having sparse cell counts in their marginal distribution (e.g., in alcohol consumption, there was only 1 person in the categories of several times a day and every day, respectively); both of the two continuous variables (age and hours sitting) deviated obviously from Gaussian distributions. Given the small sample size ($n=80$ in the training set), the GLOM is not expected to work well as there would be too many empty or sparse cells from the full cross-tabulation of the categorical variables; the SLOMAG model could generate noisy synthetic data based on its performance in the simulation study 4 where the sample size ($n=1000)$ was much higher than this case study with a much smaller cross-tabulation of the categorical variables. We also tried the second-order mixed graphical model approach on the data, and the prediction was not good even without DP perturbation. All taken together, we discretized the two continuous variables and then fitted a saturated log-linear model. The Bayesian sufficient statistics was $\mathbf{n}$. We first sanitized $\mathbf{n}$ via the Laplace mechanism with scale $\epsilon^{-1}$ to obtain $\mathbf{n}^*$. Given $\s^\ast$, we drew $\bs{\pi}^\ast$ from $f(\bs{\pi}^\ast|\mathbf{n}^\ast)=D(\bs{\alpha}+\mathbf{n}^\ast)$, then $\tilde{\x}$ from $f(\tilde{\x}|\bs{\pi}^\ast)=\text{Multinom}(n,\bs{\pi}^\ast)$. We ran 100 repetitions.

As a benchmark, we run the PrivateSVM (Algorithm 2 in \citet{rubinstein2009learning}), an approach designed specifically for releasing differentially private SVM results. PrivateSVM first applies the SVM to the original data and then returns the noisy weight vector via the Laplace mechanism. Calculating the global sensitivity of the weight vector for PrivateSVM is nontrivial. We employed the linear kernel for PrivateSVM, the global sensitivity based which is $4LC\kappa \sqrt{F}/n$ \citep{rubinstein2009learning}, where $L=1$ is the linear kernel, $C=1$ is cost of constraints violation (the $C$-constant of the regularization term in the Lagrange formulation in SVM), $\kappa$ is the upper bound for the linear kernel (9 with the normalized data in this case), $F$ is the number of features (9 in this case), and $n$ is the number of observations in the training sample (80 in this case). Thus, the total global sensitivity is 1.35. 


If the DIPS approaches performed similarly to PrivateSVM in classification accuracy, DIPS would be preferable since the data user would have the individual-level synthetic data whereas PrivateSVM only provides a differentially private SVM. When constructing the SVM on the synthetic data, we employed the \texttt{svm} command in R package \texttt{e1071} with \verb;kernel="linear"; and a 5-fold cross validation.
\begin{table}[!htb]
\centering
\resizebox{\textwidth}{!}{
\begin{tabular}{c |cc| cc |cc| cc}
\hline
& \multicolumn{8}{|c}{Predicted based on}\\
\cline{2-9}
Observed & \multicolumn{2}{c|}{Original data} 
& \multicolumn{2}{c|}{PrivateSVM} 
& \multicolumn{2}{c|}{Laplace sanitizer} 
& \multicolumn{2}{c}{MODIPS}\\
\cline{2-9}
& + & - & + & - & + & - & + & - \\
\hline
+ & 0 & 3 & 0.84 & 2.16 & 1.04 & 1.96 & 1.50 & 1.50\\
- & 0 & 17 & 4.87 & 12.13 & 5.11 & 11.89 & 8.51 & 8.49\\
\hline
\textbf{CR}$^\dag$ & \multicolumn{2}{|c|}{$\boldsymbol{17/20=85\%}$} &  \multicolumn{2}{c|}{$\boldsymbol{12.97/20=64.9\%}$} &  \multicolumn{2}{c|}{$\boldsymbol{12.93/20=64.7\%}$}  & \multicolumn{2}{c}{$\boldsymbol{9.99/20=50.0\%}$} \\
\hline
\end{tabular}}
\begin{tabular}{l}
$^\dag$\footnotesize{CR: consistency rate\hspace{4.5in}} \\
\hline
\end{tabular}
\caption{Accuracy of SVMs constructed from PrivateSVM approach and synthetic data via the Laplace Sanitizer and MODIPS approaches when $\epsilon=e$.} \label{tab:confmat}
\end{table}

Table \ref{tab:confmat} shows the averaged confusion matrices and the classification accuracy on the 20 testing cases over the 100 repeats by the SVMs constructed from PrivateSVM as well as the synthetic data from the Laplace sanitizer and MODIPS approaches. As expected, the prediction power of the SVMs constructed on the synthetic data via the Laplace sanitizer and the MODIPS approach decreased (64.7\% and 50.0\%, respectively) from the original SVM (85\%) at the privacy budget of $\epsilon=2.72$, a cost we had to pay to achieve some level of privacy. The Laplace sanitizer was no worse than PrivateSVM (64.9\%), and the data user will have access to the full synthetic data with the DIPS approach, performing any analyses as if they had the original data. The  significant decreases ($20\%$) in predictive accuracy from the original results in all 3 differentially private approaches might have something to do with the small sample size and the unbalancedness between the two categories of the outcome (88:12 normal vs. altered). On the other hand, there might exist more efficient DIPS methods that can better preserve the original info while satisfying DP, a topic we will continue to work on.

We also examined summary statistics to further assess the synthetic data quality. Table \ref{tab:dipssum} summarizes the continuous variables (\textit{Age at the time of analysis} and\textit{ Number of hours sitting per day}, which are refered to as \textit{Age} and \textit{Hours}, respectively) by the mean and SD,  and the categories variables by the averaged total variation distance in the 1-, 2-way, 3-way, and 8-way (full) cross-tabulations, respectively, constructed based on the synthetic vs the original data. In summary, \textbf{1)} for both the Laplace sanitizer and the MODIPS approach, the mean and standard deviation of the two continuous variables were close to the original data's values; \textbf{2)} the averaged total variation distances in all the  cross-tabulations were consistently smaller with the Laplace sanitizer than the MODIPS approach.

\begin{table}[!ht]
\centering 
\small \begin{tabular}{L{1in}|L{0.8in}|C{1in}| C{1.4in} | C{0.75in}  }
\hline
\textbf{statistic} & \textbf{attribute} & \textbf{original data} & \textbf{Laplace santizer} & \textbf{MODIPS}\\
\hline
mean (SD) & age & 0.67 (0.12) & 0.66 (0.14) & 0.66 (0.14) \\
          & hours & 0.41 (0.19) & 0.44 (0.21) & 0.50 (0.29) \\
\hline
\multirow{3}{1in}{total variation distance (TVD)} 
& 1-way table & --- & 0.228 & 0.250 \\
& 2-way table & --- & 0.353 & 0.379 \\
& 3-way table & --- & 0.311 & 0.330 \\
& 8-way table & --- & 0.451 & 0.483 \\
\hline
\end{tabular}
\begin{tabular}{L{5.75in}}
\footnotesize $^\dagger\; \mbox{TVD}= t^{-1}\sum_{j=1}^t|\mathbf{p}^*_j-\mathbf{p}_j|_1$, where $t$ is the number of tables, $\mathbf{p}_j$ and $\mathbf{p}^*_j$
represent the vector of cell probabilities in table $j$ constructed from the original and synthetic data, respectively.\\ 
\hline
\end{tabular}
\caption{Some summary statistics on the synthetic data} \label{tab:dipssum}
\end{table}

\section{Discussion} \label{sec:discussion}
We reviewed different DIPS methods for synthesizing differentially private individual-level data and compared some DIPS methods empirically through  simulation studies and a real-life case study on the utility and inferential properties of the synthetic data generated by the DIPS methods. To the best of our knowledge, this is the first work that compares the inferential properties of DIPS approaches across various types and sizes of data. 

The NP-DIPS approaches are robust given that they do not impose model or distribution assumptions on a given data set. However, most NP-DIPS approaches require some degree of discretization on numerical attributes. When the number of attributes $p$ is large, an important question to ask is whether there exists a ``consistent'' high-dimensional histogram density estimator $f_n$ for the underlying true density $f$ for a given sample size $n$, even before the employment of a DP technique. It is known that the number of bins of a high-dimensional histogram grows exponentially with dimension $p$, and the rate of decrease of the mean integrated squared error $\mbox{E}||(f_n-f)||_2^2$ degrades rapidly as $p$ increases compared to the ideal parametric rate $O(n)$ \citep{scott2015multivariate}. In addition, when $p$ increase, most of the hypercube bins in the high-dimensional histogram become empty and  the histogram will be rough and provides reasonable estimates only near the mode and away from the tails. When there are correlations among the variables, smaller bin widths are required to ``track'' the correlations (implying more bins), and the asymptotic mean integrated squared error is always larger than the independent case.  
In summary, in high-dimensional histograms, the variance and bias trade-off is not well accomplished unless $n$ is large. If the original high-dimensional histogram is already not a good estimator for the distribution of the data, it is not meaningful to further sanitize it. Additionally, inferences based on the synthetic data via histogram-based NP-DIPS approaches are affected by how the histogram bins are formed. There exists theoretical work in the computer science community that examines the relationship between the sample size $n$ and the accuracy of $p$ binary proportion where the accuracy is defined as how close the sanitized histogram to the original and does not involve drawing inferences about population parameters. For example, the average $l_1$ error of answering $p$ 1-way binary proportions  has a lower bound of $\Omega(p/(n\epsilon))$ \citep{geometry}; and the maximum $l_1$ error from answering $p$ binary proportion given $n$ has a upper bound of $\Omega(p\log(p)/(n\epsilon))$ \citep{Steinke}.

The P-DIPS approaches, on the other hand, often require distributional assumptions and model building, and thus are subject to appropriate model mis-specification. None of the P-DIPS procedures we have examined (except for PrivBayes in \citet{zhang2014privbayes}) have the inherent model-selection component, implying they are applied after a suitable model is identified. Broadly speaking, there are two model selection scenarios -- one costs privacy budget and the other does not. Specifically, if the model is chosen not using the knowledge in the current data, but based on previous studies and common practice, then no privacy needs to be spent. If the synthesis model is selected via a selection procedure using the data to be released, then it costs privacy and we will need to split the privacy budget between model selection and data synthesis. The current research work on  differentially private model selection focuses on feature selection the setting of a certain model type, such as \citet{Kifer2012, Smith2013, Lei2018} for linear regression; and \citet{zhang2014privbayes} for Bayesian networks. More research will be needed in differentially privately selecting among models that do not have to be of  the same type, maybe by perturbing model selection criteria such as AIC or BIC. Meanwhile, to mitigate the concern on model specification or when there are several plausible models, we incorporated the model averaging idea into the synthetic data generation, which also helps loosening the restriction the dependency of the synthetic data on a single synthesis model. 

An obvious drawback for all DIPS approaches is that the data user will not know how much the results based on the synthetic data deviate from those if they had access to the original data. If the differentially private synthetic data contain too much noise, the decisions made based on the analysis of the synthetic might be improper or wrong. \citet{barrientos2017differentially} proposed a  differentially privately mechanism to release the test statistic and p-value from testing a regression coefficient against 0 from a linear regression model. Their numerical results suggest the sign of the test statistic and the conclusion of the hypothesis test have a high probability of being consistent with the original results, it is then suggested the approach can be used to validate the linear regression analysis based on synthetic data from a DIPS method. The validation system hinted in \citet{barrientos2017differentially} is developed by   \citet{barrientos2018providing} in a more comprehensive and integrated fashion using the U.S. federal government employee longitudinal data as an example. Specifically, the system has three components: 1) release synthetic data generated from a joint distribution of the data; 2)verify /validate the statistical utility of a certain analysis (query) by comparing the results based on the synthetic data with the query result released by a differentially private mechanism; and 3) provide the raw/confidential data to approved data users via secure remote access.   In the examples given in both papers, all the given privacy budget is spent on testing or verifying  a single query result, while the reality is that a data user is often interested in estimating more than parameters. From a DP perspective, then the total privacy budget will have to be split among all queries to be verified, leading to potentially a large amount noised injected per query and diminished values of a validation system. In addition, the synthesis in the validation system mentioned in \citet{barrientos2018providing} does not have to be differentially private (and is not in the example given in the paper), that needs another disclosure risk assessment step on the released synthetic data, which replies on strong assumption and can be be ad-hoc as compared to the robustness of the DP concept. In addition, since DP aims to cover the worst-case scenario which means the statistical utility of the query result obtained via a differentially private mechanism can be further away from the original than that based on the synthetic data without DP. In other words, a significant discrepancy between the two as suggested in \citet{barrientos2018providing} does not necessarily invalidate one, or the other.  

An alternative to enhancing data users' confidence in synthetic data is to develop more efficient DP mechanisms at the same privacy cost but with less noise injected, such as taking into account the correlations among the statistics during sanitization so that the privacy budget is not spent on overlapping information, or  optimizing the privacy budget allocation scheme when the sequential composition is in effect. In all the simulation studies we conducted, statistics were sanitized independently, implying that redundant noises were introduced on correlated statistics. Accounting for the correlations among the statistics will cut the necessary noises to satisfy DP, improving the efficiency of the DIPS procedures. In addition, we could always employ a relaxed version of DP (such as aDP or pDP) to generate synthetic data as long as there is consensus the relaxed DP still provide satisfactory privacy protection. Conceptually, all the DIPS methods introduced and examined in the paper can be implemented with relaxed DP assuming  the appropriate sanitizer is employed.

Some future work could also involve developing a system that compares the various DP definitions, mechanisms, and algorithms, and recommends DP mechanisms/algorithms to users. Given the wealth of DP methods, a data user might face difficulty in selecting the most well-suited DP approach for his/her data, including considerations on the practicality and computational limitations of those DP methods.  \citet{hay2016principled} attempted to address the issue proposing DPBench as approach for standardized evaluation of privacy algorithm, as well as valuable observations and findings after comparing various data-dependent and data-independent DP methods. However, their work is limited to 1- and 2-dimensional range queries.  Motivated by DPBench, \citet{kotsogiannis2017pythia} developed Pythia, a meta-algorithm that measures the input features to select a particular DP method. Similarly, Pythia is limited to releasing certain queries such as histograms, range queries, and  Naive Bayes classifiers. 

The choice of $\epsilon$ (and the parameter that quantifies the relaxation of the strict $\epsilon$-DP if a relaxed version of DP is used) remains an open question. The concept of the  $\epsilon$-DP is abstract and does not easily relate to practically relevant measures of privacy, making the justification of a socially acceptable of $\epsilon$ difficult.  Based on the literature we have surveyed as well as the observations on the statistical utility from the simulation studies and the case study we conducted, it seems that $\epsilon$ in the neighborhood 1 (which is neither too small nor too large) seems to produce synthetic data of acceptable statistical utility. Additionally, $\epsilon=1$ has been explored frequently in experiments run in the literature.  We believe more research and further investigation on this issue will help to narrow down to a generally acceptable set of $\epsilon$ values.  

The ultimate goal of developing DIPS approaches is to employ them for public data release in practice.  The US Census Bureau aims to employ DP in major data products like the 2020 Census of Population and House, the Economic Census, and the annual American Community Surveys \citep{abowd2017privacy}.  On the other hand, real-life data can be  large in size, complex in structure, and have a large number of attributes of various types. In addition, issues such as missing data, sparse data, data entry errors, among others further complicate the application of DIPS. There is still a huge gap from the research work on DP to the wide practical application of DP. The status quo is that a large body of DP literature focus only on categorical/binary attributes and ignore missing data or data entry errors.  \citet{machanavajjhala2008privacy} demonstrated that the Multinomial-Dirichlet synthesizer  led to poor inferences due to data sparsity when releasing the  commuting patterns of the US population data and proposed combining distance-based coarsening with a probabilistic pruning algorithm and preserving $(\epsilon=8.6,\delta=10^{-5})$-pDP.  The relatively low classification accuracy based on synthetic data in our case study in Section \ref{sec:case} also suggests that direct application of a DIPS approach without any modification might not accommodate real-life situation well enough. On the other hand, local DP has been employed by big tech companies (e.g., Google and Apple) to collect users data. Though it seems to be a successful application,  multiple sources suggest the privacy budget $\epsilon$ employed by Apple to collect users data on mobile devices is too high to be acceptable for privacy protection \citep{tang2017privacy, mac}. Although Apple has provided some information about their DP approach, the information is vague. \citet{tang2017privacy} attempted to replicate the method without success and stated that ``\textit{We applaud Apple's deployment of DP for its bold demonstration of feasibility of innovation while guaranteeing rigorous privacy. However, we argue that in order to claim the full benefits of differentially private data collection, Apple must give full transparency of its implementation and privacy loss choices, enable user choice in areas related to privacy loss, and set meaningful defaults on the daily and device lifetime privacy loss permitted.}'' 

\section*{Acknowledgement}
We thank the editor, the associate editor, and the two reviewers for their valuable comments and suggestions that greatly improved the quality of the manuscript.

\bibliographystyle{agsm}
\setlength{\bibhang}{0pt}
\bibliography{reflist}

\newpage
\setcounter{figure}{0}
\setcounter{table}{0}
\begin{center}
	Supplemental Materials for ``Comparative Study of Differentially Private Data Synthesis Methods" by Claire Bowen and Fang Liu
\end{center}
This file contains the supplementary materials to accompany the paper ``Comparative Study of Differentially Private Data Synthesis Methods" with additional results from the four simulation studies.
\begin{itemize}
\item Simulation study 1: Table \ref{tab:sim1use1} shows the proportions of usable simulation repeats for all DIPS methods. A usable simulation repeat is defined as a simulated data set that leads to a synthetic data set that contains at least one of each of the two levels of the binary variable.
Figure \ref{fig:sim1p50} depicts the zoomed-in inferential results when $\pi=0.50$.
\item Simulation study 2: Table \ref{tab:histstat} shows the summary statistics for the number of histogram bins used in histogram-based DIPS methods.
Figures \ref{fig:sim2mu2} and \ref{fig:sim2sig2_cp} depict the inferential results of $\mu$ and $\sigma^2$ when the bounds of the data were $[\mu-4\sigma,\mu+4\sigma]$ (the main text contains the results on data with asymmetric bounds $[\mu-3\sigma,\mu+4\sigma]$).
\item Simulation study 3: Table \ref{tab:sumstat} contains the true values of $\bs{\mu}_1$, $\bs{\mu}_2$, and $\bs{\pi}$ for the 24 cells used for simulating the data. Tables \ref{tab:sumstatcell} and \ref{tab:histstatsim3} show the summary statistics for the number of observations in the 24 cells formed by the categorical variables $\mathbf{w}$, and summary statistics for the number of 2-dimensional histogram bins formed by the continuous variables $(\z_1,\z_2)$ in each of the 24 cell (for the histogram-based DIPS methods), respectively. Figures \ref{fig:sim3mu2rho} and \ref{fig:sim3sigma} depict the  inferential results on  $\bs{\mu}_{2}$, $\rho$, $\sigma^2_1$, and $\sigma^2_2$.
\item  Simulation study 4:  Table \ref{tab:histstatsim4} shows the summary statistics for the number of 2-dimensional histogram bins formed by the continuous variables $(\z_1,\z_2)$ in each of the 12 cells formed by the categorical variables $\w$  in the histogram-based DIPS methods. Figure \ref{Sfig:sim4trace} depicts some MCMC trace plots from the MS and the MODIPS-correct approaches using  $\boldsymbol{\beta}_1$ as example in one repeat. Figures \ref{Sfig:sim4reg1} to \ref{Sfig:sim4reg3} show the  inferential results for parameters $\mu_{1}$, $\mu_{2}$, $\sigma^2_1$, $\sigma^2_2$, $\rho$, $\bs{\beta}_1$, and $\bs{\beta}_4$.
\end{itemize}
\newpage

\noindent In Table \ref{tab:sim1use1}, an usable simulation repeat is defined as a simulated data set that leads to a synthetic data set that contains both levels of the binary variable. When a simulation repeat is unusable, the simulation is discarded since a data user would most likely not release such a data set. However, we presented Table \ref{tab:sim1use1} to demonstrate how often unusable repeats would occur for certain DIPS methods. The only DIPS method that had unusable repeats within our tested $\epsilon$ values was the Laplace Sanitizer.
\begin{table}[hbp!]
\def\arraystretch{1.5}
\centering
\caption{The proportion of usable simulation repeats (out of 5000) based on the synthetic data via the Laplace sanitizer in Simulation 1. Only the proportions $\ln(\epsilon)= -4, -2, 0$ are presented since $\ln(\epsilon)>0$ generated 100\% usable repeats.}
\label{tab:sim1use1}
\begin{tabular}{| l | c | c | c |}
  \hline
$\ln(\epsilon)$ & -4 & -2 & 0 \\
    \hline
$n=40, \pi=0.15$ & 0.9532 & 0.9852 & 0.9998 \\
$n=100, \pi=0.15$ & 0.9684 & 0.9954 & 1.0000 \\
$n=1000, \pi=0.15$ & 0.9972 & 1.0000 & 1.0000 \\

$n=40, \pi=0.25$ & 0.9566 & 0.9900 & 1.0000 \\
$n=100, \pi=0.25$ & 0.9708 & 0.9984 & 1.0000 \\
$n=1000, \pi=0.25$ & 0.9992 & 1.0000 & 1.0000 \\

$n=40, \pi=0.50$ & 0.9574 & 0.9952 & 1.0000 \\
$n=100, \pi=0.50$ & 0.9754 & 0.9998 & 1.0000 \\
$n=1000, \pi=0.50$ & 1.0000 & 1.0000 & 1.0000 \\ \hline
\end{tabular}
 \end{table}

\clearpage
\begin{figure}[!h]
\includegraphics[width=6in]{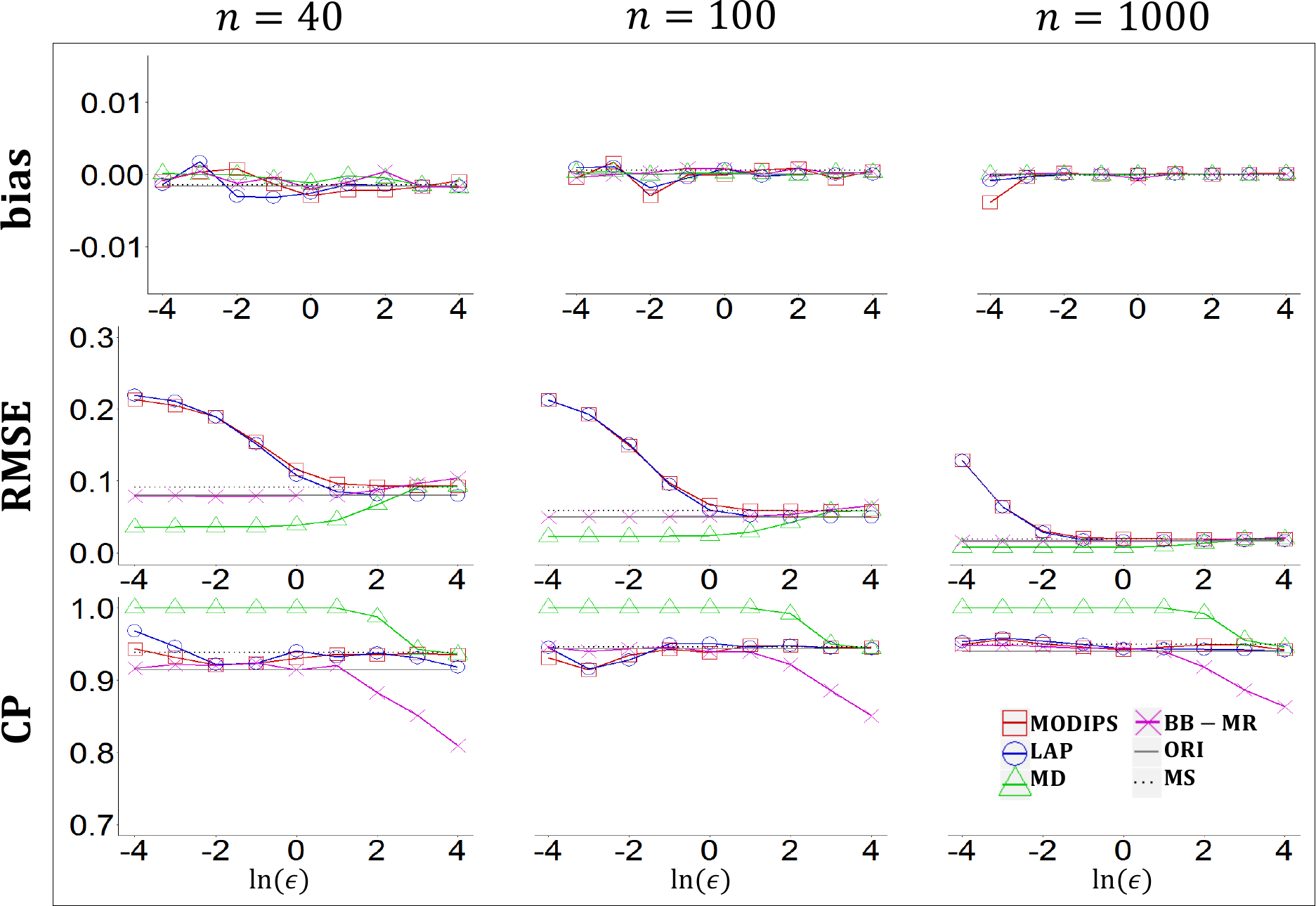}
\caption{Zoomed-in results on the bias, RMSE, and 95\% CP when $\pi=0.50$ in Simulation 1. MODIPS represents the model-based differentially private synthesis, LAP represents the Laplace synthesizer, MD represents the Multinomial-Dirichlet synthesizer, BB-MR represents the Binomial-Beta McClure-Reiter synthesizer, MS is the traditional non-DP MS, and Ori is the original results without any perturbation.
}
\label{fig:sim1p50}
\end{figure}

\clearpage
\begin{table}[!h]
\def\arraystretch{1.5}
\caption{Summary statistics for the number of the histogram bins across 5000 simulations repeats in simulation study 2.}\label{tab:histstat}
\begin{center}  \begin{tabular}{| c | c | c | c | c | c |}
  \hline
\textbf{Scenario} & \textbf{Min} & \textbf{Mean} & \textbf{Median} & \textbf{Max} \\
    \hline
$n=40, \sigma^2=1,4,9$ & 5 &  7.46 & 7 & 12\\
$n=100, \sigma^2=1,4,9$ & 8 &  9.85 & 10 & 13 \\
$n=1000, \sigma^2=1,4,9$ & 19 &  20.54 & 21 & 22 \\ \hline
\end{tabular}\end{center}
\end{table}

\clearpage
\begin{figure}[htp]
\vspace{-0.5in}
\makebox[\textwidth][c]{\includegraphics[width=6in]{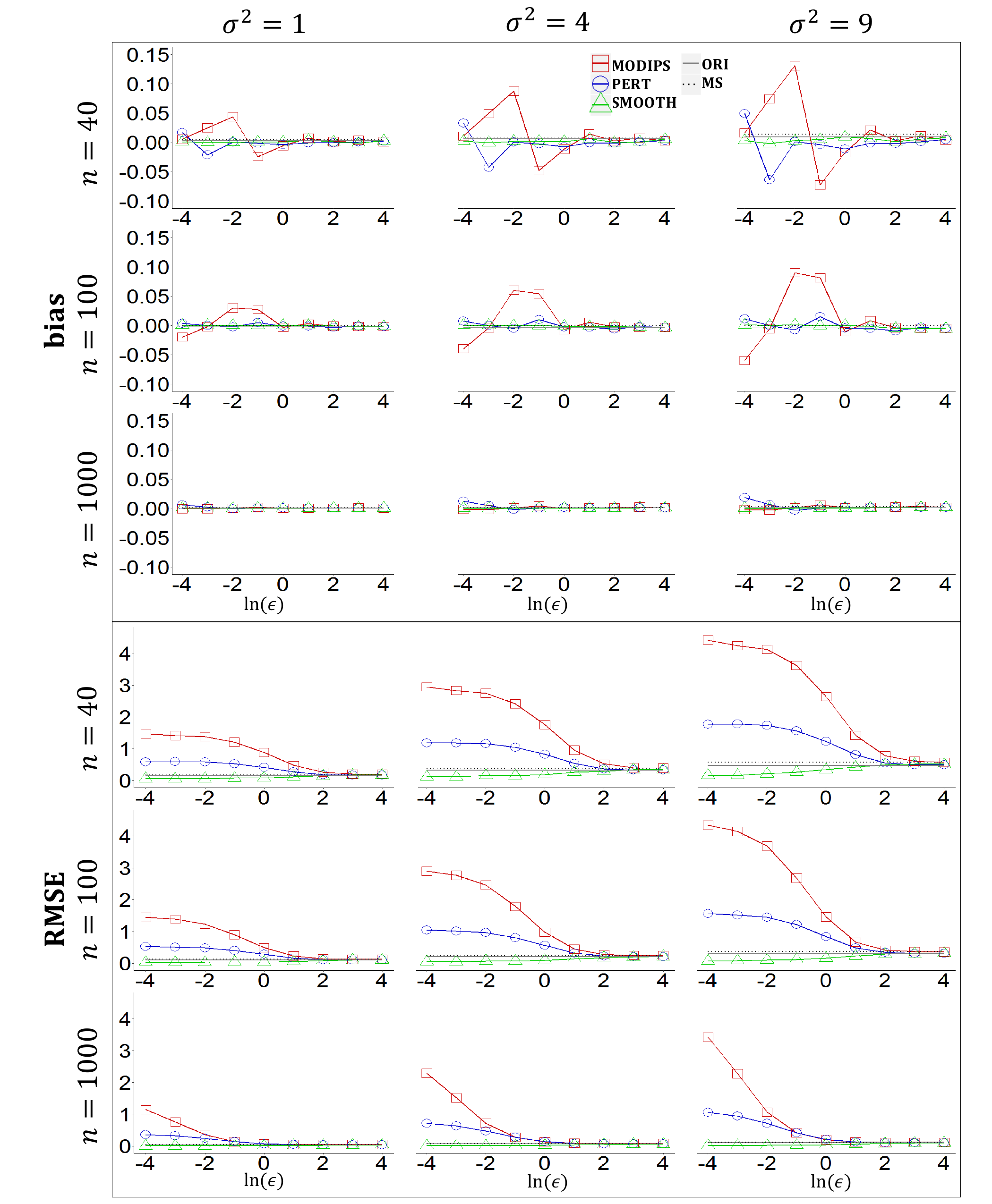}}
\vspace{12pt}
\caption{The bias and RMSE of $\mu$ in simulation study 2 when the data bounds were $[\mu-4\sigma,\mu+4\sigma]$. MODIPS represents the model-based differentially private data synthesis, PERT represents the perturbed histogram method, SMOOTH represents the smoothed histogram method, MS is the traditional MS method without DP, and Ori is the original results without perturbation.}\label{fig:sim2mu2}
\end{figure}

\vspace{-0.5in}
\begin{figure}[htp]
\makebox[\textwidth][c]{\includegraphics[width=5.5in]{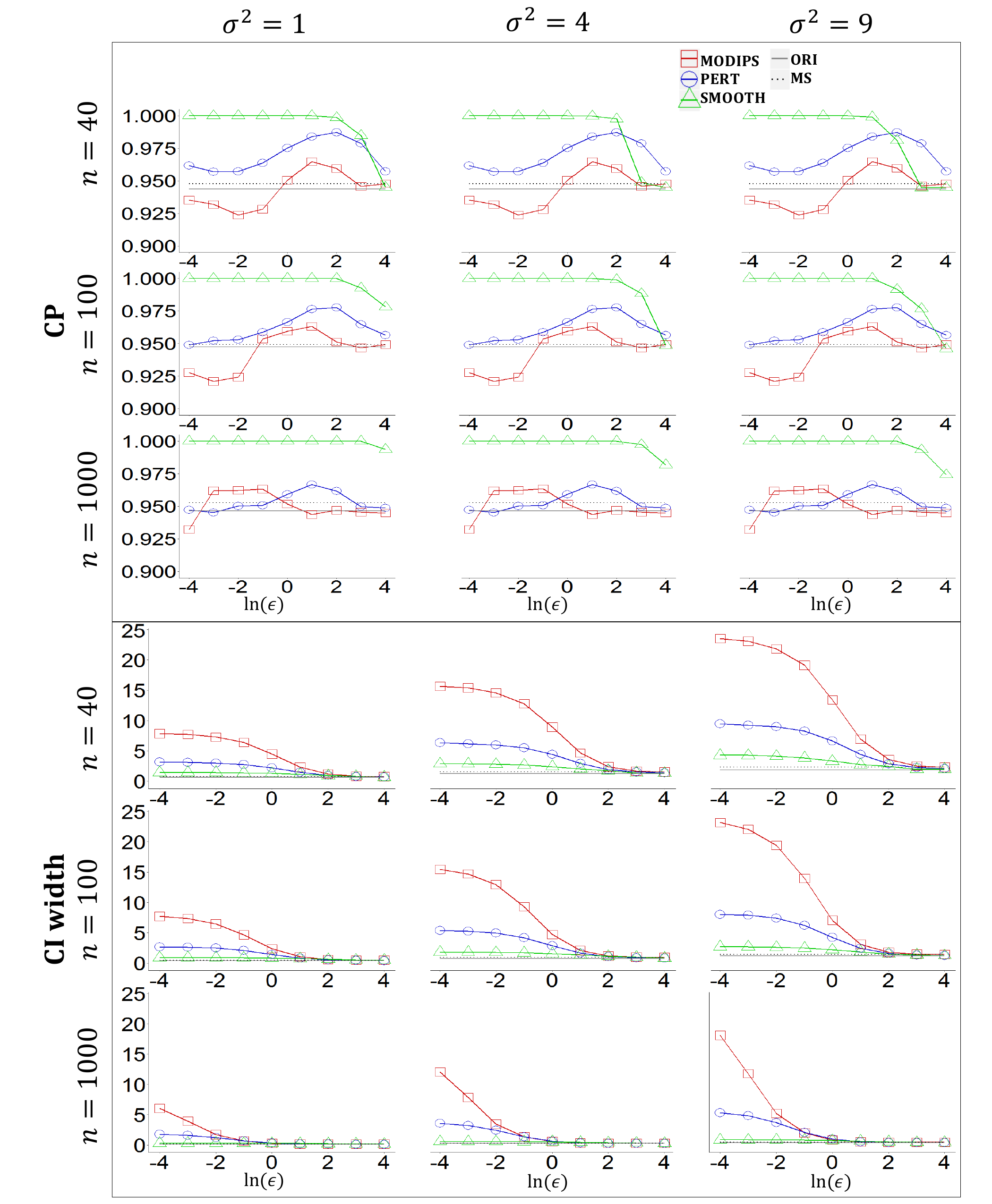}}
\vspace{12pt}
\caption{The 95\% CP and 95\% CI width of $\mu$ in simulation study 2 when the data bounds were $[\mu-4\sigma,\mu+4\sigma]$. MODIPS (with the boundary inflated truncation procedure) represents the model-based differentially private data synthesis, PERT represents the perturbed histogram method, SMOOTH represents the smoothed histogram method, MS is the traditional MS method without DP, and Ori is the original results without perturbation.}\label{fig:sim2mu2_cp}
\end{figure}

\vspace{-0.5in}
\begin{figure}[htp]
\makebox[\textwidth][c]{\includegraphics[width=5.5in]{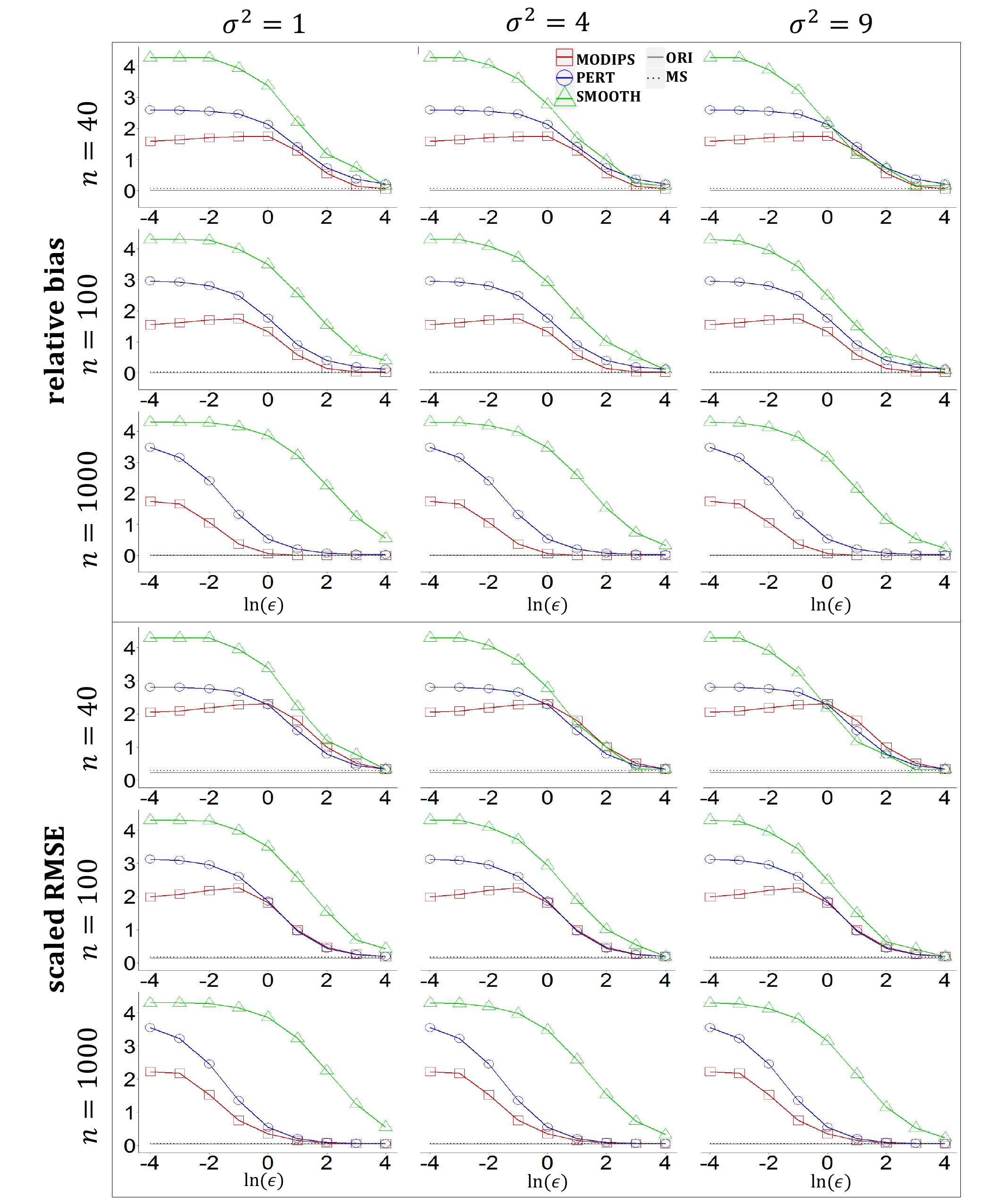}}
\vspace{12pt}
\caption{The relative bias and scaled RMSE of $\sigma^2$ in simulation study 2 when the bounds were $[\mu-4\sigma,\mu+4\sigma]$. MODIPS represents the model-based differentially private data synthesis, PERT represents the perturbed histogram method, SMOOTH represents the smoothed histogram method, MS is the traditional MS method without DP, and Ori is the original results without perturbation.}\label{fig:sim2sig2}
\end{figure}

\vspace{-0.5in}
\begin{figure}[htp]
\makebox[\textwidth][c]{\includegraphics[width=5.5in]{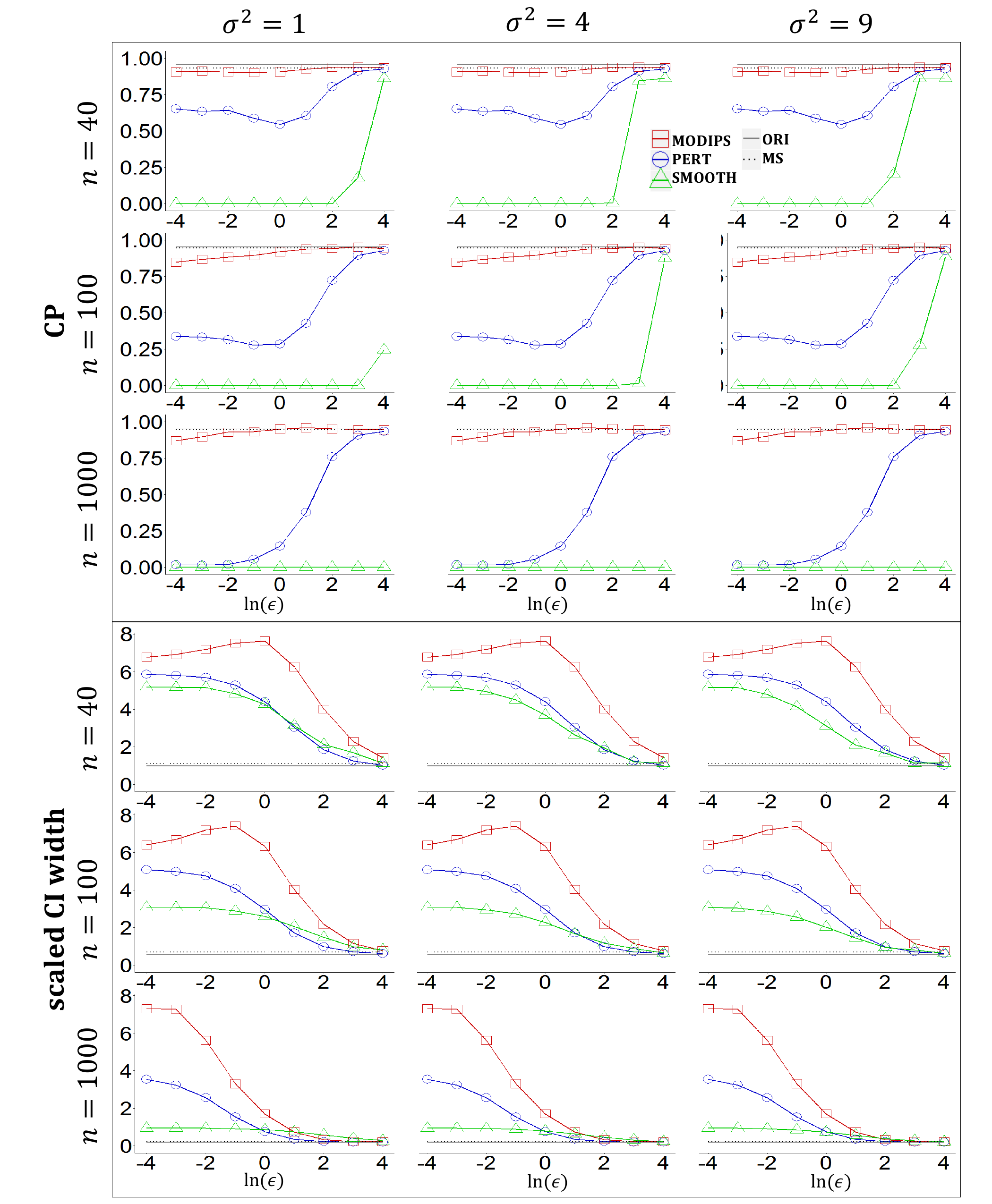}}
\vspace{12pt}
\caption{The 95\% CP and scaled 95\% CI width of $\sigma^2$ in simulation study 2 when the bounds were $[\mu-4\sigma,\mu+4\sigma]$. MODIPS (with boundary inflated truncation procedure) represents the model-based differentially private data synthesis, PERT represents the perturbed histogram method, SMOOTH represents the smoothed histogram method, MS is the traditional MS method without DP, and Ori is the original results without perturbation.}\label{fig:sim2sig2_cp}
\end{figure}

\clearpage
\begin{table}[!htb]
\def\arraystretch{1.15}
\caption{The true values of $\bs{\mu}_1$, $\bs{\mu}_2$, and $\bs{\pi}$ for the 24 cells in simulation study 3}\label{tab:sumstat}
\begin{center}  \begin{tabular}{| l | c | c | c | c |}
  \hline
\textbf{Cell} & $\bs{\mu}_1$ & $\bs{\mu}_2$ & $\bs{\pi}$ \\
    \hline
1 & 1.371 & -0.565 & 0.041 \\
  2 & 0.363 & 0.633 & 0.076 \\
  3 & 0.404 & -0.106 & 0.024 \\
  4 & 1.512 & -0.095 & 0.062 \\
  5 & 2.018 & -0.063 & 0.045 \\
  6 & 1.305 & 2.287 & 0.041 \\
  7 & -1.389 & -0.279 & 0.038 \\
  8 & -0.133 & 0.636 & 0.007 \\
  9 & -0.284 & -2.656 & 0.064 \\
  10 & -2.440 & 1.320 & 0.064 \\
  11 & -0.307 & -1.781 & 0.031 \\
  12 & -0.172 & 1.215 & 0.048 \\
  13 & 1.895 & -0.430 & 0.053 \\
  14 & -0.257 & -1.763 & 0.007 \\
  15 & 0.460 & -0.640 & 0.021 \\
  16 & 0.455 & 0.705 & 0.065 \\
  17 & 1.035 & -0.609 & 0.070 \\
  18 & 0.505 & -1.717 & 0.012 \\
  19 & -0.784 & -0.851 & 0.024 \\
  20 & -2.414 & 0.036 & 0.028 \\
  21 & 0.206 & -0.361 & 0.048 \\
  22 & 0.758 & -0.727 & 0.011 \\
  23 & -1.368 & 0.433 & 0.058 \\
  24 & -0.811 & 1.444 & 0.062 \\
\hline
\end{tabular}\end{center}
\end{table}

\begin{table}[ht]
\centering
\def\arraystretch{1.15}
\caption{Summary statistics for the number of observations in the 24 cells formed by the categorical variables across the 5000 repeats in simulation 3.}\label{tab:sumstatcell}
\begin{tabular}{| l | c | c | c | c |}
  \hline
\textbf{Cell} & \textbf{Min} & \textbf{Mean} & \textbf{Median} & \textbf{Max} \\
    \hline
      1 & 20   & 41.05   & 41   & 68   \\ 
        2 &  49   &  75.86   &  76   & 106   \\ 
        3 &  8   & 24.07   & 24   & 45   \\ 
        4 & 36   & 61.92   & 62   & 89   \\ 
        5 & 24   & 44.81   & 45   & 68   \\ 
        6 & 21   & 40.94   & 41   & 64   \\ 
        7 & 17   & 38.01   & 38   & 61   \\ 
        8 &  1   &  6.993   &  7   & 17   \\ 
        9 & 38   & 63.96   & 64   & 94   \\ 
       10 & 30   & 64.05   & 64   & 91   \\ 
       11 & 12   & 31.03   & 31   & 59   \\ 
       12 & 23   & 48.01   & 48   & 76   \\ 
       13 & 30   & 53.01   & 53   & 81   \\ 
       14 &  1   &  6.978   &  7   & 18   \\ 
       15 &  6   & 21.03   & 21   & 39   \\ 
       16 & 38   & 64.94   & 65   & 97   \\ 
       17 &  42   &  70.15   &  70   & 105   \\ 
       18 &  1   & 12   & 12   & 29   \\ 
       19 &  8   & 24.03   & 24   & 43   \\ 
       20 &  7   & 28.06   & 28   & 48   \\ 
       21 & 27   & 48.17   & 48   & 71   \\ 
       22 &  1   & 10.99   & 11   & 24   \\ 
       23 & 30   & 57.95   & 58   & 86   \\ 
       24 & 36   & 61.98   & 62   & 93   \\ 
   \hline
\end{tabular}
\end{table}

\begin{table}[!htb]
\def\arraystretch{1.15}
\caption{Summary statistics for the bin number in the 2-dimensional histogram formed by $\z_1$ and $\z_2$ in the 24 cells formed by the categorical cells (used in the NP-DIPS method) across the 5000 repeats in simulation 3.}\label{tab:histstatsim3}
\begin{center}  \begin{tabular}{| l | c | c | c | c |}
  \hline
\textbf{Cell} & \textbf{Min} & \textbf{Mean} & \textbf{Median} & \textbf{Max} \\
    \hline
1 & 36 & 37.78 & 36 & 49 \\
  2 & 49 & 51.09 & 49 & 64 \\
  3 & 25 & 34.39 & 36 & 36 \\
  4 & 42 & 49.01 & 49 & 56 \\
  5 & 36 & 42.06 & 42 & 49 \\
  6 & 36 & 37.78 & 36 & 49 \\
  7 & 36 & 36.38 & 36 & 49 \\
  8 & 16 & 16.04 & 16 & 20 \\
  9 & 49 & 49.04 & 49 & 56 \\
  10 & 49 & 49.04 & 49 & 56 \\
  11 & 36 & 36.00 & 36 & 36 \\
  12 & 36 & 45.60 & 49 & 49 \\
  13 & 36 & 48.49 & 49 & 49 \\
  14 & 16 & 16.04 & 16 & 20 \\
  15 & 25 & 28.59 & 30 & 36 \\
  16 & 49 & 49.06 & 49 & 56 \\
  17 & 49 & 49.37 & 49 & 64 \\
  18 & 20 & 25.00 & 25 & 25 \\
  19 & 25 & 34.39 & 36 & 36 \\
  20 & 25 & 35.97 & 36 & 36 \\
  21 & 36 & 45.60 & 49 & 49 \\
  22 & 16 & 24.92 & 25 & 25 \\
  23 & 36 & 48.97 & 49 & 56 \\
  24 & 42 & 49.01 & 49 & 56 \\ \hline
\end{tabular}\end{center}
\end{table}


\clearpage
\begin{figure}[htp]
\vspace{-0.5in}
\makebox[\textwidth][c]{\includegraphics[width=6.25in]{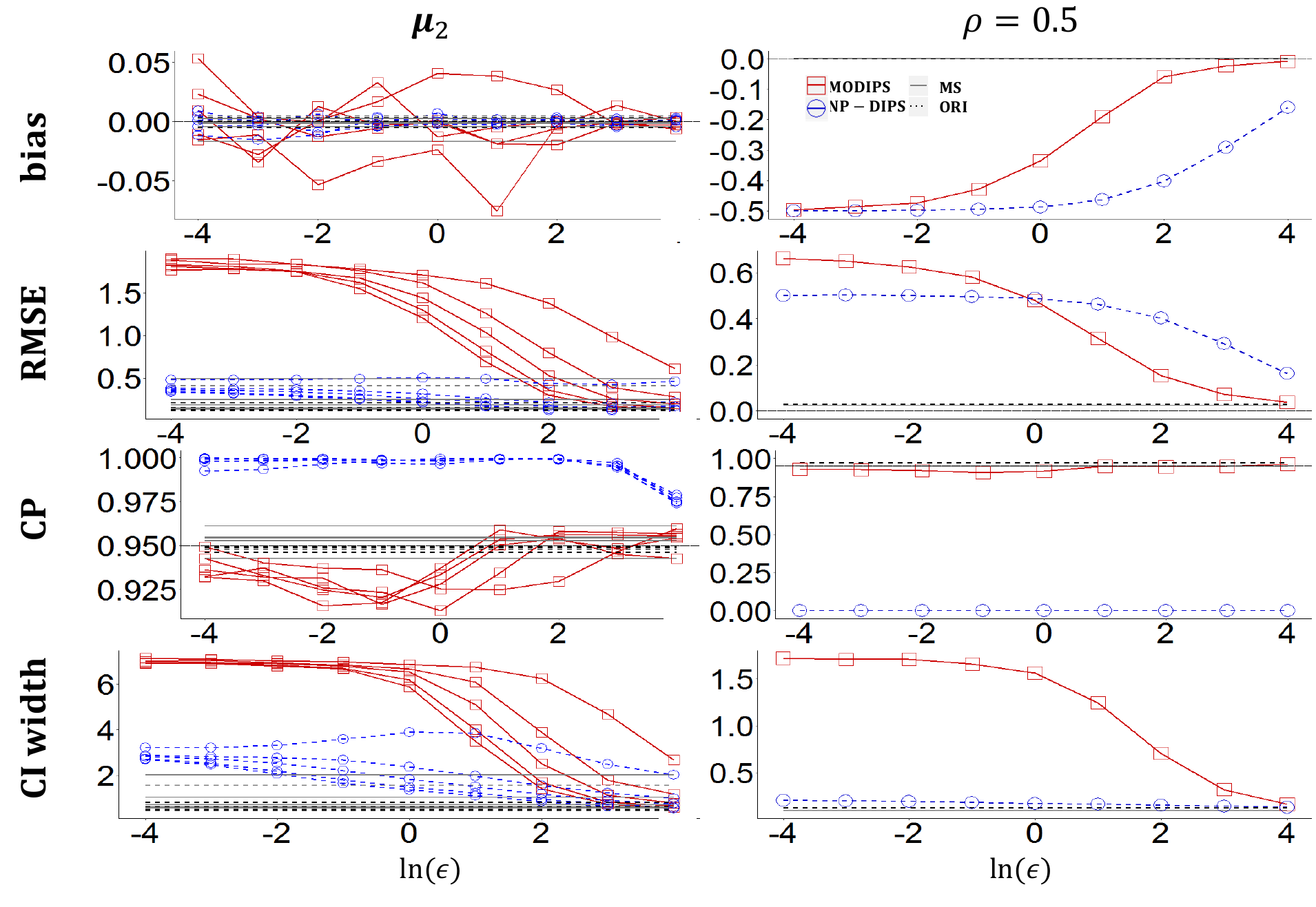}}
\caption{The bias, RMSE, 95\% CP, and 95\% CI width of $\bs{\mu}_{2}$ and $\rho$  in simulation study 3. MODIPS  represents the model-based differentially private data synthesis, NP-DIPS represents the Laplace sanitizer + perturbed histogram method, MS is the traditional MS method without DP, and Ori is the original results without  perturbation.}
\label{fig:sim3mu2rho}
\end{figure}

\begin{figure}[htp]
\vspace{-0.5in}
\makebox[\textwidth][c]{\includegraphics[width=6.25in]{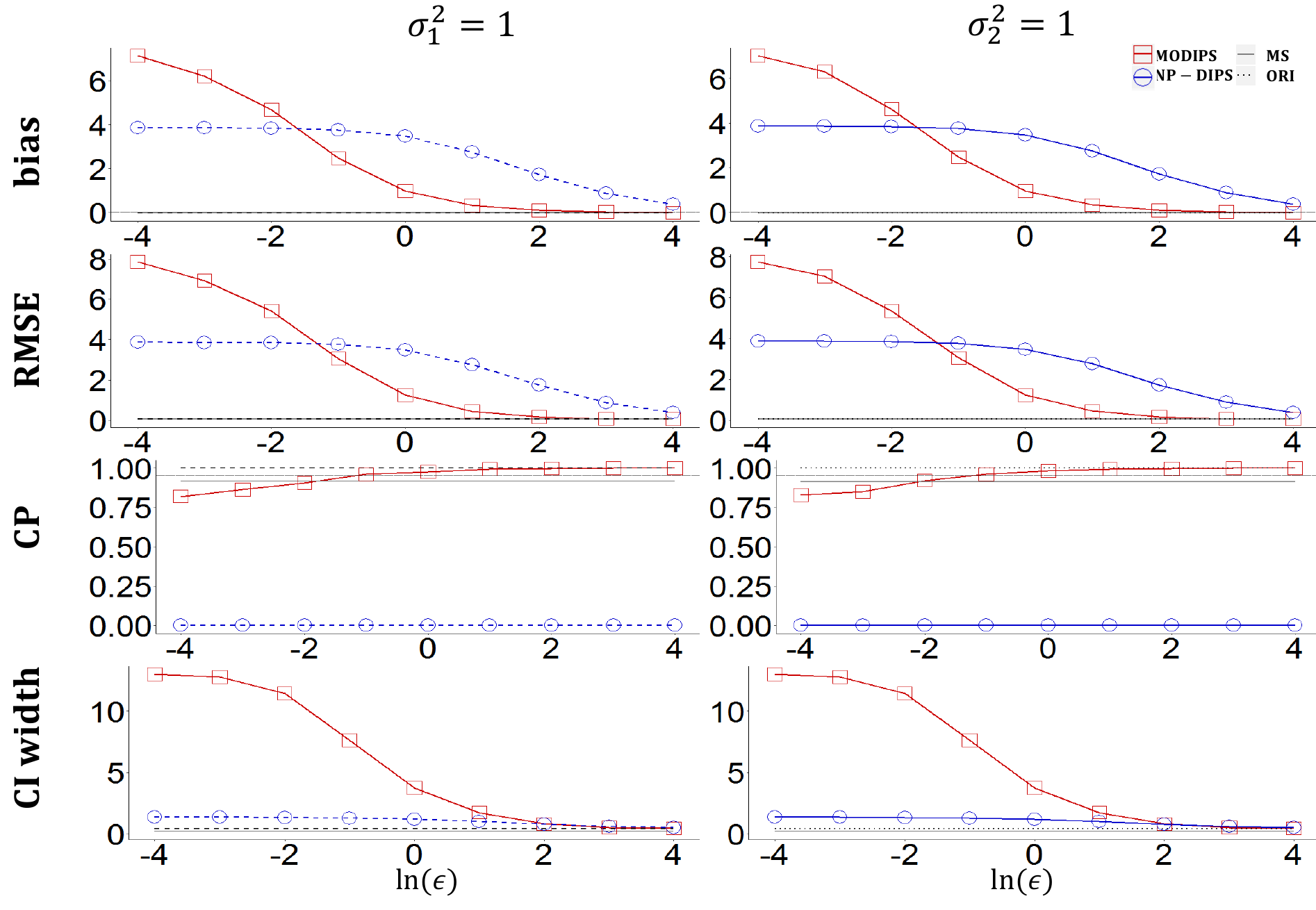}}
 \caption{The bias, RMSE, 95\% CP, and 95\% CI width of $\sigma_1^2$ and $\sigma_2^2$ in simulation study 3.  MODIPS  represents the model-based differentially private data synthesis, NP-DIPS represents the Laplace sanitizer + perturbed histogram method, MS is the traditional MS method without DP, and Ori is the original results without perturbation.}
\label{fig:sim3sigma}
\end{figure}

\clearpage
\begin{table}[ht]
\centering
\def\arraystretch{1.15}
\caption{Summary statistics for the number of observations in the 12 cells formed by the categorical cells across the 1000 repeats in simulation 4.}\label{tab:oristatsim4}
\begin{tabular}{| l | c | c | c | c |}
  \hline
\textbf{Cell} & \textbf{Min} & \textbf{Mean} & \textbf{Median} & \textbf{Max} \\
    \hline
      1 & 107   & 157.7   & 158   & 200   \\ 
        2 & 151   & 202.3   & 202   & 245   \\ 
        3 & 154   & 197   & 197   & 249   \\ 
        4 &  50   &  77.65   &  78   & 113   \\ 
        5 & 33   & 54.59   & 54   & 84   \\ 
        6 &  3   & 14.72   & 15   & 29   \\ 
        7 & 36   & 59.18   & 59   & 89   \\ 
        8 &  65   &  94.29   &  94   & 126   \\ 
        9 &  67   &  98.46   &  98   & 136   \\ 
       10 &  8   & 212   & 21   & 39   \\ 
       11 &  4   & 17.83   & 18   & 35   \\ 
       12 &  0   &  5.30   &  5   & 16   \\ 
   \hline
\end{tabular}
\end{table}

\begin{table}[!htb]
\def\arraystretch{1.15}
\caption{Summary statistics for the number of 2-dimensional histogram bins formed by the two continuous variables $(\z_1, \z_2)$ in each of the 12 cells by the categorical variables $\w$ (needed in the NP-DIPS approach) in simulation study 4.}\label{tab:histstatsim4}
\begin{center}  \begin{tabular}{| l | c | c | c | c |}
  \hline
\textbf{Cell} & \textbf{Min} & \textbf{Mean} & \textbf{Median} & \textbf{Max} \\
    \hline
  \hline
1 & 49 & 65.23 & 64 & 81 \\
  2 & 56 & 64.53 & 64 & 81 \\
  3 & 64 & 78.78 & 81 & 100 \\
  4 & 36 & 50.90 & 49 & 64 \\
  5 & 9 & 19.81 & 16 & 30 \\
  6 & 64 & 79.23 & 81 & 100 \\
  7 & 36 & 47.67 & 49 & 64 \\
  8 & 64 & 77.26 & 81 & 90 \\
  9 & 16 & 25.74 & 25 & 36 \\
  10 & 0 & 13.89 & 16 & 25 \\
  11 & 36 & 47.71 & 49 & 64 \\
  12 & 16 & 26.35 & 25 & 36 \\
   \hline
\end{tabular}\end{center}
\end{table}

\clearpage
Table \ref{tab:pval} presents the Bayesian $p$-values for the sample marginal probabilities $\w$, the sample means and  variance-covariance of $\z$  from the two models fitted to the original data over the 100 repeats. A Bayesian $p$-value that is close to 0 or 1 implies the posterior predicted value of that statistic based on the Bayesian model deviates from the observed value significantly, or the model is a not good fit for the data.  The $p$-values from the SLOMAG for all the examined statistics ranged from $\sim0.45$ to $\sim0.55$. The $p$-values from the GLOM were acceptable ($\sim 0.5$) in most of the examined statistics but not in the two sample variances $s_1^2$ and $s_2^2$ ($p$-values $>0.99$) or the sample correlation $r$ ($p$-values ranged from 0 to 0.59 with a median of 0.051). These results suggested that the SLOGMAG model was a better fit than the GLOM. 
\begin{table}[!htp]
\centering
\begin{tabular}{ll lll lll}
  \hline
model  &  statistics  & min & Q1 & mean & med & Q3 & max \\
\hline
correct   &  $\bar{z}_1$ & 0.455 & 0.491 & 0.501 & 0.501 & 0.510 & 0.544 \\
(SLOMAG)  &  $\bar{z}_2$ & 0.470 & 0.488 & 0.500 & 0.501 & 0.510 & 0.546 \\
                &  $s^2_1$ & 0.474 & 0.497 & 0.507 & 0.505 & 0.515 & 0.552 \\ 
                &  $s^2_2$ & 0.468 & 0.495 & 0.506 & 0.503 & 0.520 & 0.543 \\
                & $r$ & 0.465 & 0.492 & 0.501 & 0.500 & 0.511 & 0.540 \\
                & $\Pr(W_1=1)$ & 0.436 & 0.475 & 0.485 & 0.485 & 0.496 & 0.528 \\
                & $\Pr(W_2=1)$ & 0.451 & 0.474 & 0.483 & 0.482 & 0.493 & 0.515 \\
                & $\Pr(W_3=1)$ & 0.448 & 0.474 & 0.488 & 0.488 & 0.497 & 0.533 \\ 
                & $\Pr(W_3=2)$ & 0.452 & 0.477 & 0.486 & 0.485 & 0.495 & 0.519 \\ 
\hline
wrong &  $\bar{z}_1$ & 0.463 & 0.491 & 0.502 & 0.502 & 0.512 & 0.544 \\ 
(GLOM)  &  $\bar{z}_2$ & 0.458 & 0.490 & 0.500 & 0.500 & 0.511 & 0.543 \\ 
                &  $s^2_1$ & 0.991 & 0.998 & 0.998 & 0.999 & 1.000 & 1.000 \\ 
                &  $s^2_2$ & 0.993 & 0.998 & 0.999 & 0.999 & 1.000 & 1.000 \\
                & $r$ & 0.000 & 0.020 & 0.107 & 0.051 & 0.148 & 0.587 \\
                & $\Pr(W_1=1)$ & 0.434 & 0.479 & 0.490 & 0.490 & 0.499 & 0.527 \\ 
                & $\Pr(W_2=1)$ & 0.449 & 0.479 & 0.490 & 0.492 & 0.501 & 0.522 \\ 
                & $\Pr(W_3=1)$ & 0.446 & 0.476 & 0.488 & 0.489 & 0.498 & 0.518 \\
                & $\Pr(W_3=2)$ & 0.434 & 0.478 & 0.488 & 0.488 & 0.497 & 0.531 \\ 
  \hline
\end{tabular}
\caption{Bayesian $p$-values on selected statistics with the right and wrong model specification.}\label{tab:pval}
\end{table}

\clearpage
\begin{figure}[htp]
\vspace{-0.2in}
\makebox[\textwidth][c]{\includegraphics[width=6.25in]{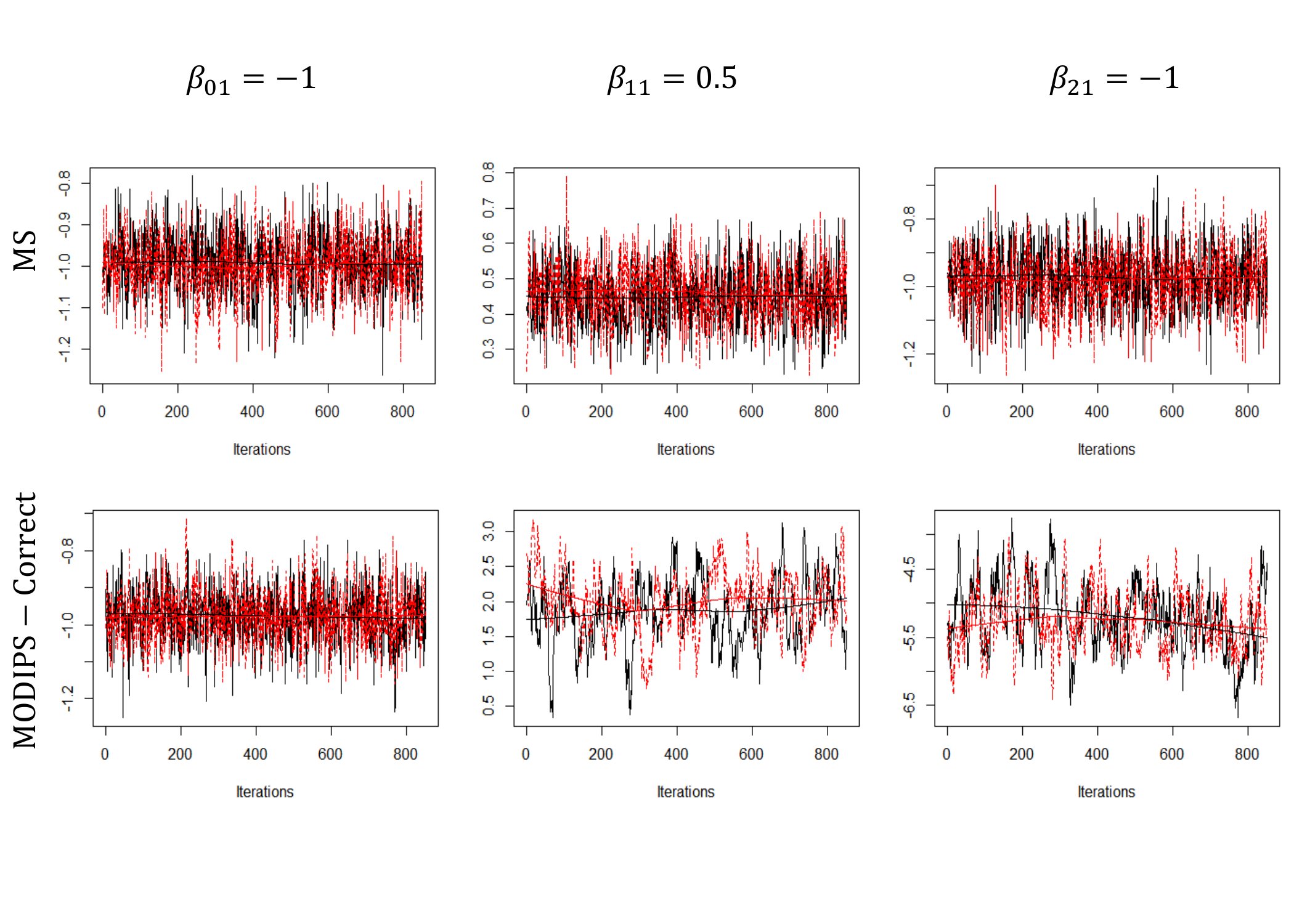}}
\caption{Sample MCMC trace plots for $\boldsymbol{\beta}_1$. MS is the traditional multiple synthesis method without DP, MODIPS-Correct is the DIPS method based on the SLOMAG model with the $l_2$ regularizer on $\boldsymbol{\beta}_1$ at $\lambda=0.50$.}
\label{Sfig:sim4trace}
\end{figure}

\begin{figure}[htp]
\vspace{-0.5in}
\makebox[\textwidth][c]{\includegraphics[width=6.5in]{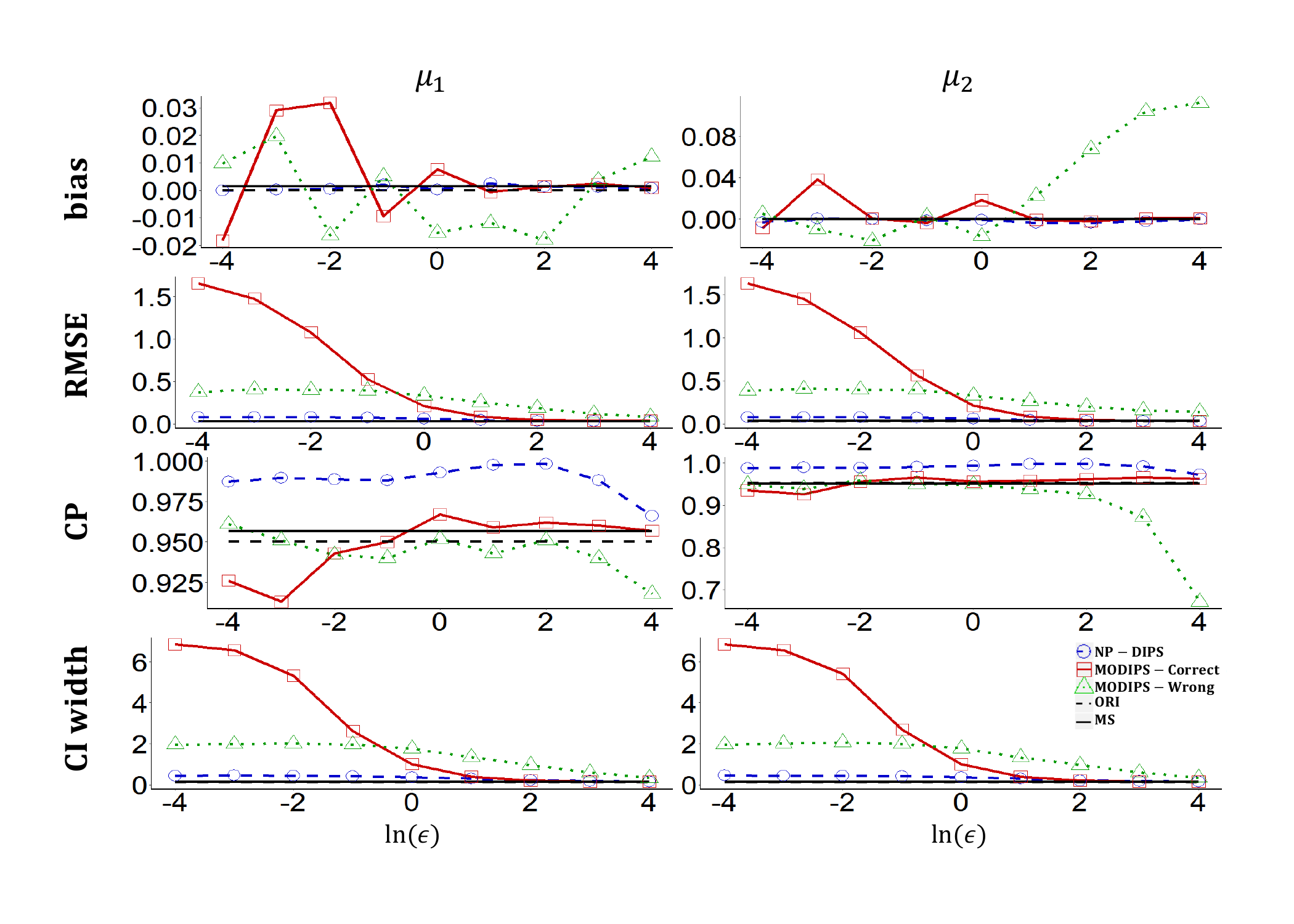}}
\caption{Bias, RMSE, CP, and 95\% CI width of $\mu_1$ and $\mu_2$ in simulation study 4. MODIPS represents the model-based differentially private data synthesis, NP-DIPS represents the Laplace sanitizer + perturbed histogram method, MS is the traditional MS method without DP, and Ori is the original results without  perturbation.}
\label{Sfig:sim4reg1}
\end{figure}

\begin{figure}[htp]
\vspace{-0.5in}
\makebox[\textwidth][c]{\includegraphics[width=6.5in]{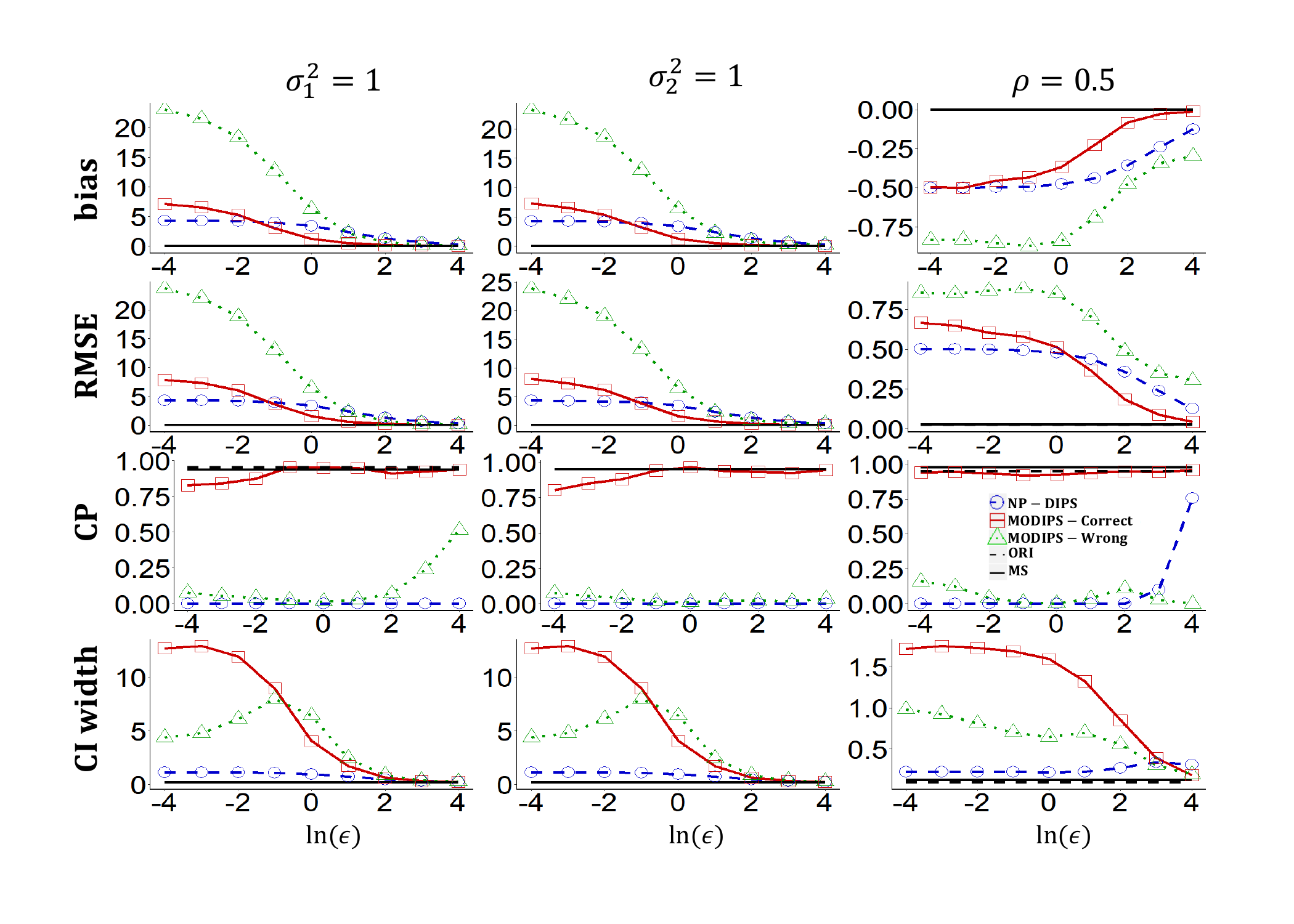}}
\caption{Bias, RMSE, 95\% CP, and 95\% CI width of $\sigma^2_1$, $\sigma^2_2$, and $\rho$ in simulation study 4.  MODIPS represents the model-based differentially private data synthesis, PERT represents the perturbed histogram method, MS is the traditional MS method without DP, and Ori is the original results without  perturbation.}
\label{Sfig:sim4reg2}
\end{figure}

\begin{figure}[htp]
\vspace{-0.5in}
\makebox[\textwidth][c]{\includegraphics[width=6.5in]{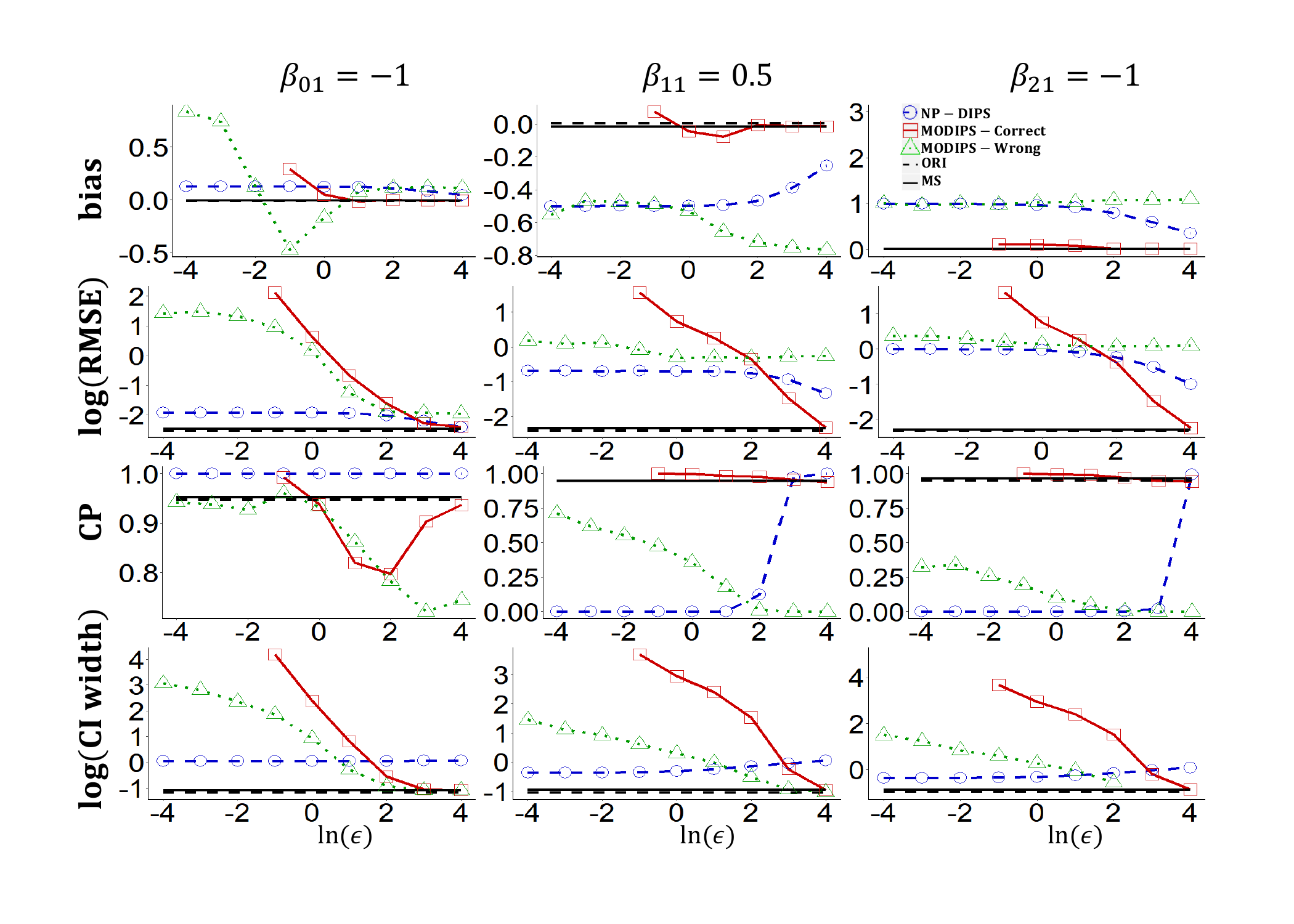}}
\caption{Bias, log of RMSE, 95\% CP, and log of the 95\% CI width for $\bs{\beta}_1$ in simulation study 4. MODIPS  represents the model-based differentially private data synthesis, PERT represents the perturbed histogram method, MS is the traditional MS method without DP, and Ori is the original results without perturbation.}
\label{Sfig:sim4reg3}
\end{figure}

\end{document}